\documentclass[journal]{IEEEtran}

\usepackage{amsmath,amssymb,amsthm,multirow,dsfont,cuted}
\usepackage{cite,caption}
\usepackage{enumitem}
\usepackage{subfigure}
\usepackage{epsfig,graphics}
\usepackage{tikz,pgfplots}
\usepackage{epsfig,graphics}
\usepackage{algorithm}
\usepackage{algpseudocode}
\usepackage{diagbox}
\usepackage{multirow}

\newtheorem{assumption}{Assumption}

\def\cblue{\textcolor{black}}

\newcommand{\cI}{\mathcal{I}}

\newcommand{\cS}{\mathcal{S}}
\newcommand{\N}{\mathcal{N}}

\newcommand{\C}{\mathcal{C}}

\newcommand{\cW}{\mathcal{W}}

\newcommand{\bA}{\boldsymbol{A}}
\newcommand{\bB}{\boldsymbol{B}}
\newcommand{\bC}{\boldsymbol{C}}
\newcommand{\bD}{\boldsymbol{D}}

\newcommand{\bI}{\boldsymbol{I}}

\newcommand{\bR}{\boldsymbol{R}}
\newcommand{\bT}{\boldsymbol{T}}
\newcommand{\bU}{\boldsymbol{U}}

\newcommand{\bW}{\boldsymbol{W}}
\newcommand{\bX}{\boldsymbol{X}}

\newcommand{\bb}{\boldsymbol{b}}
\newcommand{\bc}{\boldsymbol{c}}
\newcommand{\bd}{\boldsymbol{d}}

\newcommand{\boldf}{\boldsymbol{f}}
\newcommand{\bg}{\boldsymbol{g}}

\newcommand{\bp}{\boldsymbol{p}}
\newcommand{\br}{\boldsymbol{r}}
\newcommand{\bu}{\boldsymbol{u}}

\newcommand{\bw}{\boldsymbol{w}}
\newcommand{\bx}{\boldsymbol{x}}
\newcommand{\by}{\boldsymbol{y}}

\newcommand{\bpsi}{\boldsymbol{\psi}}
\newcommand{\bphi}{\boldsymbol{\phi}}

\newcommand{\bGamma}{\boldsymbol{\Gamma}}

\newcommand{\bSig}{\boldsymbol{\Sigma}}
\newcommand{\bsig}{\boldsymbol{\sigma}}

\def\One{\mathds{1}}
\def\Zero{\boldsymbol{0}}

\newcommand{\cA}{\boldsymbol{\cal{A}}}
\newcommand{\cB}{\boldsymbol{\cal{B}}}

\newcommand{\cD}{\boldsymbol{\cal{D}}}
\newcommand{\cF}{\boldsymbol{\cal{F}}}
\newcommand{\cG}{\boldsymbol{\cal{G}}}
\newcommand{\cH}{\boldsymbol{\cal{H}}}
\newcommand{\cK}{\boldsymbol{\cal{K}}}

\newcommand{\cP}{\boldsymbol{\cal{P}}}
\newcommand{\cR}{\boldsymbol{\cal{R}}}
\newcommand{\cT}{\boldsymbol{\cal{T}}}

\newcommand{\cY}{\boldsymbol{\cal{Y}}}
\newcommand{\cZ}{\boldsymbol{\cal{Z}}}

\newcommand{\bwt}{\widetilde\bw}
\newcommand{\bpsit}{\widetilde\bpsi}
\newcommand{\bphit}{\widetilde\bphi}

\newcommand{\tr}{\text{tr}}
\newcommand{\vc}{\text{vec}}
\newcommand{\bvc}{\text{bvec}}
\newcommand{\col}{\text{col}}

\newcommand{\diag}{\text{diag}}

\newcommand{\proj}{\text{P}}

\newcommand{\expec}{\mathbb{E}}

\DeclareMathOperator*{\minimize}{minimize}
\DeclareMathOperator*{\st}{subject\,to}

\def\R{\ensuremath{{\mathrm{I\!R}}}}

\begin{document}
\title{Diffusion LMS for {Multitask Problems\\ with Local Linear Equality Constraints}}

\author{Roula Nassif, C{\'e}dric Richard, \IEEEmembership{Senior Member, IEEE} \\
Andr{\'e} Ferrari, \IEEEmembership{Member, IEEE}, Ali H. Sayed, \IEEEmembership{Fellow Member, IEEE}
\thanks{The work of C. Richard and A. Ferrari was partly supported by ANR and DGA grant ANR-13-ASTR-0030 (ODISSEE project). The work of A. H. Sayed was supported in part by NSF grants CIF-1524250 and ECCS-1407712.

R. Nassif, C. Richard, and A. Ferrari are with the Universit{\'e} de Nice Sophia-Antipolis, France (email: roula.nassif@oca.eu; cedric.richard@unice.fr; andre.ferrari@unice.fr).

A. H. Sayed is with the department of electrical engineering, University of California, Los Angeles, USA (email: sayed@ee.ucla.edu).
}}
\maketitle

\begin{abstract}
We consider distributed multitask learning problems over a network of agents where each agent is interested in estimating its own parameter vector, {also called task, and where the tasks at neighboring agents are related according to a set of linear equality constraints}. Each agent possesses its own convex cost function of its parameter vector and a set of linear equality constraints {involving its own parameter vector and the parameter vectors of its neighboring agents.} We propose an adaptive stochastic algorithm based on the projection gradient method and diffusion strategies {in order to allow} the network to optimize the individual costs subject to all constraints. Although the derivation is carried out for linear equality constraints, the technique can be applied to other {forms} of convex constraints. We conduct a detailed mean-square-error analysis of the proposed algorithm and derive closed-form expressions to predict its learning behavior. We provide simulations to illustrate the theoretical findings. {Finally, the algorithm is employed for solving two problems in a distributed manner: a minimum-cost flow problem over a network and a space-time varying field reconstruction problem.}
\end{abstract}





\section{Introduction}

Single-task distributed optimization over networks allows to minimize the aggregate sum of convex cost functions, each available at an agent, subject to convex constraints that are also distributed across the agents. Each learner seeks to estimate the minimizer through local computations and communications among neighboring agents without the need to know any of the constraints or costs besides their own. Several useful strategies have been proposed to solve constrained and unconstrained versions of this problem in a fully decentralized \cblue{manner~\cite{bertsekas1997new,nedic2001incremental,lopes2008diffusion,ram2010distributed,srivastava2011distributed,mota2013D-ADMM,lee2013distributedDRP,chen2013distributed,sayed2014adaptation,towfic2014adaptive,sayed2014diffusion,towfic2015stability,zhang2016decentralized}}. Diffusion strategies~\cite{lopes2008diffusion,chen2013distributed,sayed2014adaptation,towfic2014adaptive,sayed2014diffusion,towfic2015stability} are attractive since they are scalable, robust, {and} enable continuous learning and adaptation in response to drifts in the location of the minimizer due to changes in the costs or in the constraints.

Multitask distributed learning over networks {is particularly well-suited for applications} where several parameter vectors need to be estimated simultaneously from successive noisy measurements using in-network \cblue{processing~\cite{bertrand2010distributed,bertrand2012distributed,eksin2012distributed,kekatos2013distributed,bogdanovic2013distributed,bogdanovic2014distributed,platachaves2015distributed,mota2015distributed,chen2013distributed,chen2014multitask,nassif2016multitask,nassif2015proximal,nassif2015multitask,chen2014diffusion,zhao2012clustering,chen2015diffusion,zhao2015distributed,chen2016group,platachaves2016unsupervised,abdolee2014estimation,chen2015dictionary}.} The network is decomposed into clusters of agents and each cluster {estimates} its own parameter vector~\cite{chen2014multitask}. Distributed strategies for {solving} multitask problems have been addressed in two main ways. In a first scenario, agents do not know the cluster they belong to and no prior information on possible relationships between tasks is assumed. In this case, all agents cooperate with each other as dictated by the network topology. It is shown in~\cite{chen2013distributed} that, {in this case,} the diffusion iterates {will converge} to a Pareto optimal solution corresponding to a multi-objective optimization problem. To avoid cooperation with neighbors seeking different objectives, automatic clustering techniques based on diffusion strategies have {also been proposed}~\cite{zhao2012clustering,chen2015diffusion,zhao2015distributed}.
In a second scenario, it is assumed that agents know which cluster they belong to. Multitask diffusion strategies {are then derived} by exploiting prior information about relationships among the tasks. For example, one way to model and exploit relationships among tasks is to formulate convex optimization problems with appropriate co-regularizers between \cblue{the agents~\cite{eksin2012distributed,chen2014multitask,nassif2016multitask,nassif2015proximal,nassif2015multitask}. While~\cite{eksin2012distributed} deals with deterministic optimization problems, \cite{chen2014multitask,nassif2016multitask,nassif2015proximal,nassif2015multitask} are concerned with adaptive estimation problems.} In~\cite{bertrand2012distributed}, distributed algorithms are derived to estimate node-specific signals that lie in a common latent signal subspace in the presence of node-specific linear equality constraints. \cblue{Several useful works consider stochastic~\cite{kekatos2013distributed,bogdanovic2013distributed,bogdanovic2014distributed,platachaves2015distributed} and deterministic~\cite{mota2015distributed} multitask estimation problems with overlapping parameter vectors. They assume that each agent is interested in estimating its own parameter vector, and that the local parameter vectors at neighboring agents have some entries that are equal. Unsupervised strategies are also considered in~\cite{chen2016group,platachaves2016unsupervised} to address multitask overlapping problems.} 
In~\cite{chen2014diffusion}, a diffusion algorithm is derived to solve multitask estimation problem where the parameter space is decomposed into two orthogonal subspaces, with one of the subspaces being common to all agents. 

\cblue{In some applications, it happens} that the optimum parameter vectors to be estimated {at neighboring agents} are related according to a set of constraints. {This observation motivates us to consider in this work} multitask \cblue{estimation} problems subject to linear equality constraints of the form:
\begin{subequations}
	\label{eq: problem in the introduction}
	\begin{align}
	&\minimize\limits_{\bw_1,\ldots,\bw_N}~J^{\text{glob}}(\bw_1,\ldots,\bw_N)\triangleq\sum_{k=1}^N J_k(\bw_k),\\
	&\st~\sum_{\ell\in\cI_p}\bD_{p\ell}\,\bw_\ell+\bb_p=\Zero, ~p=1,\ldots,P.\label{eq: multitask problem (b)}
	\end{align}
\end{subequations}
Each agent $k$ in the network seeks to estimate its own $M_k \times 1$ parameter vector $\bw_k$, and {has knowledge of} its cost function $J_k(\cdot)$ and the set of linear equality {constraints that agent $k$ is involved in}. Each constraint is indexed by~$p$, and defined by the $L_p\times M_\ell$ matrices $\bD_{p\ell}$, the $L_p\times1$ vector $\bb_p$, and the set $\cI_p$ of agent indices involved in this constraint. \cblue{Note that, by properly selecting the matrices $\bD_{p\ell}$ and setting the vectors~$\bb_p$ to $\Zero$ in~\eqref{eq: problem in the introduction}, the \textit{single-task} estimation problem~\cite{nedic2001incremental,lopes2008diffusion,ram2010distributed} and the \textit{multitask overlapping} estimation problem~\cite{kekatos2013distributed,bogdanovic2013distributed,bogdanovic2014distributed,platachaves2015distributed,mota2015distributed} can be recast as problem~\eqref{eq: problem in the introduction}. }

\textbf{Assumption.} In the current work, it is assumed that each agent $k$ in $\cI_p$ can collect estimates from all agents in $\cI_p$ in order to satisfy the $p$-th constraint, i.e., $\cI_p\subseteq\N_k$ for all $k\in\cI_p$ where $\N_k$ denotes the neighborhood of agent~$k$. This assumption is reasonable in many applications, for instance, in remote monitoring of physical phenomena involving discretization of spatial differential equations~\cite{bertsekas1989parallel}, and in network monitoring involving conservation laws at each junction~\cite{ahuja1993network}. 

\begin{figure}[t]
	{\centering
	\includegraphics[scale=0.3]{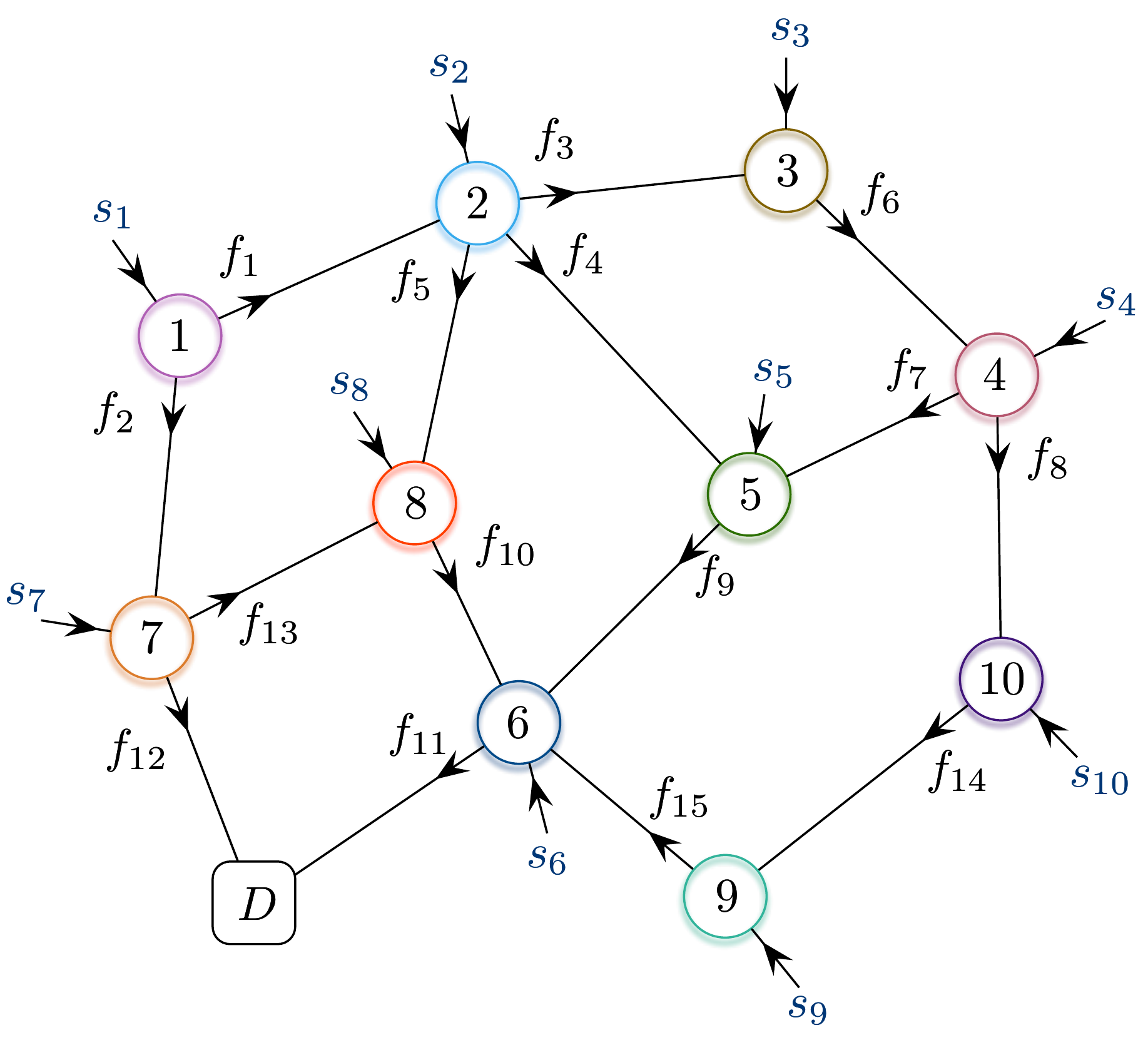}
	\caption{Flow network topology with $10$ nodes, $1$ destination sink $D$, and $15$ communication links.}
	\label{fig: network flow problem - intro}}
\end{figure}

{For illustration purposes, consider a minimum-cost flow problem over the network shown in Fig.~\ref{fig: network flow problem - intro}. This network consists of $10$ nodes, $1$ destination sink $D$, and $15$ communication links. With each link $j$, we associate a directed arc and we let $f_j$ denote the flow or traffic on this link, with $f_j>0$ meaning that the flow is in the direction of the arc, and $f_j<0$ otherwise. At each node $k$, an external source flow $s_k$ enters and flows through the network to the destination sink. The flow must satisfy a conservation equation, which states that at each node $k$, the sum of flows entering the node, plus the external source $s_k$, is equal to the sum of flows leaving node~$k$. Given the external sources $s_k$ and the network topology, a number of studies have been devoted to finding the optimal flows $f_j^\star$ that minimize a total flow transmission cost and satisfy the conservation equations~\cite{ahuja1993network,boyd2004convex,ventura1991computational}. Problems of this type arise in applications such as electrical networks, telecommunication networks, pipeline networks~\cite{ahuja1993network}. In some of these applications, it happens that node~$k$ has only access to noisy measurements $s_k(i)$ of the external source at each time instant~$i$.  \cblue{For example, in electrical networks, the agents may not be able to collect the exact values of the current sources (or the current demands).} Denoting by $\bw_k$ the $M_k\times 1$ vector containing the flows $f_j$ entering and leaving node~$k$, we are interested in distributed online learning settings where each node~$k$ seeks to estimate $\bw_k$ from noisy measurements $s_k(i)$ by relying only on local computations and communications with its neighbors. This problem can be recast in the form~(\ref{eq: problem in the introduction}a)--(\ref{eq: problem in the introduction}b) and addressed with the multitask strategy proposed in this paper. This example will be considered further in the numerical experiments section.} 

\cblue{{We shall} propose a primal adaptive technique (based on propagating and estimating the primal variable) for solving {problem}~\eqref{eq: problem in the introduction} in a distributed manner.}  {The technique relies on combining diffusion adaptation with a stochastic gradient projection step, and} on the use of constant {step-sizes} to enable continuous adaptation and learning from streaming data. Since we are learning from streaming data, the dual function cannot be computed exactly and the use of primal-dual methods may result in stability problems as {already shown} in~\cite{towfic2015stability}. For this reason, we focus on primal techniques. {Our} current work is able to cope with the following two scenarios: 1) multitask problems with prior information on linear relationships between tasks, {and} 2) constrained multitask problems with distributed information access. We analyze the behavior of our algorithm in the mean and mean-square-error sense (w.r.t. the minimizers of the local costs and w.r.t. the solution of the constrained multitask problem) and we derive expressions to predict its transient and steady-state behavior. \cblue{Some simulation results show} that, for small constant step-sizes, the expected distance between the estimates at each agent and the optimal value can be made arbitrarily small.  

\textbf{Notation.} Normal font letters denote scalars, boldface lowercase letters denote column vectors, and boldface uppercase letters denote matrices. We use the symbol $(\cdot)^\top$ to denote matrix transpose, the symbol $(\cdot)^{-1}$ to denote matrix inverse, the symbol $(\cdot)^{\dagger}$ to denote the pseudo-inverse of a full {row-rank} matrix, and the symbol $\tr(\cdot)$ to denote the trace operator. 
The symbol $\diag\{\cdot\}$ forms a matrix from block arguments by placing each block immediately below and to the right of its predecessor. The operator $\col\{\cdot\}$ stacks the column vector entries on top of each other. The symbols~$\otimes$ and $\otimes_b$ denote the Kronecker product and the block Kronecker product, respectively. The symbol $\vc(\cdot)$ refers to the standard vectorization operator that stacks the columns of a matrix on top of each other and the symbol $\bvc(\cdot)$ refers to the block vectorization operation that vectorizes each block of a matrix and stacks the vectors on top of each other. The identity matrix of size $N\times N$ is denoted by $\bI_N$. The $N\times 1$ vector of ones is denoted by $\One_{N\times 1}$. {For a $P\times N$ block matrix $\cA$, the $1\times N$ $k$-th block row is denoted by $[\cA]_{k,\bullet}$ and the $P\times 1$ $k$-th block column is denoted by $[\cA]_{\bullet,k}$.} The notation $\proj_\Omega(\bw)$ denotes the projection of $\bw$ onto the manifold~$\Omega$.
\section{Problem formulation and centralized solution}

\subsection{Problem formulation and assumptions}

{Consider a network of $N$ agents, labeled $k=1,\ldots, N$.} At each time instant $i$, each agent $k$ has access to a zero-mean real-valued observation {$d_k(i)$,}  and a zero-mean real-valued $M_k\times 1$ regression vector {$\bx_k(i)$, with positive covariance matrix $\bR_{x,k}=\expec\{\bx_k(i)\bx^\top_k(i)\}>0$.} We assume the data to be related via the linear data model:
\begin{equation}
	\label{eq: linear data model}
	d_k(i)=\bx_k^\top(i)\bw^o_k+z_k(i),
\end{equation}
where $\bw^o_k$ is an $M_k\times 1$ unknown parameter vector, and $z_k(i)$ is a zero-mean measurement noise of variance $\sigma^2_{z,k}$, independent of $\bx_{\ell}(j)$ for all $\ell$ and $j$, and independent of $z_{\ell}(j)$ for $\ell\neq k$ or $i\neq j$. We {let} $\br_{dx,k}\triangleq\expec\{d_k(i)\bx_k(i)\}$ {and} $\sigma^2_{d,k}\triangleq\expec|d_k(i)|^2$.

Let $\bw_k$ denote {some generic} $M_k\times 1$ vector {that is associated} with agent $k$. The objective {of agent $k$} is to find an estimate for $\bw^o_k$, and we associate with this agent the mean-square-error criterion:
\begin{equation}
	\label{eq: MSE at each agent}
	J_k(\bw_k)=\expec\,|d_k(i)-\bx^\top_k(i)\bw_k|^2,
\end{equation}
which is strongly convex, second-order differentiable, and minimized at $\bw^o_k$. In addition, $P$ linear equality constraints of the form~\eqref{eq: multitask problem (b)} are imposed on the {parameter vectors $\{\bw_k\}$} at each time instant $i$. Let us collect the parameter vectors $\{\bw_k\}$ and  $\{\bw^o_k\}$ from across the network into $N\times 1$ block column {vectors} $\bw$ and $\bw^o$, respectively:
\begin{equation}
	\bw\triangleq\col\{\bw_1,\ldots,\bw_N\},\qquad
	\bw^o\triangleq\col\{\bw^o_1,\ldots,\bw^o_N\}\label{eq: network block column vector w0},
	\end{equation}
and {let us write the} $P$ linear equality constraints in~\eqref{eq: multitask problem (b)} more compactly as:
\begin{equation}
	\label{eq: constraint}
	\cD\bw+\bb=\Zero,
\end{equation}
where $\cD$ {is a} $P\times N$ block matrix, with each block $\bD_{p\ell}$ having dimensions $L_p\times M_{\ell}$, and $\bb$ {is a} {$P\times 1$} block column vector where each {block $\bb_p$ has dimensions $L_p\times 1$.} Combining~\eqref{eq: constraint} and~\eqref{eq: MSE at each agent}, the network optimization problem becomes:
\begin{equation}
	\label{eq: network optimization problem}
	\begin{split}
		\minimize_{\bw}\quad&\sum_{k=1}^N\expec\,|d_k(i)-\bx^\top_k(i)\bw_k|^2,\\
		\st\quad&\cD\bw+\bb=\Zero,
	\end{split}
\end{equation}
where each agent $k$ is in charge of estimating the $k$-th sub-vector $\bw_k$ of $\bw$. Since the mean-square-error criterion in~\eqref{eq: network optimization problem} is separable, we shall assume without loss of generality that each parameter vector $\bw_k$ is involved in at least one constraint so {that cooperation is justified}. We shall also assume that $\cD$ is full row-rank to ensure that equation $\cD\bw+\bb=\Zero$ has at least one solution. {We also introduce an assumption on the availability of the constraints. Let $\cI_p $ be the set of agent indices involved in the $p$-th constraint. We shall assume that every agent $k$ in $\cI_p$ is aware of the $p$-th constraint, and that the network topology permits this agent to collect estimates from all agents in $\cI_p$, that is, $\cI_p\subseteq\N_k$, so it can apply this constraint to its own estimate. This assumption is reasonable in many applications, for instance, in remote monitoring of physical phenomena~\cite{bertsekas1989parallel}, and in network distribution system monitoring (as described in the introduction)~\cite{ahuja1993network}. These examples will be considered in numerical experiments section.}

Before proceeding, note that problem~\eqref{eq: network optimization problem} can be recast as a quadratic program (QP)~\cite{boyd2004convex}, {and any} algorithm that solves QPs can solve it. We are {interested instead} in distributed adaptive {solutions that can operate in real-time on streaming data}. As we will see later, the {traditional constrained LMS} algorithm~\cite{frost1972algorithm} can solve~\eqref{eq: network optimization problem} in a centralized manner. In {this} centralized solution, each agent at each iteration sends its data to a fusion center{, which in turn} processes {the} data and sends the estimates {back to} the agents. The entire matrix $\cD$ and the entire vector $\bb$ then need to be available at the fusion center. \cblue{While centralized solutions can be powerful, decentralized solutions are more attractive since they are more robust and respect the privacy policy of each agent~\cite{estrin2001instrumenting,sayed2014adaptation,harrane2016toward}.} 
\subsection{Centralized solution}
\label{subsec: Centralized solution}
{Let us first describe the centralized solution. We} assume that the agents transmit the collected data $\{d_k(i), \bx_k(i)\}$ to a fusion center for processing. Problem~\eqref{eq: network optimization problem} can be written equivalently as:
\vspace{0mm}
\begin{equation}
	\label{eq: network optimization problem centralized}
	\begin{split}
	\minimize_{\bw}\quad&\bw^\top\cR_x\bw-2\br_{dx}^\top\bw+\br_{d}^\top\One_{N\times 1},\\
	\st\quad&\cD\bw+\bb=\Zero,
	\end{split}
\end{equation}
where the $N\times N$ block diagonal matrix $\cR_x$, the $N\times 1$ block column vector $\br_{dx}$, and the $N\times 1$ column vector $\br_d$ are given by:
\begin{align}
	\cR_x&\triangleq\diag\{\bR_{x,1},\ldots,\bR_{x,N}\},\\
	\br_{dx}&\triangleq\col\{\br_{dx,1},\ldots,\br_{dx,N}\},\\
	\br_{d}&\triangleq\col\{\sigma^2_{d,1},\ldots,\sigma^2_{d,N}\}.
\end{align}  
Since $\cR_x$ is positive definite, problem~\eqref{eq: network optimization problem centralized} is a positive definite quadratic program with equality constraints. It has a unique global minimum given by: 
\begin{equation}
	\label{eq: closed form solution}
	\bw^\star=\bw^o-\cR_x^{-1}\cD^\top(\cD\cR_x^{-1}\cD^\top)^{-1}(\cD\bw^o+\bb).
\end{equation}
Let $\Omega$ {denote} the linear manifold:
\begin{equation}
	\Omega\triangleq\{\bw\!:\,\cD\bw+\bb=\Zero\}.
\end{equation}
If $\bw^o\in\Omega$, the optimum $\bw^{\star}$ coincides with $\bw^o$. In this case, {the constrained optimization problem~\eqref{eq: network optimization problem} can be thought as estimating the unknown parameter vectors $\bw^o_k$ given prior information about relationships between tasks of the form~\eqref{eq: multitask problem (b)}. Exploiting such prior information} may improve the estimation as we will see in the experiments. Let $M$ denote the dimension of the network parameter vector $\bw$, i.e., $M=\sum_{k=1}^N M_k$. The projection of any vector~$\cblue{\by}\in\R^M$ onto $\Omega$ is given by:
\begin{equation}
	\label{eq: projection onto linear manifolds}
	\proj_{\Omega}(\cblue{\by})=\cP\cblue{\by}-\boldf,
\end{equation}
where
\begin{equation}
	\cP\triangleq\bI_M-\cD^\dagger\cD,\qquad
	\boldf\triangleq\cD^\dagger\bb.
\end{equation}
Let $\bw(i)$ denotes the estimate of $\bw^\star$ at iteration $i$. In order to solve~\eqref{eq: network optimization problem centralized} iteratively, the gradient projection method~\cite{bertsekas1999nonlinear} can be {applied on top of a gradient-descent iteration:}
\begin{equation}
\label{eq: offline centralized solution}
	\bw(i+1)=\proj_{\Omega}\big(\bw(i)+\mu[\br_{dx}-\cR_x\bw(i)]\big),\qquad i\geq 0.
\end{equation}
In order to run recursion~\eqref{eq: offline centralized solution}, we need to have access to the second-order moments $\{\bR_{x,k},\br_{dx,k}\}$. Since these moments are rarely available beforehand, the agents use their instantaneous data $\{d_k(i),\bx_k(i)\}$ to approximate the second-order moments, namely, $\bR_{x,k}\approx\bx_k(i)\bx_k^\top(i)$ and $\br_{dx,k}\approx d_k(i)\bx_k(i)$. Doing so and replacing $\proj_{\Omega}(\cdot)$ by~\eqref{eq: projection onto linear manifolds}, we obtain the following stochastic-gradient algorithm in lieu of~\eqref{eq: offline centralized solution}:
\begin{equation}
	\label{eq: online centralized solution}
	\bw(i+1)=\cP\cdot\col\big\{\bw_k(i)+\mu\bx_k(i)[d_k(i)-\bx_k^\top(i)\bw_k(i)]\big\}_{k=1}^N-\boldf.
\end{equation}
Collecting the regression vectors into {the $M\times N$ matrix $\bX(i)\triangleq\diag\{\bx_k(i)\}_{k=1}^N$ and the observations into the $N\times 1$ vector $\bd(i)\triangleq\col\{d_k(i)\}_{k=1}^N$}, algorithm~\eqref{eq: online centralized solution} becomes the Constrained {Least-Mean-Squares} (CLMS) algorithm:
\begin{equation}
	\label{eq: CLMS}
	\bw(i+1)=\cP\big(\bw(i)+\mu\bX(i)[\bd(i)-\bX^\top(i)\bw(i)]\big)-\boldf.
\end{equation}
{This procedure} was originally proposed in~\cite{frost1972algorithm} as an online linearly constrained minimum variance (LCMV) {filter for solving mean-square-error estimation problems subject to linear constraints; the motivation there was not concerned with multi-task problems. In this section, we showed that the centralized multitask constrained problem reduces to a similar problem, for which algorithm~\eqref{eq: CLMS} can be applied. The performance of such stand-alone centralized solutions} was studied {in~\cite{frost1972algorithm,arablouei2015mean,sayed2011adaptive}}.

\section{Problem reformulation and Distributed solution}

\subsection{Problem reformulation}
{We move on to develop a distributed solution with a continuous adaptation mechanism. First, note that several} works for solving problems of the form~\eqref{eq: network optimization problem} with possible distributed information access {already exist} in the \cblue{literature~\cite{mota2012distributed,mota2013D-ADMM,chen2015dictionary,mota2015distributed,towfic2014adaptive,lee2013distributedDRP,ram2010distributed,yuan2013convergence,zhang2016decentralized}}. However, \cblue{except for \cite{chen2015dictionary,mota2015distributed}}, these {other works} solve single-task estimation problems where the entire network is employed to estimate the minimizer of~\eqref{eq: network optimization problem}. Furthermore, \cblue{compared to~\cite{mota2012distributed,mota2013D-ADMM,chen2015dictionary,yuan2013convergence,mota2015distributed}}, we {shall assume} stochastic errors in the evaluation of the gradients of local cost functions.

{To proceed with the analysis, one of the challenges we now face is} that {any} given agent $k$ may be involved in several constraints. Our strategy is to transform~\eqref{eq: network optimization problem} into an equivalent optimization problem exhibiting structure amenable to distributed optimization with separable constraints. Let $j_k$ denote the number of constraints {that} agent~$k$ is involved in. We expand each node $k$ into a cluster $\C_k$ of $j_k$ {\emph{virtual}} sub-nodes, namely, $\C_k\triangleq\{k_m\}_{m=1}^{j_k}$. Each one of these sub-nodes is involved in a single constraint. Let $\bw_{k_m}$ denote the $M_k\times 1$ auxiliary vector associated with sub-node $k_m$. In order to ensure that agent $k$ satisfies simultaneously all the constraints at convergence, we {will allow} all sub-nodes at agent $k$ {to run} diffusion {learning to reach agreement on their estimates $\{\bw_{k_m}\}$ asymptotically.} We denote by $\cI_{e,p}$ the set of sub-nodes which are involved {in} the $p$-th constraint.

\begin{figure}[t]
	{\centering
	\includegraphics[scale=0.34]{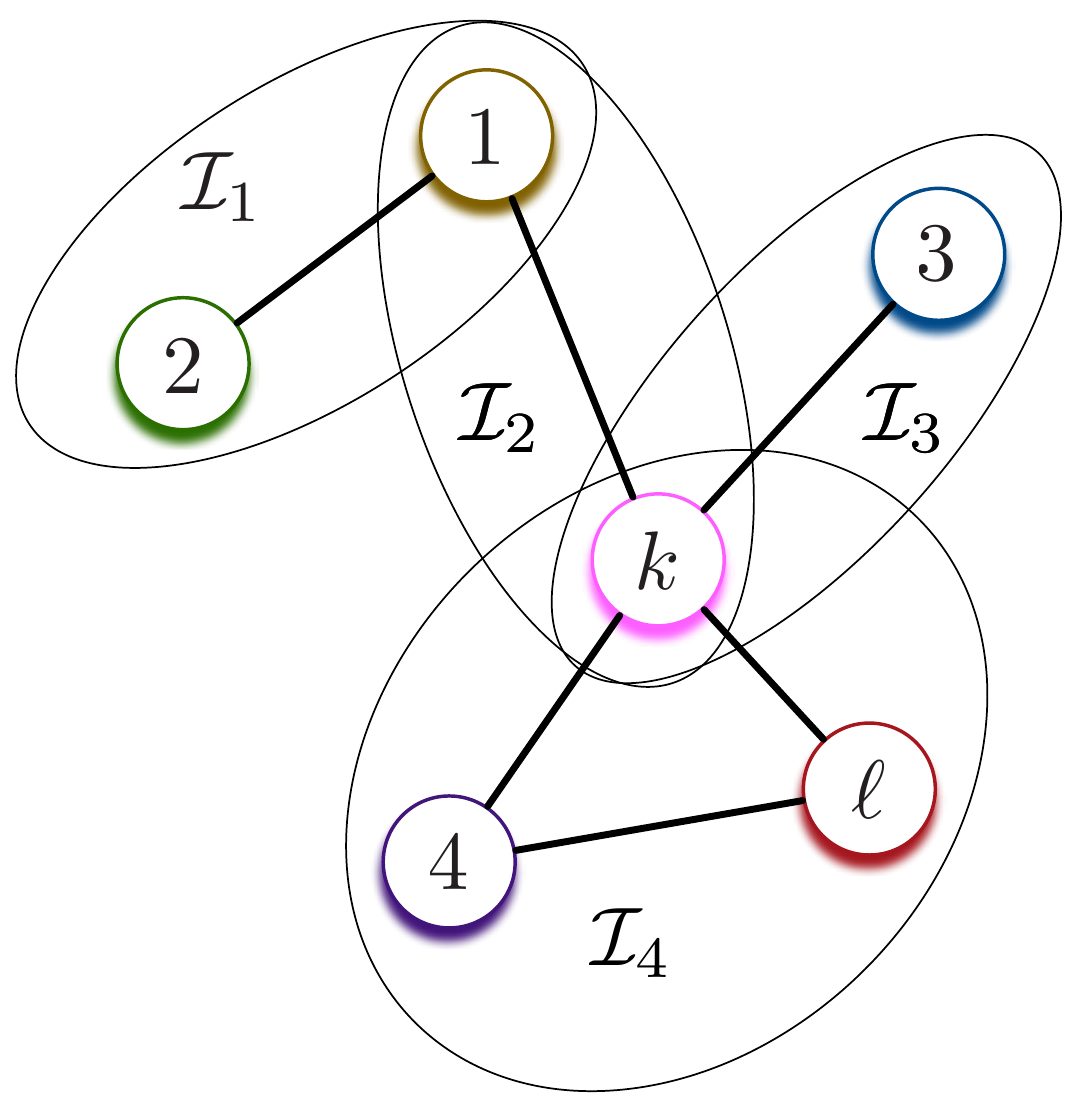}
	\includegraphics[scale=0.3]{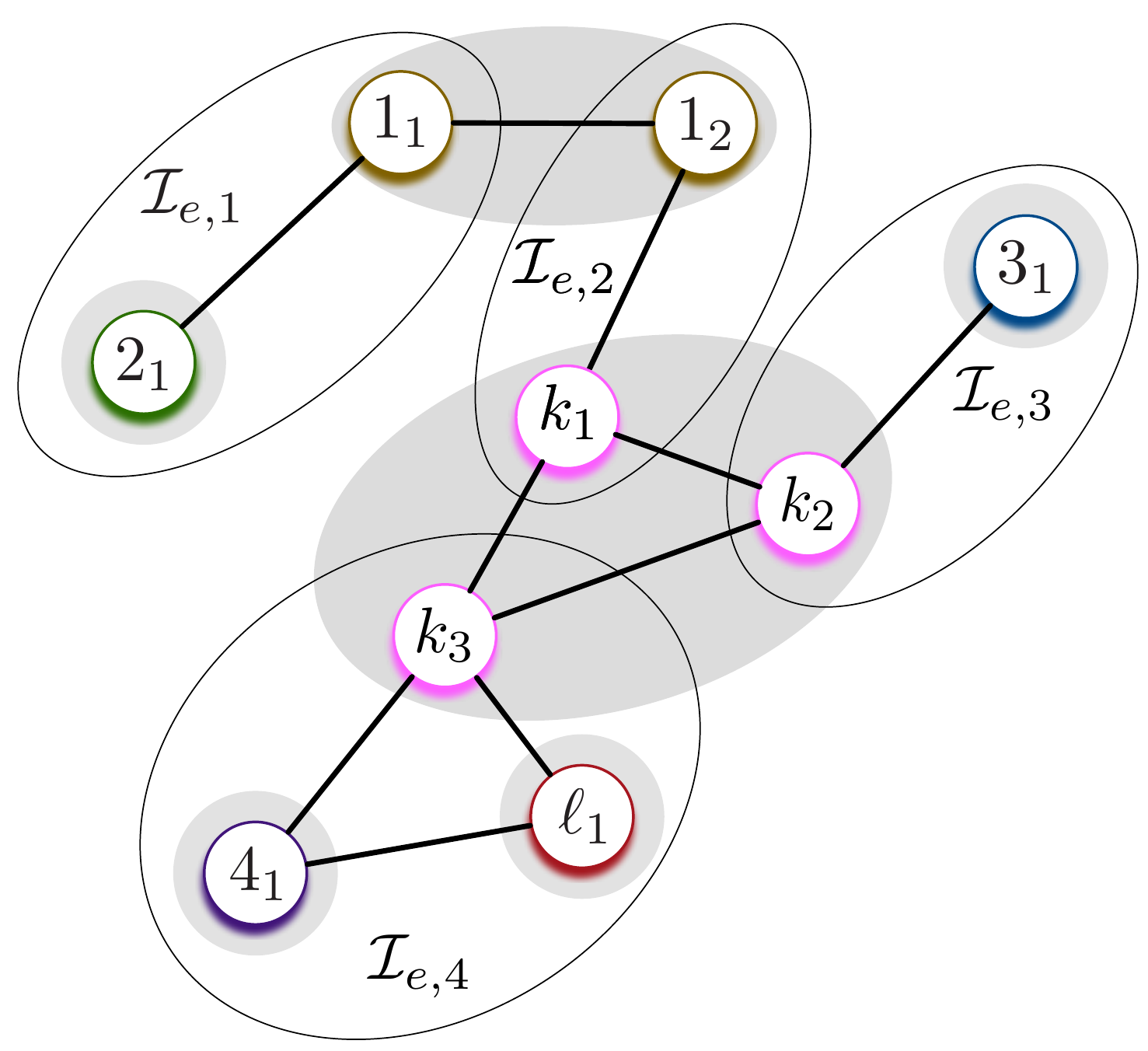}
	\caption{({\em Left}) Network topology with constraints identified by the subsets of nodes $\cI_1$, $\cI_2$, $\cI_3$, \cblue{and $\cI_4$}. 
	({\em Right}) Network topology model with \cblue{fully-connected} clusters shown in grey color and constraints now identified by the subsets of 
	sub-nodes $\cI_{e,1}$, $\cI_{e,2}$, $\cI_{e,3}$, \cblue{and $\cI_{e,4}$}. All sub-nodes in this model are involved in a single constraint. Diffusion learning is run in clusters with more than one sub-node to reach agreement on local estimates while satisfying their respective constraints. 
	}
	\label{fig: illustrative example 2}}
\end{figure}

 {In order to clarify the presentation, an illustrative example is provided in Fig.~\ref{fig: illustrative example 2}. On the left of this panel is the original network topology} with $N=6$ agents and $P=3$ constraints. On the right is the network topology model with clusters of sub-nodes shown in grey color. Observe that \cblue{$\,\cI_2=\{1,k\}$, $\,\cI_3=\{3,k\}$ and $\,\cI_4=\{4,k,\ell\}$}, which means that agent $k$ is involved in \cblue{constraints~$2$,~$3$, and~$4$}. Agent~$k$ is thus expanded into a cluster \cblue{$\C_k=\{k_1,k_2,k_3\}$ of $3$ sub-nodes}. \cblue{Sub-nodes $k_1$, $k_2$, and $k_3$ are assigned to constraints $2$, $3$, and $4$, respectively.} Each other agent, say $\ell$, involved in a single constraint is renamed $\ell_1$ and assigned to a single-node cluster $\C_\ell=\{\ell_1\}$ for consistency of notation. \cblue{This leads to the sets $\,\cI_{e,2}=\{1_2,k_1\}$, $\,\cI_{e,3}=\{3_1,k_2\}$ and $\,\cI_{e,4}=\{4_1,\ell_1,k_3\}$} where all sub-nodes are involved in a single constraint. 
\cblue{All sub-nodes $k_m$ in cluster~$\C_k$ can share data since they refer to the same agent $k$. In the sequel, we shall propose a general algorithm for strongly-connected clusters (see~\eqref{eq: multitask algorithm} below) and show how the designer can simplify the algorithm by choosing fully-connected clusters (see~\eqref{eq: multitask algorithm reduced} below).}

Accordingly, we {can} now reformulate problem~\eqref{eq: network optimization problem}. We start by collecting the vectors $\bw_{k_m}$ into the $N_e\times 1$ network block column vector:
\begin{equation}
\label{eq: vector extended}
	\bw_e\triangleq\col\Big\{\col\big\{\bw_{k_m}\big\}_{m=1}^{j_k}\Big\}_{k=1}^N,
\end{equation}
where {$N_e\triangleq\sum_{k=1}^Nj_k$}. \cblue{Throughout this work, a subscript ``$_e$'' below a symbol indicates an extended version  associated with sub-nodes (auxiliary variables). For example, while the symbol $N$ represents the number of nodes, the symbol $N_e$ represents the number of sub-nodes. Likewise, the vector $\bw_e$ in~\eqref{eq: vector extended} corresponds to the extended version of the vector $\bw$ in~\eqref{eq: network block column vector w0}.} We introduce for each agent $k$ a set of $j_k$ coefficients $\{c_{k_m}\}$ that satisfy two conditions:
\begin{equation}
	\label{eq: conditions on c}
	c_{k_m}>0,\text{ for }m=1,\ldots, j_k, \quad\text{and}\quad \sum_{m=1}^{j_k}c_{k_m}=1.
\end{equation}
The coefficients $\{c_{k_m}\}$ are free parameters that are chosen by the user. A natural choice is $c_{k_m}=\frac{1}{j_k}$ for all $m$. The global cost in~(\ref{eq: problem in the introduction}a) can be written as:
\begin{equation}
	J^{\text{glob}}(\bw_1,\ldots,\bw_N)\triangleq\sum_{k=1}^N J_k(\bw_k)=\sum_{k=1}^N\sum_{m=1}^{j_k} c_{k_m}J_k(\bw_k).
\end{equation}
We reformulate problem~{\eqref{eq: problem in the introduction}} in the following equivalent form by introducing the auxiliary variables {$\{\bw_{k_m}\}$}:
\begin{subequations}
	\label{eq: problem reformulation}
	\begin{align}
	&\minimize\limits_{\bw_e}\quad\sum\limits_{k=1}^N\sum\limits_{m=1}^{j_k}{c_{k_m}J_{k}(\bw_{k_m})}\label{eq: cost function of the problem}\\
	&\st\quad {\sum_{\ell_n\in\cI_{e,p}}\bD_{p\ell_n}\bw_{\ell_n}+\bb_p=\Zero,\quad p=1,\ldots,P,}\label{eq: constraint set 1}\\ 
	&\qquad\qquad\qquad\bw_{k_1}=\ldots=\bw_{k_{j_k}},\quad {k=1,\ldots,N}\label{eq: constraint set 2}.
	\end{align}
\end{subequations}
In the following, we shall address the equality constraints~\eqref{eq: constraint set 2} with a diffusion algorithm within each cluster of sub-nodes with the objective of reaching {an agreement} within each cluster (all sub-nodes converge to the same estimate). {Since the} diffusion strategy in a single-task network allows the agents to converge to the same limit point asymptotically for sufficiently small constant step-sizes when the network is strongly connected~\cite{sayed2014adaptation}, we {allow the} sub-nodes in cluster $\C_k$ to be connected such that the resultant cluster $\C_k$ is strongly connected. This does not lead to a change in the network topology since each sub-node in a cluster refers to {the} same agent. We refer to the \textit{virtual} set of neighboring sub-nodes of $k_m$ in $\C_k$ by {$\N_{k_m}\!\!\cap\C_k$}. 

The cost function in~\eqref{eq: cost function of the problem} can be written as:
\begin{equation}
	\sum\limits_{k=1}^N\sum\limits_{m=1}^{j_k}{c_{k_m}J_{k}(\bw_{k_m})}=\bw_e^\top\cR_{x,e}\bw_e-2\br_{dx,e}^\top\bw_e+\br_{d,e}^\top\One_{N_e\times1},
\end{equation}
where the $N_e\times N_e$ block diagonal matrix $\cR_{x,e}$, the $N_e\times 1$ block column vector $\br_{dx,e}$, and the $N_e\times 1$ column vector $\br_{d,e}$ are given by: 
\begin{align}
	\cR_{x,e}&\triangleq\diag\big\{\bC_k\otimes\bR_{x,k}\big\}_{k=1}^N,\label{eq: R_x e}\\
	\br_{dx,e}&\triangleq\col\big\{\bc_k\otimes\br_{dx,k}\big\}_{k=1}^N,\label{eq: r_dx e}\\
	\br_{d,e}&\triangleq\col\big\{\sigma^2_{d,k}\bc_k\big\}_{k=1}^N,\label{eq: r_d e}
\end{align}
with $\bC_k\triangleq\diag\{c_{k_m}\}_{m=1}^{j_k}$ and $\bc_k\triangleq\col\{c_{k_m}\}_{m=1}^{j_k}$.

The {equality} constraints in~\eqref{eq: constraint set 1}--\eqref{eq: constraint set 2} can be written more compactly as:
\begin{equation}
	\cD_e'\bw_e+\bb' = \Zero
\end{equation}
with
\begin{equation}
\label{eq: D'_e and b'}
	\cD'_e=\left[
	\begin{array}{lr}
	\cD_e\\
	\cH
	\end{array}
	\right],\quad
	\bb'=\left[
	\begin{array}{lr}
	\bb\\
	\Zero
	\end{array}
	\right],
\end{equation}
where $\cD_e$ is a $P\times N_e$ block matrix constructed according to~\eqref{eq: constraint set 1} which can be viewed as an expanded form of the $P\times N$ block matrix $\cD$, and $\cH$ is a $\sum_{k=1}^N(j_k-1)\times N_e$ block matrix constructed according to~\eqref{eq: constraint set 2}.

Using similar arguments as in Section~\ref{subsec: Centralized solution}, we find that the solution of~\eqref{eq: problem reformulation} is given by:
\begin{equation}
	\label{eq: extended closed form solution 1}
	\bw^\star_e=\bw^o_e-\cR_{x,e}^{-1}\cD_e'^\top(\cD_e'\cR_{x,e}^{-1}\cD_e'^\top)^{-1}(\cD_e'\bw^o_e+\bb'){,}
\end{equation}
where the $N_e\times 1$ block column vector $\bw^o_e$ is given by:
\begin{equation}
	\bw^o_e\triangleq\col\big\{\One_{{j_k}\times1}\otimes\bw^o_k\big\}_{k=1}^N\label{eq: expanded w0}.
\end{equation}
Let $\bw^\star_k$ {denote} the $k$-th block of $\bw^\star$ in~\eqref{eq: closed form solution}. The optimum vector $\bw^\star_e$ can be written alternatively as:
\begin{equation}
	\label{eq: extended closed form solution 2}
	\bw^\star_e=\col\big\{\One_{{j_k}\times1}\otimes\bw^\star_k\big\}_{k=1}^N.
\end{equation}

\subsection{Distributed solution}
\label{subsec: Distributed solution}

To solve problem~\eqref{eq: problem reformulation} with distributed information access, we propose an iterative algorithm based on diffusion strategies
and gradient-projection principle. First, we present the algorithm when the second order moments of the observations are assumed to be known by each sub-node. Although cluster $\C_k$ and agent $k$ refer to the same entity, we shall use the notion of cluster and sub-nodes in order to simplify the presentation.

Let $\bw_{e,p}$ {denote} the $i_p\times 1$ block column vector given by $\bw_{e,p}=\col\{\bw_{\ell_n}\}_{\ell_n\in\cI_{e,p}}$ where $i_p$ is the number of nodes involved in the $p$-th constraint. {Also, note that $i_p$ is the cardinality of $\cI_{p}$ and $\cI_{e,p}$. Let $\Omega_{p}$ denote the linear manifold corresponding to the $p$-th constraint in~\eqref{eq: constraint set 1}, namely, $\Omega_{p}\triangleq\{\cD_{p}\bw_{e,p}+\bb_p=\Zero\}$ where $\cD_{p}$ is a $1\times i_p$ block matrix.} Let $\bw_{k_m}(i)$ be the estimate of $\bw^\star_k$ at sub-node $k_m$ and time instant $i$. We assume that $k_m\in\cI_{e,p}$. Following the {same line of reasoning as}~\cite{sayed2014diffusion} in the single-task case, and extending the argument to our multitask problem, {we arrive at} the following diffusion algorithm consisting of three steps:
\begin{subequations}
	\label{eq: diffusion iterates in the absence of gradient noise}
	\begin{align}
		&\bpsi_{k_m}(i+1)=\bw_{k_m}(i)+\mu\,c_{k_m}[\br_{dx,k}-\bR_{x,k}\bw_{k_m}(i)]\label{eq: step 1}\\
		&\bphi_{k_m}(i+1)=\sum_{k_n\in{\N_{k_m}\!\cap\,\C_k}}a_{k_n,k_m}\bpsi_{k_n}(i+1)\label{eq: step 2}\\
		&\bw_{k_m}(i+1)=\left[\proj_{\Omega_{p}}\big(\col\big\{\bphi_{\ell_n}(i+1)\big\}_{\ell_n\in\cI_{e,p}}\big)\right]_{k_m}\label{eq: step 3}
	\end{align}
\end{subequations}
where $\mu>0$ is a constant step-size parameter, $[\bx]_{k_m}$ is the block of $\bx$ corresponding to {sub-node} $k_m$, {and $\bw_{k_m}(0)=\bw_k(0)$ for all $m$.} In the first step~\eqref{eq: step 1}, also called adaptation step, each sub-node $k_m$ in the network adapts its estimate $\bw_{k_m}(i)$ via gradient descent on $c_{k_m}J_{k}(\cdot)$. This step results in the intermediate estimate $\bpsi_{k_m}(i+1)$. 

In the combination step~\eqref{eq: step 2}, each sub-node $k_m$ combines its estimate $\bpsi_{k_m}(i+1)$ with the estimates $\bpsi_{k_n}(i+1)$ of its intra-cluster neighbors {$\N_{k_m}\!\!\cap\C_k$}. This step results in the intermediate estimate $\bphi_{k_m}(i+1)$. The nonnegative coefficients $\{a_{k_n,k_m}\}$ are chosen to satisfy the following conditions:
\begin{equation}
\begin{split}
	\label{eq: conditions on the combination coefficients}
	a_{k_n,k_m} \geq 0,&
	\sum_{k_m\in{\N_{k_n}\cap\,\C_k}}a_{k_n,k_m}=1,
	\sum_{k_n\in{\N_{k_m}\cap\,\C_k}}a_{k_n,k_m}=1,\\
	&\text{and } a_{k_n,k_m}=0 ~~\text{if}~~k_n\notin{\N_{k_m}\!\!\cap\C_k}.
\end{split}
\end{equation}
Collecting these coefficients into a $j_k\times j_k$ matrix $\bA_k$ for each cluster $\C_k$, it follows that $\bA_k$ is doubly stochastic. 

{Let $M_p$ denote the dimension of the vector $\bw_{e,p}$, i.e., $M_p=\sum_{\ell_n\in\cI_{e,p}}M_{\ell}$.} Before describing the third step, we recall that the projection of any point $\cblue{\by}$ onto $\Omega_{p}$ has the form:
\begin{equation}
	\label{eq: projection onto the local manifolds}
	\proj_{\Omega_{p}}\big(\cblue{\by}\big)=\cP_{p}\,\cblue{\by}-\boldf_{p}
\end{equation}
where
\begin{equation}
	\label{eq: cP r p}
	\cP_{p}\triangleq\bI_{M_p}-\cD_{p}^\dagger\cD_{p}\qquad\text{and}\qquad \boldf_{p}\triangleq\cD_{p}^\dagger\bb_p.
\end{equation}
To evaluate the block $\big[\proj_{\Omega_{p}}(\cblue{\by})\big]_{k_m}$, even if sub-node $k_m$ is only in charge of estimating $\bw_{k_m}$, it needs the entire vector~$\cblue{\by}$, the $M_k\times M_p$ matrix $[\cP_{p}]_{k_m,\bullet}$, and the $M_k\times 1$ vector $[\boldf_{p}]_{k_m}$. In the projection step~\eqref{eq: step 3}, each sub-node $k_m\in\cI_{e,p}$ collects the intermediate estimates $\bphi_{\ell_n}(i+1)$ from all sub-nodes $\ell_n\in\cI_{e,p}$ and combines them according to~\eqref{eq: step 3}. This step results in the estimate $\bw_{k_m}(i+1)$ of $\bw^\star_k$ at sub-node $k_m$ and iteration $i+1$. 

The adaptation step~\eqref{eq: step 1} requires knowledge of the second-order moments of data. Proceeding as in the centralized {case, and replacing the moments by instantaneous approximations,} we obtain algorithm~\eqref{eq: multitask algorithm} for solving~\eqref{eq: problem reformulation} in a distributed way:
 	\begin{subequations}
		\label{eq: multitask algorithm}
		\begin{align}
			&\bpsi_{k_m}(i+1)=\bw_{k_m}(i)+\mu\,c_{k_m}\bx_k(i)[d_k(i)-\bx_k^\top(i)\bw_{k_m}(i)],\label{eq: 1 adaptation step}\\
			&\bphi_{k_m}(i+1)= [\cP_{p}]_{k_m,\bullet}\cdot\col\big\{\bpsi_{\ell_n}(i+1)\big\}_{\ell_n\in\cI_{e,p}}\hspace{-3mm}-[\boldf_{p}]_{k_m},\label{eq: 3 projection step}\\
			&\bw_{k_m}(i+1)=\sum_{k_n\in{\N_{k_m}\cap\,\C_k}}a_{k_n,k_m}\bphi_{k_n}(i+1).\label{eq: 2 combination step}
		\end{align}
	\end{subequations}

Compared to~\eqref{eq: diffusion iterates in the absence of gradient noise}, observe in~\eqref{eq: multitask algorithm} that each sub-node~$k_m$ projects its intermediate estimate before combining it. We recommend this permutation since it allows, with the appropriate parameter settings described below, to reduce the algorithm complexity without compromising its convergence, as confirmed in the sequel. Consider any agent $k$. By setting factors $c_{k_m}$ to $\frac{1}{j_k}$ for all $m=1,\ldots,j_k$, and combining the intermediate estimate $\bphi_{k_m}(i+1)$ at each sub-node $k_m$ with the estimates of all other sub-nodes available at node~$k$ using uniform combination coefficients, i.e., $\N_{k_m}\cap\C_k=\C_k$ and $a_{k_n,k_m}=\frac{1}{j_k}$ for $n=1,\ldots,j_k$, \eqref{eq: 1 adaptation step} and~\eqref{eq: 2 combination step} reduce to:
\begin{align}
	&\bpsi_{k_m}(i+1)=\bpsi_k(i+1), \quad \text{for } m=1,\ldots,j_k,\label{eq: 36}\\
	&\bw_{k_m}(i+1)=\bw_k(i+1), \quad \text{for } m=1,\ldots,j_k,
\end{align}
where $\bpsi_k(i+1)$ and $\bw_k(i+1)$ are given by:
\begin{align}
	&\bpsi_k(i+1)=\bw_k(i)+\frac{\mu}{j_k}\bx_k(i)\big[d_k(i)-\bx_k^\top(i)\bw_k(i)\big],\\
	&\bw_k(i+1)=\frac{1}{j_k}\cblue{\sum_{m=1}^{j_k}\bphi_{k_m}(i+1)}.\label{eq: 39}
\end{align}
\cblue{In this case, at each agent $k$, the algorithm~\eqref{eq: multitask algorithm} becomes:
	\begin{subequations}
		\label{eq: multitask algorithm reduced}
		\begin{align}
			&\bpsi_{k}(i+1)=\bw_k(i)+\frac{\mu}{j_k}\bx_k(i)\big[d_k(i)-\bx_k^\top(i)\bw_k(i)\big]\\
			&\bphi_{k_m}(i+1)= [\cP_{p}]_{k_m,\bullet}\cdot\col\big\{\bpsi_{\ell}(i+1)\big\}_{\ell\in\cI_{p}}-[\boldf_{p}]_{k_m},\nonumber\\
			&\qquad\qquad\qquad k_m\in\cI_{e,p}, \quad m=1,\ldots,j_k,\\
			&\bw_{k}(i+1)=\frac{1}{j_k}\sum_{m=1}^{j_k}\bphi_{k_m}(i+1).
		\end{align}
	\end{subequations}
Instead of maintaining and updating $j_k$ coefficient vectors $\bpsi_{k_m}(i+1)$, agent $k$ maintains and updates only one parameter vector $\bpsi_k(i+1)$. Then, it transmits the vector $\bpsi_k(i+1)$ to its neighbors, receives $\{\bpsi_\ell(i+1)\}$ from its neighborhood, and generates $j_k$ parameter vectors $\bphi_{k_m}(i+1)$ by projecting onto its constraints. Finally, it combines these  vectors to obtain $\bw_k(i+1)$, i.e., the estimate of $\bw^\star_k$ at iteration $i+1$. Therefore, with this setting, the computational and communication complexity of our distributed algorithm is significantly reduced.}

\vspace{-0.3cm}



\section{Performance analysis Relative to $\bw_e^o$}
\label{sec: Performance analysis I}

\subsection{Network error vector recursion}
\label{subsec: network error vector recursion}
We shall first study the stochastic behavior of algorithm~\eqref{eq: multitask algorithm} with respect to the optimal parameter vector $\bw_e^o$. 
We introduce the error vector $\bwt_{k_m}(i)\triangleq\bw^o_k-\bw_{k_m}(i)$ and the intermediate error vectors $\bpsit_{k_m}(i)\triangleq\bw^o_k-\bpsi_{k_m}(i)$ and $\bphit_{k_m}(i)\triangleq\bw^o_k-\bphi_{k_m}(i)$. We further introduce the $N_e\times 1$ block network error vector:
\begin{equation}
\label{eq: definition of wt_e}
	\bwt_e(i)\triangleq\col\left\{\col\big\{\bwt_{k_m}(i)\big\}_{m=1}^{{j_k}}\right\}_{k=1}^N.
\end{equation}
{Let $M_e$ denote the length of the network error vector $\bwt_e(i)$, that is, $M_e\triangleq\sum_{k=1}^Nj_kM_k$.} Using the linear model~\eqref{eq: linear data model}, the estimation error in the adaptation step~\eqref{eq: 1 adaptation step} can be written as:
\begin{equation}
	\label{eq: estimation error 1}
	d_k(i)-\bx_k^\top(i)\bw_{k_m}(i)=\bx_k^{\top}(i)\bwt_{k_m}(i)+z_k(i).
\end{equation}
Subtracting $\bw^o_k$ from both sides of the adaptation step~\eqref{eq: 1 adaptation step}, using~\eqref{eq: estimation error 1}, and collecting the error vectors $\bpsit_{k_m}(i)$ into the $N_e\times 1$ block vector $\bpsit_e(i)\triangleq\col\left\{\col\big\{\bpsit_{k_m}(i)\big\}_{m=1}^{{j_k}}\right\}_{k=1}^N$, we obtain:
\begin{equation}
	\label{eq: bpsit(i+1)}
	\bpsit_e(i+1)=\big[\bI_{M_e}-\mu\cR_{x,e}(i)\big]\bwt_e(i)-\mu\,\bp_{zx,e}(i),
\end{equation}
where 
\begin{align}
	&\cR_{x,e}(i)\triangleq\diag\Big\{\bC_k\otimes\bx_k(i)\bx^\top_k(i)\Big\}_{k=1}^N \label{eq: R x},\\
	&\bp_{zx,e}(i)\triangleq\col\Big\{\bc_k\otimes \bx_k(i)z_k(i)\Big\}_{k=1}^N \label{eq: p zx}.
\end{align} 
{Projecting $\bpsi_e(i+1)$ onto the sets $\Omega_{p}$ in~\eqref{eq: projection onto the local manifolds}, we obtain {from~\eqref{eq: 3 projection step}}:
\begin{equation}
	\label{eq: recursion w}
	\bphi_e(i+1)=\cP_e\bpsi_e(i+1)-\boldf_e,
\end{equation}
where 
\begin{align}
        \bpsi_e(i)\triangleq\col\Big\{\col\big\{\bpsi_{k_m}(i)\big\}_{m=1}^{j_k}\Big\}_{k=1}^N,\\
	\quad \bphi_e(i)\triangleq\col\Big\{\col\big\{\bphi_{k_m}(i)\big\}_{m=1}^{j_k}\Big\}_{k=1}^N, 
\end{align}
$\cP_e$ is an $M_e\times M_e$ orthogonal projection matrix, and $\boldf_e$ is an $M_e\times 1$ vector given by (see~Appendix~A):
\begin{align}
	\cP_e&\triangleq\bI_{M_e}-\cD_e^\dagger\cD_e=\bI_{M_e}-\cD_e^\top(\cD_e\cD_e^\top)^{-1}\cD_e\label{eq: cP e}\\
	\boldf_e&\triangleq\cD_e^\dagger\bb=\cD_e^\top(\cD_e\cD_e^\top)^{-1}\bb.\label{eq: boldf e}
\end{align}
Subtracting $\bw^o_e$ in~\eqref{eq: expanded w0} from both sides of recursion~\eqref{eq: recursion w}, we obtain:
\begin{align}
	\bphit_e(i+1)&\triangleq\col\Big\{\col\big\{\bphit_{k_m}(i+1)\big\}_{m=1}^{j_k}\Big\}_{k=1}^N\notag\\
	&=\cP_e\bpsit_e(i+1)+\big(\bI_{M_e}-\cP_e\big)\bw^o_e+\boldf_e.
\end{align}
Subtracting $\bw^o_k$ from both sides of the combination step~\eqref{eq: 2 combination step} and using~\eqref{eq: conditions on the combination coefficients}, we obtain that the network error vector for the diffusion strategy~\eqref{eq: multitask algorithm} evolves according to the following recursion:
\begin{equation}
	\label{eq: weight error vector recursion 0}
	\boxed{
	\begin{aligned}
	&\bwt_e(i+1)=\cA^\top\cP_e\big[\bI_{M_e}-\mu\cR_{x,e}(i)\big]\bwt_e(i)-\\ 
	&\qquad\mu\cA^\top\cP_e\bp_{zx,e}(i)+\cA^\top(\bI_{M_e}-\cP_e)\bw^o_e+\cA^\top\boldf_e,
	\end{aligned}
	}
\end{equation}
where $\cA\triangleq\diag\{\bA_k\otimes\bI_{M_k}\}_{k=1}^N$.} Before proceeding, let us introduce the following assumption on the regression data.

\begin{assumption}
\label{assump: independent regressors}
{\emph{(Independent regressors)}} The regression vectors $\bx_k(i)$ arise from a zero-mean random process that is temporally white and spatially independent.
\end{assumption}

Under this assumption, $\bx_k(i)$ is independent of $\bw_{\ell_m}(j)$ for $i\geq j$ and for all $\ell_m$. This assumption is commonly used in the adaptive filtering literature since it helps simplify the analysis, and the performance results obtained under this assumption match well the actual performance of stand-alone filters for sufficiently small step-sizes~\cite{sayed2011adaptive}.

\subsection{Mean behavior analysis}
\label{subsec: Mean behavior analysis I}

Recursion~\eqref{eq: weight error vector recursion 0} can be rewritten in a more compact form:
\begin{equation}
	\label{eq: weight error vector recursion 1}
	\bwt_e(i+1)=\cB(i)\bwt_e(i)-\mu\bg(i)+\br{,}
\end{equation}
where we introduced the following notations:
{\begin{align}
	\cB(i)&\triangleq\cA^\top\cP_e\big[\bI_{M_e}-\mu\cR_{x,e}(i)\big],\label{eq: B(i)}\\
	\bg(i)&\triangleq\cA^\top\cP_e\bp_{zx,e}(i),\label{eq: g(i)}\\
	\br&\triangleq\cA^\top(\bI_{M_e}-\cP_e)\bw^o_e+\cA^\top\boldf_e.\label{eq: br}
\end{align}}
Taking the expectation of both sides of recursion~\eqref{eq: weight error vector recursion 1}, using Assumption~\ref{assump: independent regressors}, and  $\expec\,\bg(i)=\Zero$, we find that the mean error vector evolves according to the recursion:
\begin{equation}
	\label{eq: mean error recursion}
	\expec\,\bwt_e(i+1)=\cB\,\expec\,\bwt_e(i)+\br,
\end{equation}
\cblue{where 
\begin{equation}
	\cblue{\cB\triangleq\expec\,\cB(i)}=\cA^\top\cP_e\big(\bI_{M_e}-\mu\cR_{x,e}\big)\label{eq: B},
\end{equation}
with $\cR_{x,e}=\expec\cR_{x,e}(i)$ given in~\eqref{eq: R_x e}\footnote{\cblue{If $\bU(i)$ is a random matrix, its expected value $\expec\,\bU(i)$ is denoted by $\bU$.} }.} Recursion~\eqref{eq: mean error recursion} converges as $i\rightarrow\infty$ if the matrix $\cB$ is stable. If we let $i\rightarrow\infty$ on both sides of~\eqref{eq: mean error recursion}, we find that the asymptotic mean bias is given by:
\begin{equation}
	\label{eq: bias}
	\expec\,\bwt_e(\infty)=\lim_{i\rightarrow\infty}\expec\,\bwt_e(i)=(\bI_{M_e}-\cB)^{-1}\br.
\end{equation}
{It is known that any induced matrix norm is lower bounded by the spectral radius of the matrix. We can thus write in terms of the $2$-induced matrix norm:
\begin{equation}
	\rho(\cB)
	\leq\|\cA^\top\|_2\cdot\|\cP_e\|_2\cdot\|\bI_{M_e}-\mu\cR_{x,e}\|_2,
\end{equation}
where we used the sub-multiplicative property of the $2$-induced norm. Since $\cP_e$ is an orthogonal projection matrix and $\cA^\top$ is a doubly-stochastic matrix, their $2$-induced norms are equal to one. Since the matrix $\bI_{M_e}-\mu\cR_{x,e}$ is a symmetric block diagonal matrix, its $2$-induced norm agrees with its spectral radius:
\begin{equation}
	\begin{split}
	\|\bI_{M_e}-\mu\cR_{x,e}\|_2&=\rho(\bI_{M_e}-\mu\cR_{x,e})\\
	&=\max_{1\leq k \leq N}\max_{1\leq m\leq j_k}\rho(\bI_{M_k}-\mu\, c_{k_m}\bR_{x,k}).
\end{split}
\end{equation}
}
Thus, the stability of $\cB$ is ensured by choosing $\mu$ such that:
\begin{equation}
	\label{eq: stability conditions}
	0<\mu<\frac{2}{c_{k,\max}\cdot\lambda_{\max}(\bR_{x,k})},\quad\forall k=1,\ldots,N.
\end{equation}
where $c_{k,\max}\triangleq\max\limits_{ 1\leq m\leq j_k}c_{k_m}$. We observe that when $\bw^\star_e=\bw^o_e$, i.e., perfect model scenario where $\bw^o$ satisfies the linear equality constraints, the bias reduces to $\Zero$.

\subsection{Mean-square-error behavior analysis}
{To perform the mean-square-error analysis, we shall use the block Kronecker product {operator~\cite{Koning1991nlockkronecker}} and the block vectorization operator $\bvc(\cdot)$. As explained in~\cite{sayed2014adaptation}, these block operators preserve the locality of the blocks in the original matrix arguments. To analyze the convergence in mean-square-error sense, we consider the variance of the weight error vector $\bwt_e(i)${,} weighted by any positive-definite matrix $\bSig$, that is, $\expec\,\|\bwt_e(i)\|^2_{\bSig}$, where $\|\bwt_e(i)\|^2_{\bSig}\triangleq\bwt_e^\top(i)\bSig\bwt_e(i)$. The freedom in selecting $\bSig$ allows us to extract various types of information about the network and the sub-nodes. From~\eqref{eq: weight error vector recursion 1} and Assumption~\ref{assump: independent regressors}, we obtain:}
\begin{equation}
\begin{split}
	\label{eq: variance relation 1}
	\expec\{\|\bwt_e(i+1)\|^2_{\bSig}\}&=\expec\{\|\bwt_e(i)\|^2_{\bSig'}\}+\mu^2\expec\{\|\bg(i)\|^2_{\bSig}\}+\\
	&\qquad\|\br\|^2_{\bSig}+2\br^\top\bSig\cB\expec\,\bwt_e(i),
\end{split}
\end{equation}
where matrix $\bSig'$ is given by:
\begin{equation}
\label{eq: sigma'}
\bSig'\triangleq\expec\{\cB^\top(i)\bSig\cB(i)\}.
\end{equation}

Let $\bsig$ denotes the $M_e^2\times 1$ vector representation of $\bSig$ that is obtained by the block vectorization operator, namely, $\bsig\triangleq\bvc(\bSig)$. In the sequel, it will be more convenient to work with $\bsig$ than with $\bSig$ itself. We will use the notations $\|\bx\|^2_{\bSig}$ and $\|\bx\|^2_{\bsig}$ to denote the same quantity $\bx^\top\bSig\bx$. Let $\bsig'=\bvc(\bSig')$. Using the property $\bvc(\bU\bSig\bW)=(\bW^\top\otimes_b\bU)\bsig$, the vector $\bsig'$ can be related to $\bsig$:
\begin{equation}
	\bsig'=\cF\bsig,
\end{equation}
where $\cF$ is an $M_e^2\times M_e^2$ matrix given by:
\begin{equation}
	\label{eq: matrix F}
	\cF\triangleq\expec\{\cB^\top(i)\otimes_b\cB^\top(i)\}.
\end{equation}
The evaluation of the matrix $\cF$ requires knowledge of the {fourth-order} moments of the regression vectors. \cblue{In practice, when $\expec\{\cR_{x,e}(i)\otimes_b\cR_{x,e}(i)\}$ can be computed, as in the case of zero-mean Gaussian regressors (see~Appendix~B), the matrix $\cF$ can be calculated in closed form and its stability can be checked for a given $\mu$.\footnote{\cblue{When $\expec\{\cR_{x,e}(i)\otimes_b\cR_{x,e}(i)\}$ cannot be evaluated, a common alternative is {to use the approximation} $\cF\approx\cB^\top\!\otimes_b\cB^\top$ for sufficiently small step-sizes (see \cite{sayed2014adaptation,sayed2014diffusion}). In this case, we have $\rho(\cF)\approx\rho(\cB^\top\otimes_b\cB^\top)=\rho(\cB)^2$. As long as this approximation is reasonable, the stability of $\cF$ is ensured if $\rho(\cB)<1$, i.e., if the step-size is chosen according to condition~\eqref{eq: stability conditions}.} }}

The second term on the RHS of relation~\eqref{eq: variance relation 1} can be written as:
\begin{align}
	\mu^2\expec\{\|\bg(i)\|^2_{\bSig}\}&=\mu^2\expec\{\bg^\top(i)\bSig\bg(i)\}\notag\\
	&=\mu^2\tr(\bSig\cG)=\mu^2\big[\bvc(\cG^\top)\big]^\top\bsig,
\end{align}
where $\cG$ is the $M_e\times M_e$ matrix given by:
{\begin{equation}
	\cG\triangleq\expec\{\bg(i)\bg^\top(i)\}=\cA^\top\cP_e\diag\big\{\bc_k\bc_k^\top\otimes\sigma^2_{z,k}\bR_{x,k}\big\}_{k=1}^N\cP_e\cA.
\end{equation} }
{Similarly}, the third term on the RHS of relation~\eqref{eq: variance relation 1} can be written as: 
\begin{equation}
	\|\br\|^2_{\bSig}=\big[\bvc(\br\br^\top)\big]^\top\bsig,
\end{equation}
{and} the fourth term can be written as:
\begin{align}
	2\br^\top\bSig\cB\expec\,\bwt_e(i)&=2\tr(\br^\top\bSig\cB\expec\,\bwt_e(i))\notag\\
	&=2\big[\bvc(\br\expec\{\bwt_e^\top(i)\}\cB^\top)\big]^\top\bsig.
\end{align}
Let us define the $M_e\times M_e$ time dependent matrix $\cY(i)$ given by:
\begin{equation}
	\label{eq: Y(i)}
	\cY(i)\triangleq\mu^2\cG^\top+\br\br^\top+2\br\expec\{\bwt_e(i)\}^\top\cB^\top.
\end{equation}
{Then}, the variance relation~\eqref{eq: variance relation 1} can be expressed as:
\begin{equation}
\label{eq: variance relation 2}
	\expec\{\|\bwt_e(i+1)\|^2_{\bsig}\}=\expec\{\|\bwt_e(i)\|^2_{\cF\bsig}\}+\big[\bvc(\cY(i))\big]^\top\bsig.
\end{equation}

Provided that $\cF$ is stable, recursion~\eqref{eq: variance relation 2} is stable. Since $\cG$, $\br$, $\cB$, $\bsig$, and $\mu$ are constant and finite terms, the boundedness of $\big[\bvc(\cY(i))\big]^\top\bsig$ depends on $\expec\,\bwt_e(i)$ being bounded. We know from~\eqref{eq: mean error recursion} that $\expec\,\bwt_e(i)$ is bounded \cblue{if the step-size~$\mu$ is chosen according to condition~\eqref{eq: stability conditions}} because~\eqref{eq: mean error recursion} is a {Bounded-Input Bounded-Output} (BIBO) stable recursion with a bounded driving term $\br$. It follows that $\big[\bvc(\cY(i))\big]^\top\bsig$ is uniformly bounded. As a result, \cblue{the algorithm is mean-square-error stable, i.e., $\expec\{\|\bwt_e(i+1)\|_{\bsig}^2\}$ converges to a bounded value as $i\rightarrow \infty$, if $\mu$ is chosen such that  $\cF$ in~\eqref{eq: matrix F} is stable in addition to condition~\eqref{eq: stability conditions} that ensures mean stability. As explained above, step-sizes that ensure stability in the mean and that are sufficiently small will also ensure stability in the mean-square.}

Following similar arguments as in~\cite {chen2014multitask,nassif2016multitask,chen2014diffusion} and doing the required adjustments, we find that the weighted variance $\expec\{\|\bwt_e(i+1)\|^2_{\bsig}\}$ evolves according to the following recursion:
\begin{equation}
\begin{split}
	\label{eq: MSD transient curve}
	\expec&\{\|\bwt_e(i+1)\|^2_{\bsig}\}=\expec\{\|\bwt_e(i)\|^2_{\bsig}\}+\\
	&\big[\bvc(\expec\{\bwt_e(0)\bwt_e^\top(0)\})\big]^\top(\cF-\bI_{M_e^2})\cF^{i}\bsig+\\
	&\big[\bvc(\cY(i))\big]^\top\bsig+\bGamma(i)\bsig,
	\end{split}
\end{equation} 
where $\bwt_e(0)$ is the initial condition and $\bGamma(i+1)$ is a $1\times M_e^2$ vector that can be evaluated from $\bGamma(i)$ according to:
\begin{equation}
	\bGamma(i+1)=\bGamma(i)\cF+\big[\bvc(\cY(i))\big]^\top(\cF-\bI_{M_e^2}),
\end{equation}
with $\bGamma(0)=\Zero$. 

The steady-state network performance with metric $\bsig_{ss}$ is defined as:
\begin{equation}
\label{eq: zeta star}
	\zeta^\star=\lim_{i\rightarrow\infty}\expec\|\bwt_e(i)\|^2_{\bsig_{ss}}.
\end{equation}
If the matrix $\cF$ is stable, from the recursive expression~\eqref{eq: variance relation 2}, we obtain as $i\rightarrow\infty$:
\begin{equation}
	\label{eq: steady-state variance}
	\lim_{i\rightarrow\infty}\expec\{\|\bwt_e(i)\|^2_{(\bI_{M_e^2}-\cF)\bsig}\}=[\bvc(\cY(\infty))]^\top\bsig,
\end{equation}
where, from~\eqref{eq: Y(i)} and~\eqref{eq: bias}, we have:
\begin{equation}
\label{eq: Yinf}
\cY(\infty)=\mu^2\cG^\top+\br\br^\top+2\br\expec\{\bwt_e(\infty)\}^\top\cB^\top.
\end{equation}
To obtain~\eqref{eq: zeta star}, we replace $\bsig$ in~\eqref{eq: steady-state variance} by $(\bI_{M_e^2}-\cF)^{-1}\bsig_{ss}$. The theoretical findings~\eqref{eq: mean error recursion},~\eqref{eq: bias},~\eqref{eq: MSD transient curve}, and~\eqref{eq: steady-state variance} allow us to predict the behavior in the mean and in the mean-square-error sense of the stochastic algorithm~\eqref{eq: multitask algorithm} w.r.t. the parameter vector $\bw_e^o$. {Note that, the network MSD w.r.t. $\bw^o_e$ given by:
\begin{equation}
\text{MSD}_{\text{net}}(i)\triangleq
\frac{1}{N}\sum_{k=1}^N\left(\frac{1}{j_k}\sum_{m=1}^{j_k}\expec\|\bwt_{k_m}(i)\|^2\right),
\end{equation}
can be obtained by setting $\bSig=\frac{1}{N}\,\diag\left\{\frac{1}{j_k}\bI_{j_k\cdot M_k}\right\}_{k=1}^N$.
}


\section{Performance Analysis Relative to $\bw^\star_e$}
\label{sec: Performance analysis II}
We shall now study the convergence behavior of algorithm~\eqref{eq: multitask algorithm} toward the solution $\bw^\star_e$ of the optimization problem with constraints~\eqref{eq: problem reformulation}. To this end, we introduce for each sub-node $k_m$ the weight error vector:
\begin{equation}
	\bwt'_{k_m}(i)\triangleq\bw^\star_k-\bw_{k_m}(i),\qquad
\end{equation} 
and the $N_e\times 1$ network block error vector:
\begin{equation}
\label{eq: definition of wt'_e}
	\bwt_e'(i)\triangleq\col\Big\{\col\big\{\bwt'_{k_m}(i)\big\}_{m=1}^{{j_k}}\Big\}_{k=1}^N
\end{equation}
We note that the behavior of algorithm~\eqref{eq: multitask algorithm} with respect to $\bw_e^\star$ can be deduced from its behavior with respect to $\bw_e^o$ using the following relation:
\begin{equation}
	\label{eq: relation between wt and wt'}
	\bwt_e'(i+1)=\bwt_e(i+1)-\bw^\delta_e.
\end{equation}
where $\bw^\delta_e\triangleq\bw^o_e-\bw^\star_e$. Using \eqref{eq: relation between wt and wt'} with \eqref{eq: weight error vector recursion 0}, the fact that $\bw^\star_e$ verifies the constraints $\{\cD_e\bw_e+\bb=\Zero\}$, namely,
\begin{equation}
\cP_e\bw^\star_e-\boldf_e=\bw^\star_e,
\end{equation} 
and the fact that $\cA^\top\One=\One$, we obtain that $\bwt_e'(i+1)$ evolves according to the following recursion:
{\begin{equation}
	\label{eq: weight error vector recursion 0'}
	\boxed{
	\begin{aligned}
	&\bwt_e'(i+1)=\cA^\top\cP_e\big[\bI_{M_e}-\mu\cR_{x,e}(i)\big]\bwt_e'(i)-\\
	&\qquad\mu\cA^\top\cP_e\bp_{zx,e}(i)-\mu\cA^\top\cP_e\cR_{x,e}(i)\bw^\delta_e.
	\end{aligned}}
\end{equation}}
Taking the expectation of both sides of recursion~\eqref{eq: weight error vector recursion 0'}, using Assumption~\ref{assump: independent regressors}, and $\expec\,\bp_{zx,e}(i)=\Zero$, the mean error vector evolves according to:
\begin{equation}
	\label{eq: mean error recursion'}
	\expec\,\bwt_e'(i+1)=\cB\,\expec\,\bwt_e'(i)-\mu\br',
\end{equation}
where $\cB$ is given by~\eqref{eq: B} and
{\begin{equation}
	\label{eq: r'}
	\br'\triangleq\cA^\top\cP_e\cR_{x,e}\bw^\delta_e.
\end{equation}}
Using arguments similar to subsection~\ref{subsec: Mean behavior analysis I}, we find that the multitask diffusion algorithm~\eqref{eq: multitask algorithm} is stable in the mean if the step-size is chosen such that the matrix $\cB$ is stable. The asymptotic mean bias is given by:
\begin{equation}
	\label{eq: bias 2}
	\lim_{i\rightarrow\infty}\expec\,\bwt_e'(i)=-\mu[\bI_{M_e}-\cB]^{-1}\br'.
\end{equation}
Note that the bias depends on the step-size $\mu$ and the vector $\bw^\delta_e=\bw^o_e-\bw^\star_e$. In the next section, we shall {illustrate} with simulation results that $\lim_{i\rightarrow\infty}\|\expec\,\bwt_e'(i)\|^2$ is on the order of $\mu^2$. The bias \eqref{eq: bias 2} is $\Zero$ in two cases: 1) in the perfect model scenario where $\bw^o_k$ satisfy the constraints ($\bw_e^\delta=\Zero$); 2) if each agent is involved in at most one constraint ($\cD_e=\cD'_e=\cD$). In this second case, consider~\eqref{eq: r'} and observe that $\cA=\bI_{M_e}$. Replacing $\bw_e^\delta$ by its expression obtained from \eqref{eq: extended closed form solution 1}, and $\cP_e$ by \eqref{eq: cP e}, yields $\br' =\Zero$.

To obtain the behavior of algorithm~\eqref{eq: multitask algorithm} toward $\bw^\star_e$ in the mean-square sense, we use~\eqref{eq: relation between wt and wt'} to write:
\begin{equation}
\begin{split}
	\label{eq: relation between the variances}
	&\expec\{\|\bwt_e'(i+1)\|^2_{\bSig}\}=\expec\{\|\bwt_e(i+1)\|^2_{\bSig}\}-\\
	&\qquad2\,\expec\{\bwt_e^\top(i+1)\}\bSig\bw_e^\delta+\|\bw^\delta_e\|^2_{\bSig}.
	\end{split}
\end{equation}
The transient and steady-state behaviors of $\expec\{\|\bwt_e'(i)\|^2_{\bSig}\}$ can be derived from the models derived for $\bwt_e(i)$ in the mean and mean-square sense. We shall show with simulation results that the steady-state $\lim_{i\rightarrow\infty}\expec\,\|\bwt_e'(i)\|^2$ is {on} the order of $\mu$. We observed experimentally that modeling the behavior of $\expec\{\|\bwt_e'(i)\|^2_{\bSig}\}$ accurately needs the exact expression of $\cF$. {For zero-mean real valued regressors with $M_k=M_0$ $\forall~k$, the evaluation of $\cF$ leads to (see~Appendix~B):
\begin{equation}
\begin{split}
&\cF=\cB^\top\otimes_b\cB^\top+\\
&\mu^2\sum_{k=1}^N\Big[\left(\cS_k^\top(\bI_{N_e}\otimes\bR_{x,k})\otimes_b\cS_k(\bI_{N_e}\otimes\bR_{x,k})\right)+\\
&\qquad\left(\cS_k^\top\otimes_b\left(\cS_k(\bI_{N_e}\otimes\bR_{x,k})\right)\right)\\
&\left(\bI_{N_e^2}\otimes\vc(\bI_{M_0})\otimes[\vc(\bR_{x,k})]^\top\right)\Big](\cP_e\cA\otimes_b\cP_e\cA),
\end{split}
\end{equation}
where $\cS_k$ is the $N\times N$ block diagonal matrix whose $(k,k)$-th block is equal to $\bC_k\otimes\bI_{M_0}$. }

\section{Simulation results}
\label{sec: Simulation results}
{Throughout this section, the factors $c_{k_m}$ were set to $\frac{1}{j_k}$, and $\N_{k_m}\cap\C_k=\C_k$ for all $m$. We run algorithm~\eqref{eq: multitask algorithm} with uniform combination coefficients $a_{k_n,k_m}=\frac{1}{j_k}$ for all $n$.}

\subsection{Theoretical model validation}

\begin{figure*}[t]
{\centering
	\includegraphics[scale=0.28]{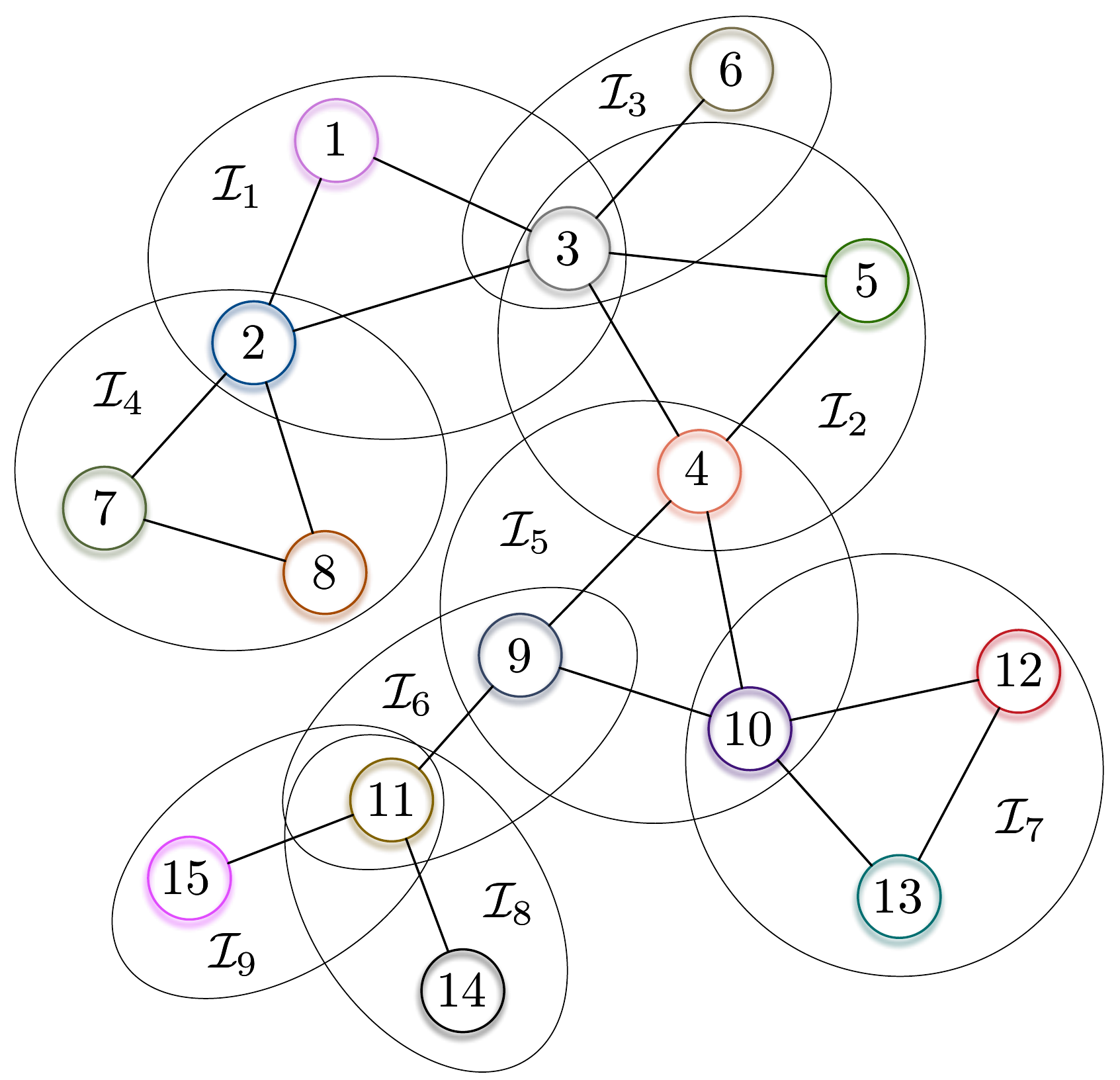}
	\qquad\qquad
	\includegraphics[scale=0.35]{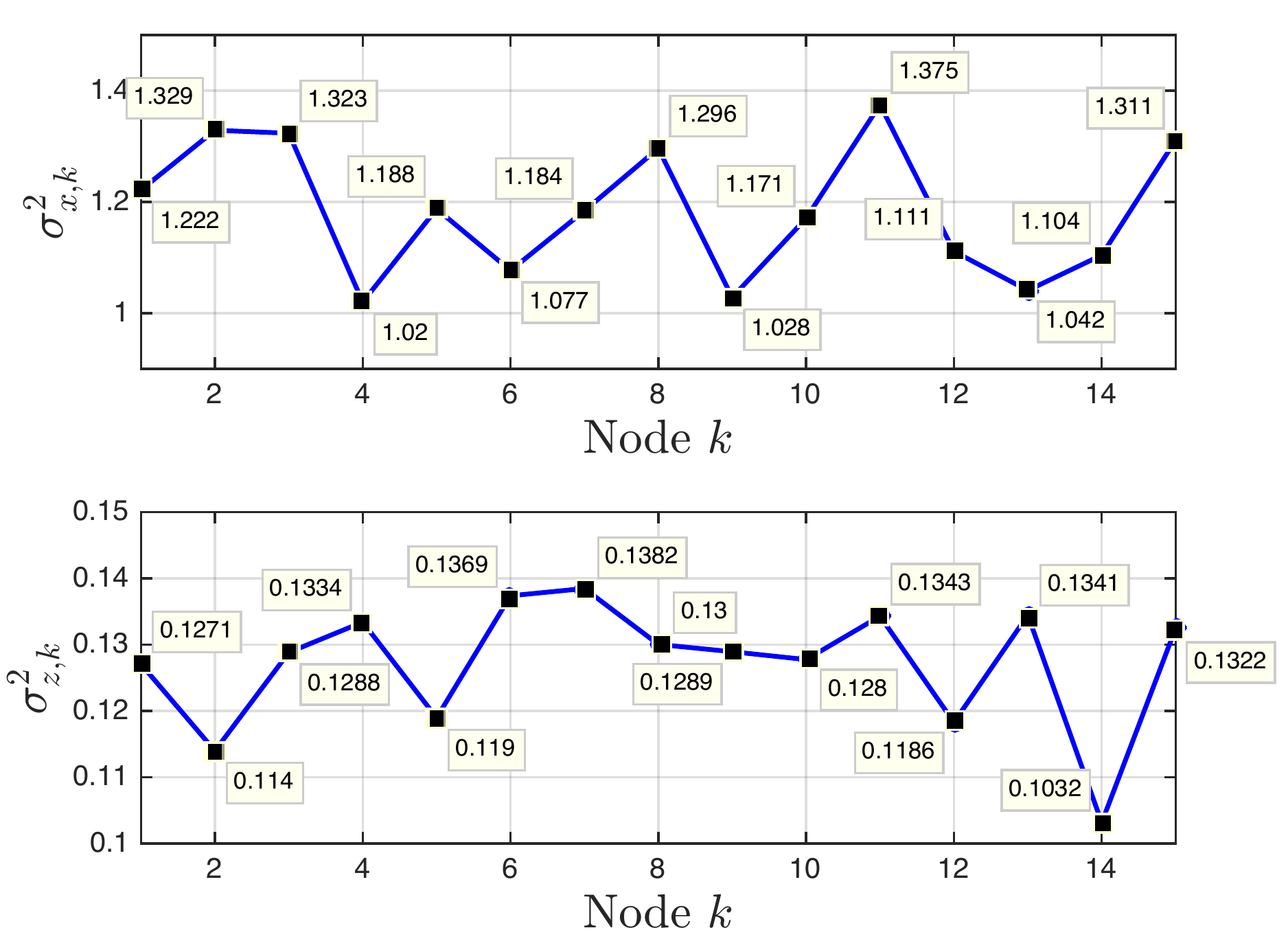}
	\caption{Experimental setup. ({\em Left}) Network topology with constraints. ({\em Right}) Regression and noise variances.}
	\label{fig: experimental setup}}
\end{figure*}

We considered a network consisting of $15$ agents with the topology shown in Fig.~\ref{fig: experimental setup}. The regression vectors $\bx_k(i)$ were $2\times 1$ zero-mean Gaussian distributed with covariance matrices $\bR_{x,k}=\sigma^2_{x,k}\bI_2$. The noises $z_k(i)$ were zero-mean i.i.d. Gaussian random variables, independent of any other signal with variances $\sigma^2_{z,k}$. The variances $\sigma^2_{x,k}$ and $\sigma^2_{z,k}$ are shown in Fig.~\ref{fig: experimental setup}. We randomly sampled $9$ linear equality constraints of the form:
\begin{equation}
	\label{eq:constraints-ref}
	\sum_{\ell\in\cI_p}d_{p\ell}\bw_{\ell}=b_p\cdot\One_{2\times 1},
\end{equation}
where the scalars $d_{p\ell}$ and $b_p$ were randomly chosen from the set $\{-3,-2,-1,1,2,3\}$.  We used a constant step-size $\mu=0.025$ for all agents. The results were averaged over 200 Monte-Carlo runs. 

\begin{figure*}[t]
	\includegraphics[scale=0.333]{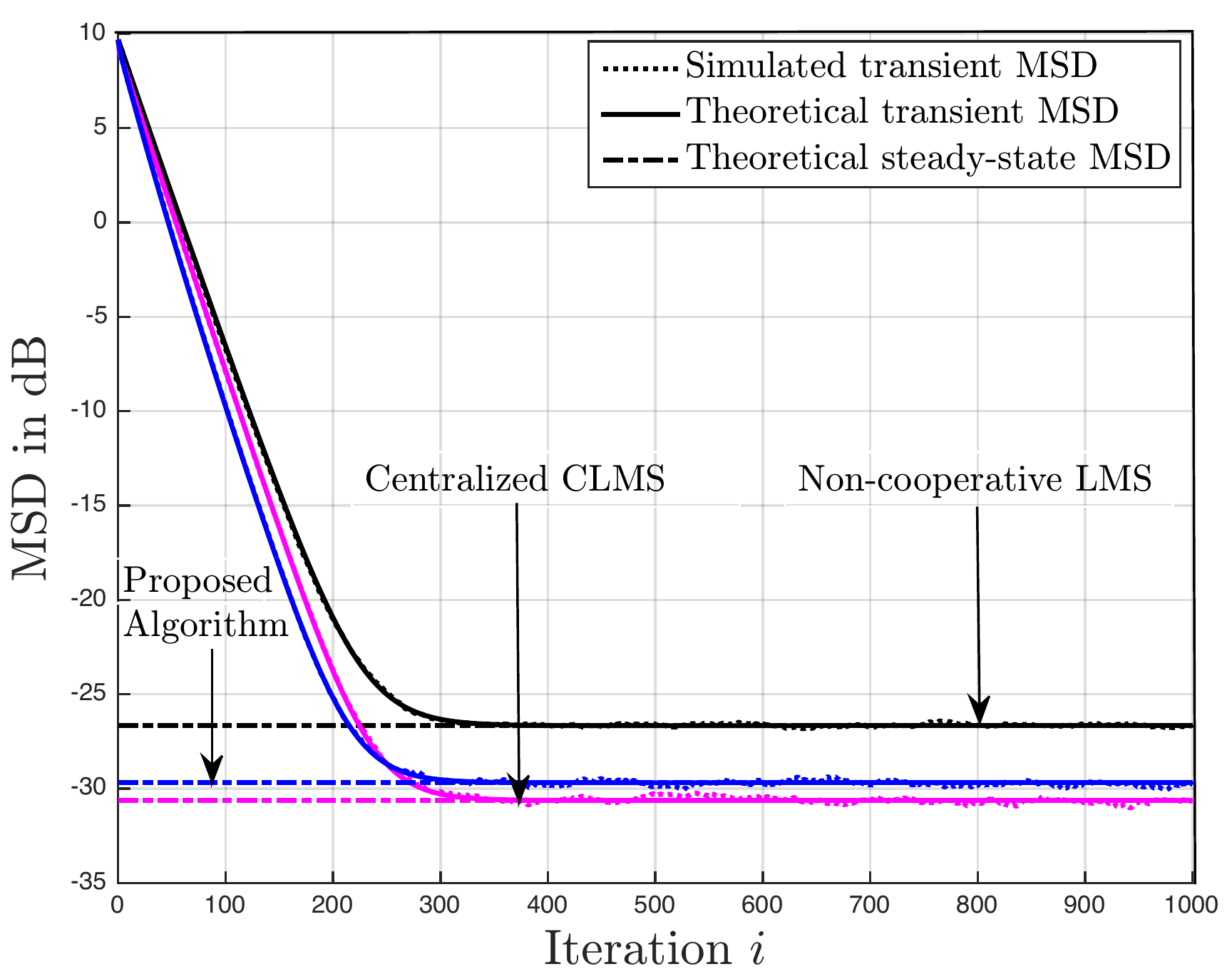}
	\includegraphics[scale=0.333]{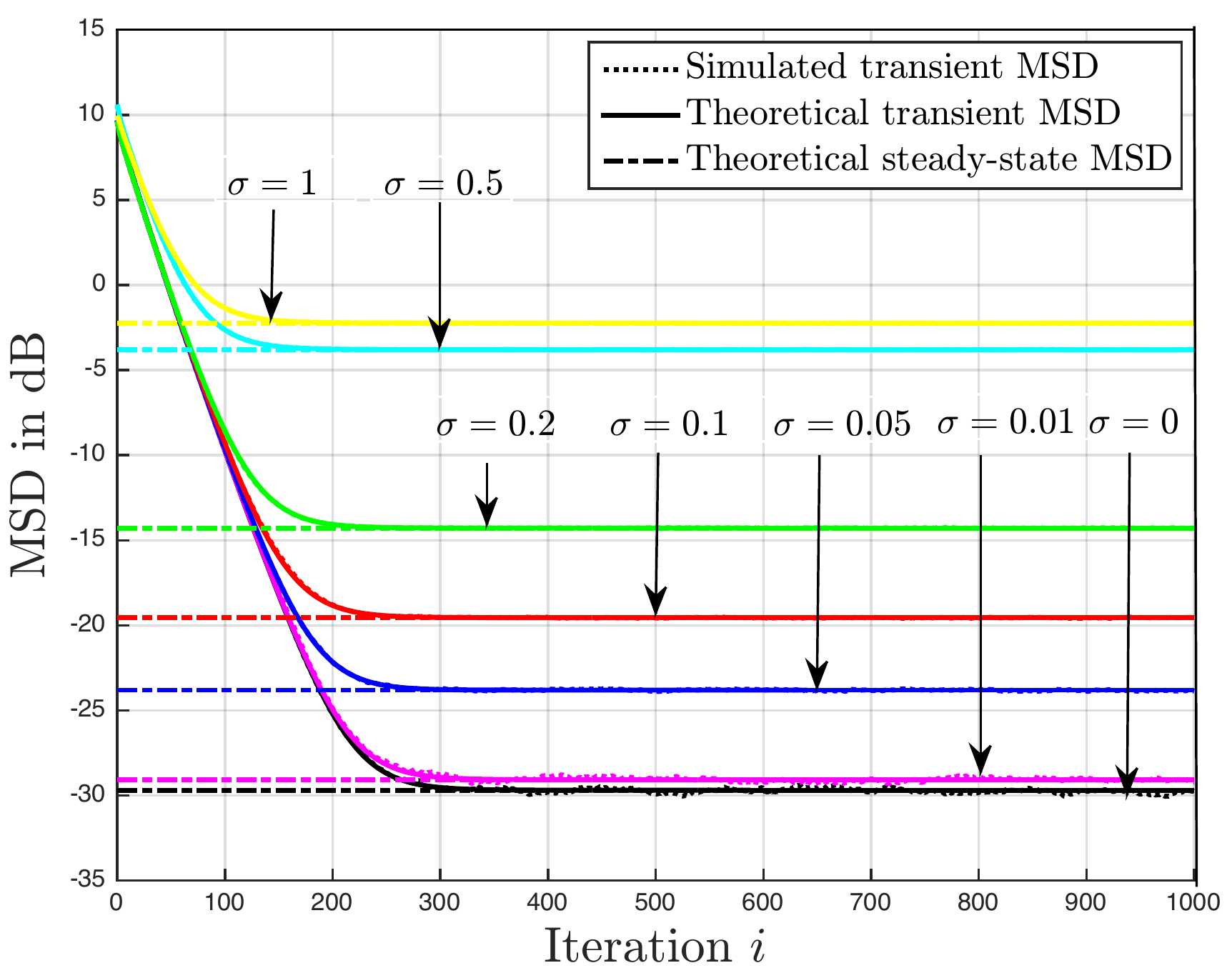}
	\includegraphics[scale=0.333]{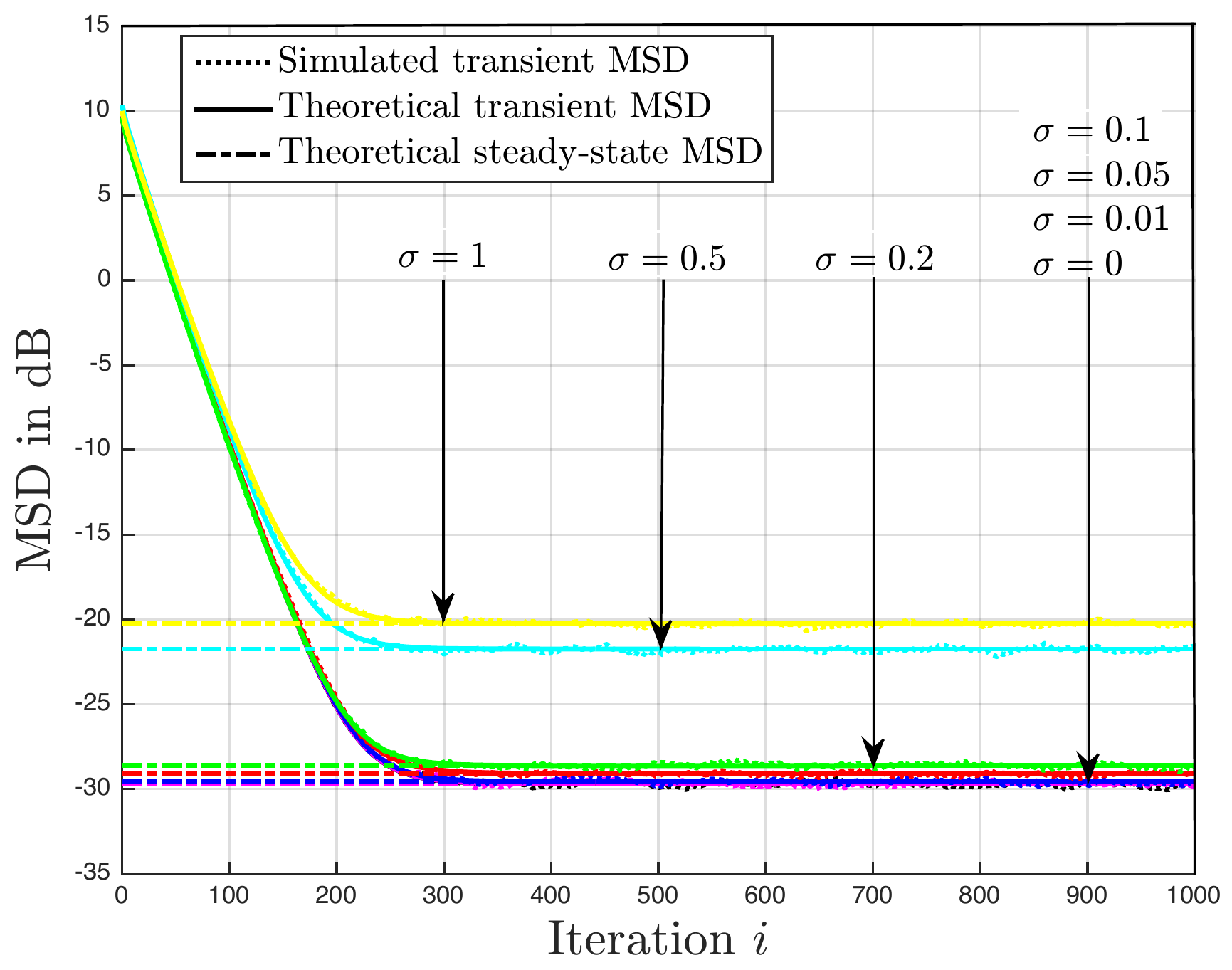}
\caption{({\em Left}) MSD comparison of the non-cooperative LMS, the centralized CLMS, and our multitask algorithm for the perfect model scenario. Learning curves of algorithm~\eqref{eq: multitask algorithm} with respect to $\bw^o_e$ ({\em middle}) and $\bw^\star_e$ ({\em right}) for $6$ different values of $\sigma$.}
\label{fig: MSD comparisons}
\end{figure*}

First, we considered the case of a perfect model scenario where the observation parameter vector $\bw^o$ satisfies the equality constraints, i.e., $\bw^\star=\bw^o$.
In Fig.~\ref{fig: MSD comparisons} (left), we compare three algorithms: the non-cooperative LMS algorithm (obtained from~\eqref{eq: CLMS} by setting $\cP=\bI_{M}$ and $\boldf=\Zero$), the centralized CLMS algorithm~\eqref{eq: CLMS} which assumes that the constraints are available at the fusion center, and algorithm~\eqref{eq: multitask algorithm}. For each algorithm, we report the theoretical transient MSD, the theoretical steady-state MSD, and the simulated MSD. We observe that the simulation results match well the actual performance. Furthermore, the network MSD is improved by promoting relationships between tasks. Finally, our algorithm performs well compared to the centralized solution.

{Next,} we perturbed the optimum parameter vector $\bw^o$ as follows:
\begin{equation}
	\label{eq: optimum parameter vector perturbation}
	\bw^o_\text{pert}=\bw^o+\bu^o,
\end{equation}
so $\bw^o_\text{pert}$ does not satisfy the constraints \eqref{eq:constraints-ref}. The entries of $\bu^o$ were sampled from Gaussian distribution $\N(0,\sigma^2)$. We evaluated algorithm~\eqref{eq: multitask algorithm} on $6$ different setups characterized by $\sigma\in\{0,0.01,0.05,0.1,0.2,0.5,1\}$. The theoretical and simulated learning curves with respect to $\bw^o_e$ and $\bw^\star_e$ are reported in Fig.~\ref{fig: MSD comparisons}. Observe that the performance with respect to $\bw^o_e$ highly deteriorates when $\sigma$ increases. However, even for the largest values of $\sigma=1$, algorithm~\eqref{eq: multitask algorithm} still performs well with respect to the solution $\bw^\star_e$ of the optimization problem with constraints.

\begin{figure*}
{\centering
	\includegraphics[scale=0.325]{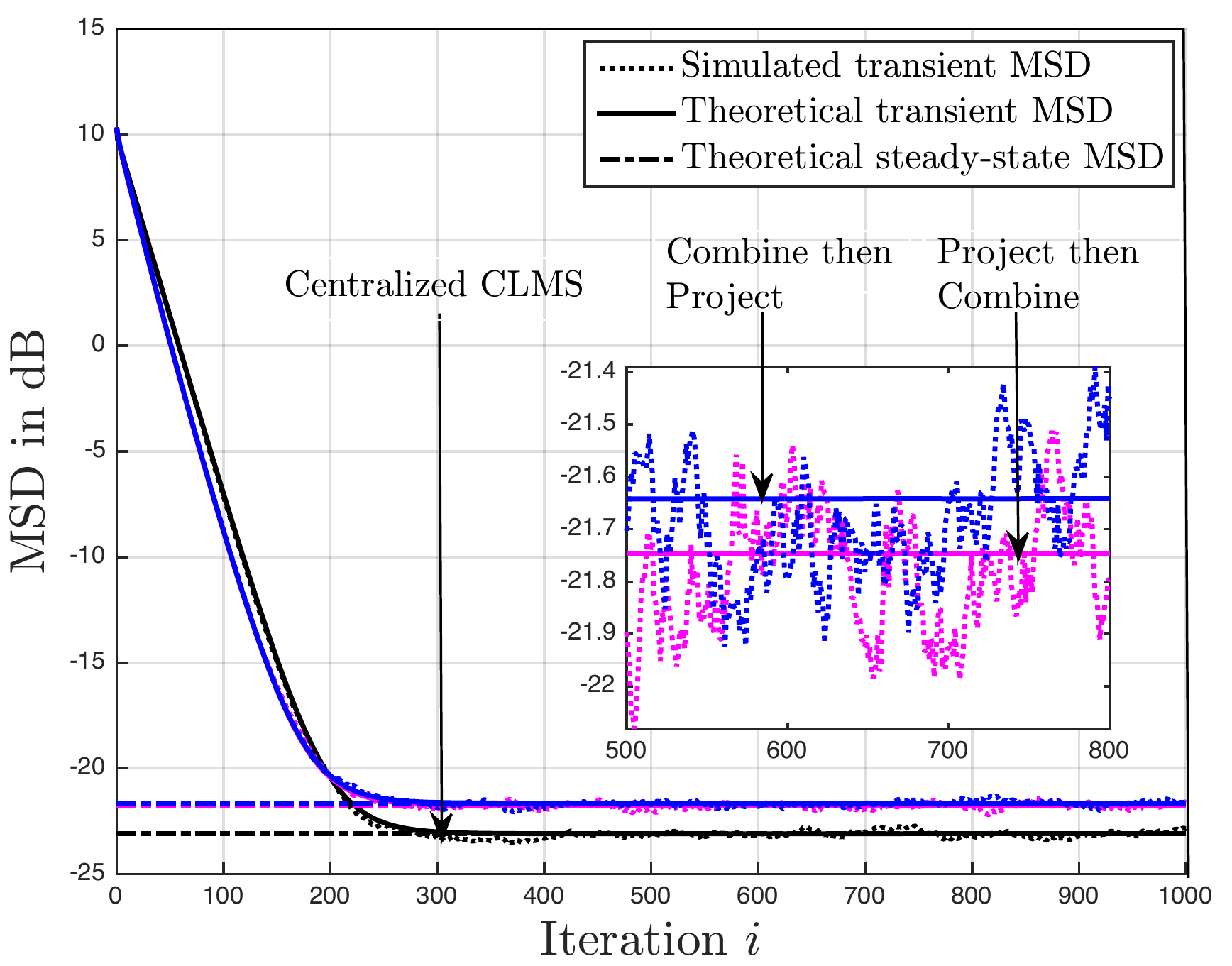}
	\qquad\qquad\includegraphics[scale=0.325]{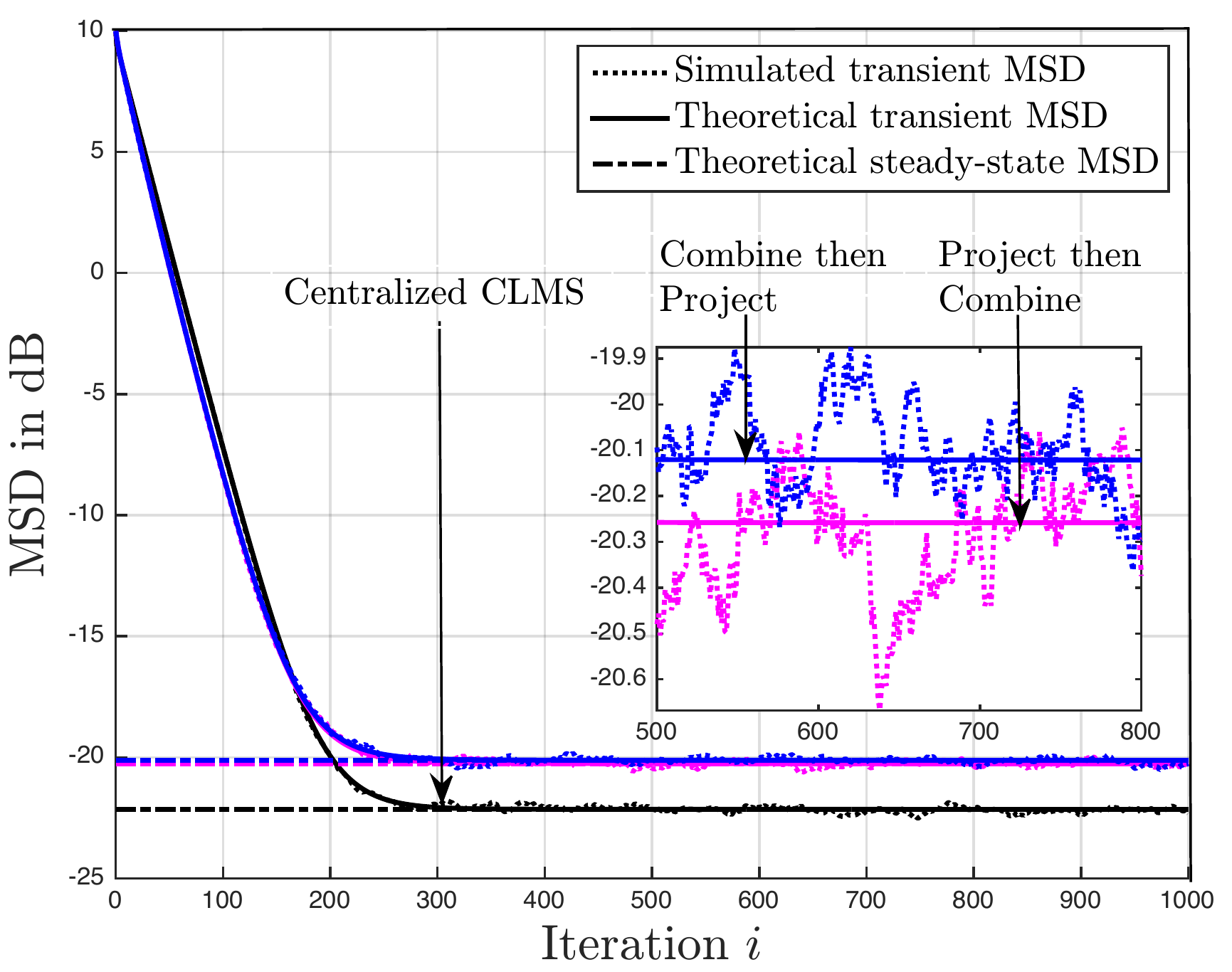}
\caption{Network MSD comparison with respect to $\bw^\star$, $\bw^\star_e$ of the centralized CLMS~\eqref{eq: CLMS}, algorithm~\eqref{eq: multitask algorithm}, and stochastic version of algorithm~\eqref{eq: diffusion iterates in the absence of gradient noise} for $\sigma=0.5$ (\emph{left}) and $\sigma=1$ (\emph{right}).}
\label{fig: MSD comparisons CLMS}}
\end{figure*}

For comparison purposes, we illustrate in Fig.~\ref{fig: MSD comparisons CLMS} the theoretical and simulated learning curves with respect to $\bw^\star$ for the settings where $\sigma=0.5$ (left) and $\sigma=1$ (right) of the centralized CLMS algorithm~\eqref{eq: CLMS}, algorithm~\eqref{eq: multitask algorithm} where the sub-nodes ``project-then-combine'', and the stochastic version (obtained by replacing the moments by instantaneous approximations) of algorithm~\eqref{eq: diffusion iterates in the absence of gradient noise} where the sub-nodes ``combine-then-project" (Appendix~C explains how the performance of this algorithm can be obtained). Observe that both algorithms~\eqref{eq: multitask algorithm} and the stochastic version of~\eqref{eq: diffusion iterates in the absence of gradient noise} have approximately the same performance. However, with the settings considered in this section, algorithm~\eqref{eq: multitask algorithm} is less complex than algorithm~\eqref{eq: diffusion iterates in the absence of gradient noise} as explained in subsection~\ref{subsec: Distributed solution}. Furthermore, we observe that the larger the vector $\bw^\delta_e$ is, the larger the performance gap between the centralized solution and the distributed solutions is. This is due to the bias~\eqref{eq: bias 2} induced in the distributed solution which does not exist in the centralized CLMS algorithm (see~Appendix~C). In order to characterize the constraints violation at the sub-nodes for the setting where $\sigma=0.5$, we evaluate the steady-state quantity $\|\cD'_e\bw_e(\infty)+\bb'\|^2$ where $\cD'_e$ and $\bb'$ are given by~\eqref{eq: D'_e and b'}  and $\bw_e(\infty)\triangleq\lim_{i\rightarrow\infty}\bw_e(i)$. When the sub-nodes project first and then combine, we obtain $\|\cD'_e\bw_e(\infty)+\bb'\|^2=\|\cD_e\bw_e(\infty)+\bb\|^2=0.0264$. On the contrary, when the sub-nodes combine first and then project, we obtain $\|\cD'_e\bw_e(\infty)+\bb'\|^2=\|\cH\bw_e(\infty)\|^2=0.0072$. Thus, at the expense of a higher computational complexity, the constraints violation, measured by $\|\cD'_e\bw_e(\infty)+\bb'\|^2$, is smaller when the projection step is performed after the combination step\footnote{We show in Appendix~D that, for the perfect model scenario, the steady-state MSD is lower when the combination step is the last step in the algorithm.}. 

\begin{figure*}[h]
{\centering
	\includegraphics[scale=0.325]{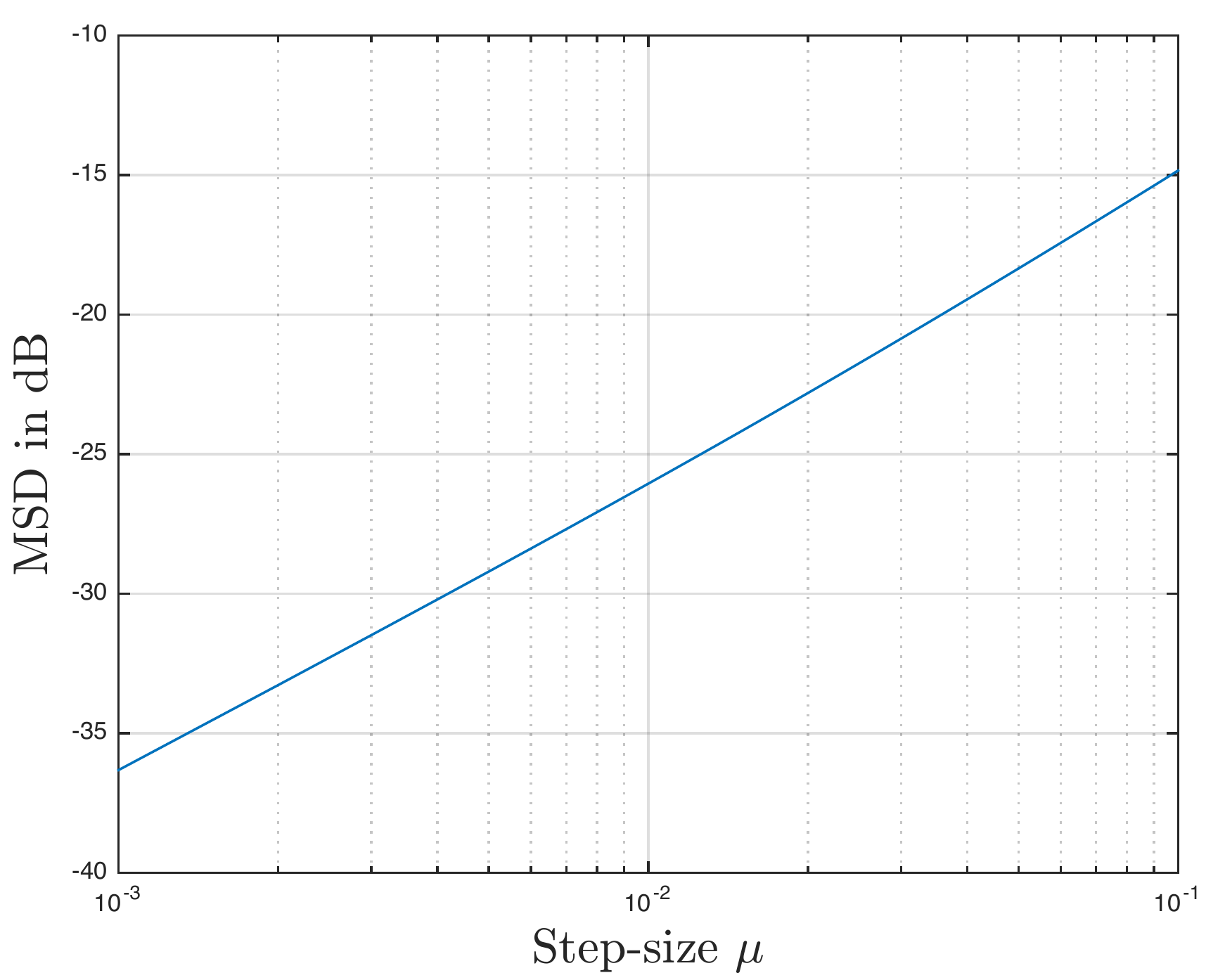}
	\qquad\qquad
	\includegraphics[scale=0.325]{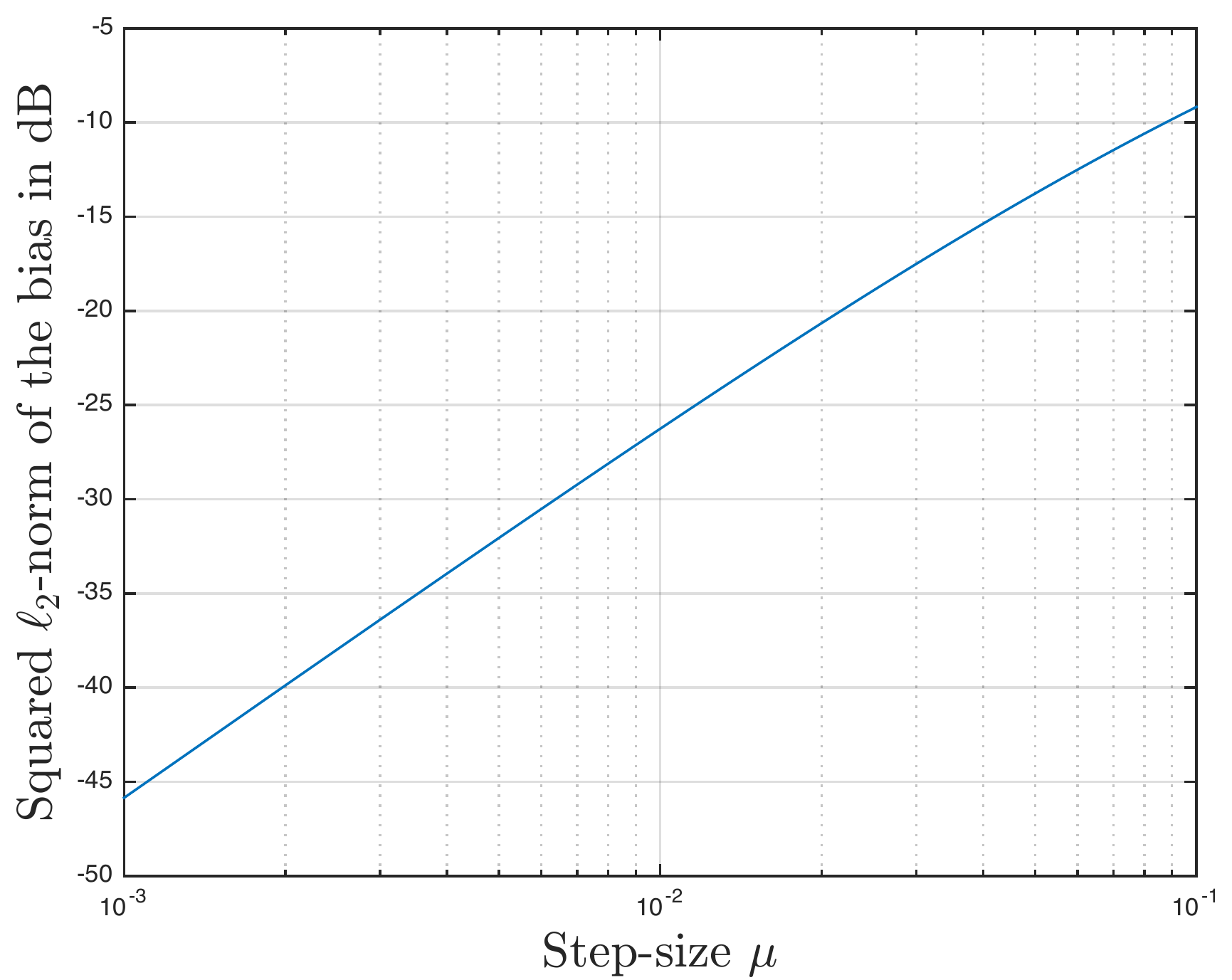}
\caption{Influence of the step-size $\mu$ on the performance of the algorithm. ({\em Left}) Network steady-state MSD for different values of $\mu$. ({\em Right}) Squared norm of the bias, i.e, $\lim\limits_{i\rightarrow\infty}\|\expec\,\bwt'(i)\|^2$, for different values of $\mu$.}
\label{fig: variable step-size}}
\end{figure*}

In order to characterize the influence of the step-size $\mu$ on the performance of algorithm~\eqref{eq: multitask algorithm}, Fig.~\ref{fig: variable step-size} (left) reports the theoretical steady-state MSD with respect to $\bw^\star_e$ for different values of $\mu$. We observe that the network MSD increases $10$ dB per decade (when the step-size goes from $\mu_1$ to $10\mu_1$). This means that the steady-state MSD is {on} the order of $\mu$. Fig.~\ref{fig: variable step-size} (right) reports the squared norm of the bias~\eqref{eq: bias 2} for different values of $\mu$. We note that it increases approximately $20$ dB per decade. This shows that, as expected, this quantity is {on} the order of $\mu^2$.

\begin{figure*}[h]
{\centering
\includegraphics[scale=0.33]{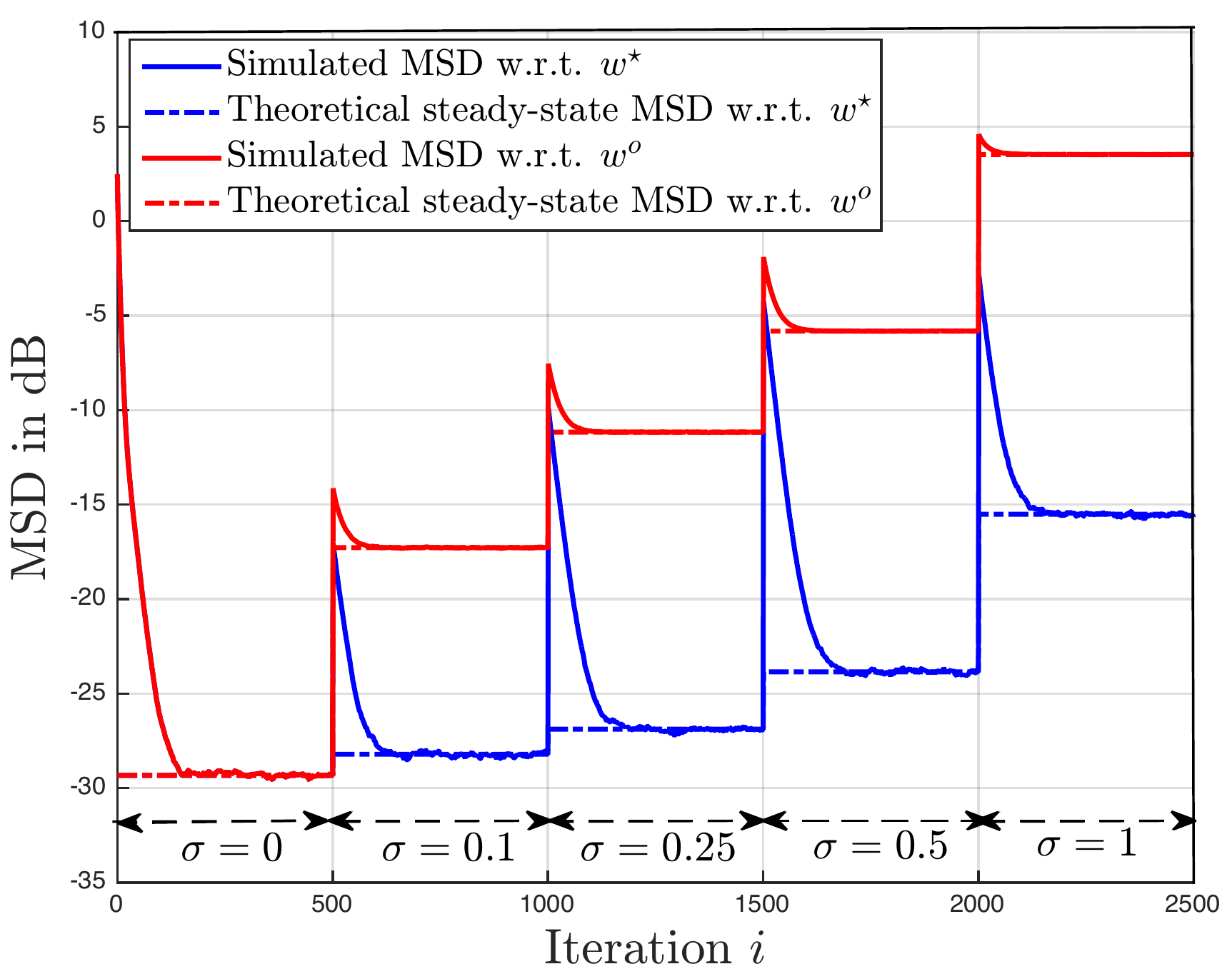}
\qquad\qquad
\includegraphics[scale=0.33]{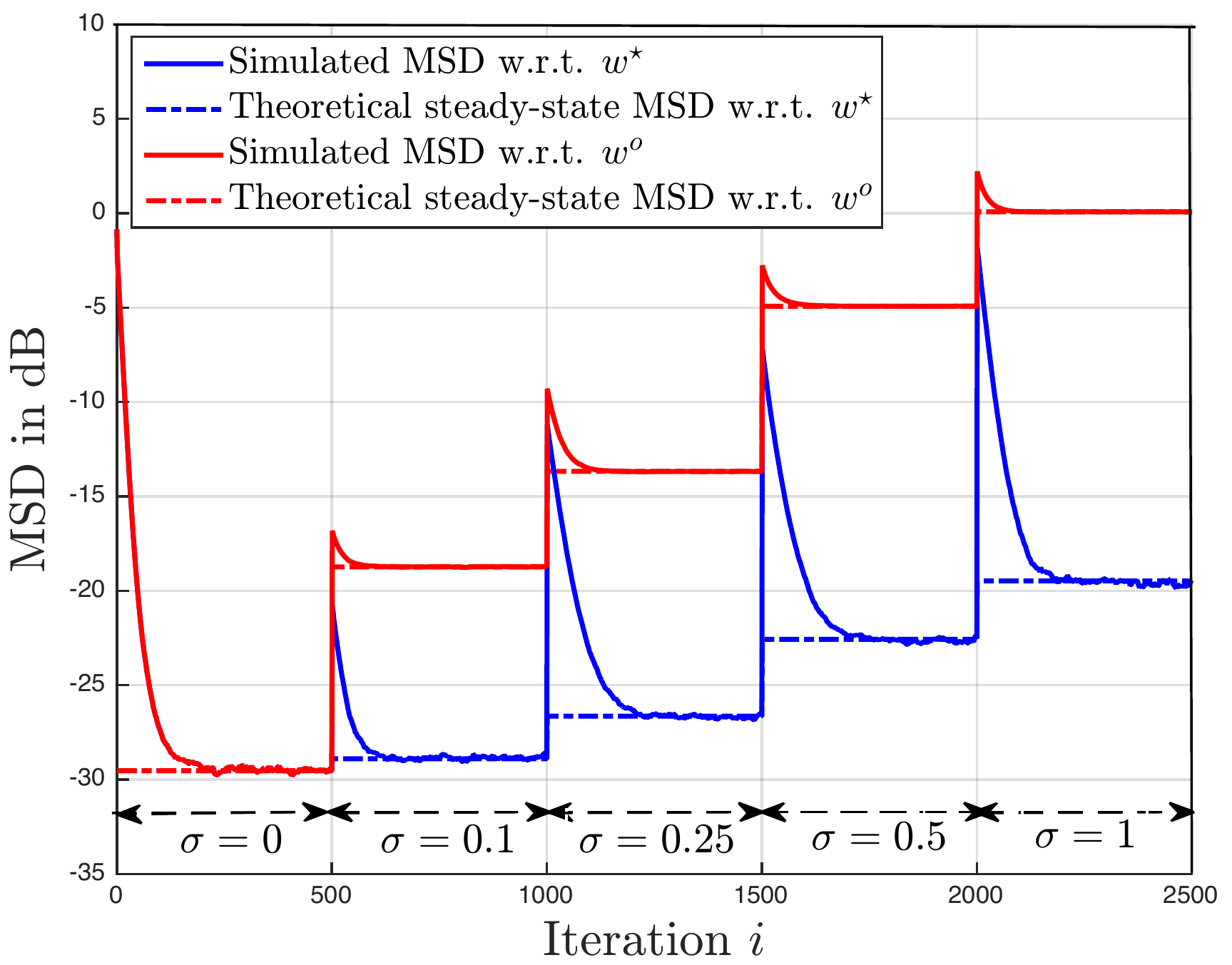}
\caption{Tracking ability of the algorithm for two sets of linear equality constraints. ({\em Left}) $\sigma^2_D=0.01$. ({\em Right}) $\sigma^2_D=1$.}
\label{fig: perturbed D}}
\end{figure*}

Next, we considered the case of non-diagonal matrices $\bD_{p\ell}$ defined as:
\begin{equation}
	\bD_{p\ell}=d_{p\ell}\bI_2+\boldsymbol{\Delta}_{p\ell}
\end{equation}
Parameters $d_{p\ell}$ were randomly selected as in~\eqref{eq:constraints-ref}. The entries of the $2\times 2$ matrix $\boldsymbol{\Delta}_{p\ell}$ were sampled from Gaussian distribution $\N(0,\sigma^2_{D})$. As shown in Fig.~\ref{fig: perturbed D}, the variance $\sigma^2_{D}$ was set to $0.01$ (left) and $1$ (right). To test the tracking ability of algorithm~\eqref{eq: multitask algorithm}, we also perturbed the parameter vector $\bw^o$ as in~\eqref{eq: optimum parameter vector perturbation} by increasing $\sigma^2$ every 500 iterations. In both cases, i.e., $\sigma^2_{D}=0.01$ and $\sigma^2_{D}=1$, $\bw^o$ in~\eqref{eq: optimum parameter vector perturbation} was set to satisfy the equality constraints defined by $\bD_{p\ell}$. We observe that the theoretical models match well the actual performance whatever the constraints are. Furthermore, algorithm~\eqref{eq: multitask algorithm} adapts its response to drifts in the location of $\bw^\star$ when $\bw^o$ changes over time. 

\subsection{Optimal network flow}

As briefly discussed in the Introduction, we shall now consider the minimum-cost flow problem over the network with topology shown in Fig.~\ref{fig: network flow problem - intro}. We are interested in online distributed learning where each node~$k$ seeks to estimate the entering and leaving flows $f_j$ from noisy measurement $s_k(i)$ of the external source, by relying only on local computations and communications with its neighbors.

Let $M_k$ be the number of flows to be estimated at node $k$. We denote by $\bw_k$ the $M_k\times 1$ parameter vector containing the flows $f_j$ entering and leaving node $k$, negatively and positively signed, respectively. For instance, for nodes $1$ and $2$, we have:
\begin{equation}
	\bw_1\triangleq[f_1~f_2]^\top \quad \bw_2\triangleq[-f_1~f_3~f_4~f_5]^\top
\end{equation}
From the flow conservation principle, the noisy measurement $s_k(i)$ can be related to $\bw_k(i)$ as follows:
\begin{equation}
	s_k(i)={\bf{1}}_{M_k \times 1}^\top\bw_k+z_k(i),
\end{equation}
with $z_k(i)$ a zero-mean measurement noise, and ${\bf{1}}_{M_k \times 1}$ an $M_k \times 1$ vector of ones. We consider the bi-objective problem consisting of minimizing $\expec\,|z_k(i)|^2$ and the cost network flow. We shall assume that the cost for flow through an arc is quadratic in the flow, as in applications such as electrical network monitoring and urban traffic control~\cite{ahuja1993network,ventura1991computational}. We formulate the estimation problem as follows:
\begin{equation}
	\label{eq: optimization problem of the network flow}
	\begin{split}
		&\minimize\limits_{\bw_1,\ldots,\bw_N}~
		\sum_{k=1}^N \Big(\expec~|s_k(i)-{\bf{1}}_{M_k \times 1}^\top\bw_k|^2+\frac{\eta}{2}\|\bw_k\|^2\Big),\\
		&\st~[\bw_k]_{f({k,\ell})}+[\bw_\ell]_{f({\ell,k})}=0,~ \ell\in\N_k,~\text{for all } k,
	\end{split}
\end{equation}
where $[\bw_p]_{f(p,q)}$ returns the flow entry in $\bw_p$ that node $p$ has in common with node $q$, and $\eta$ is a tuning parameter to trade off between both objectives.

\begin{figure}[h]
	{\centering
	\includegraphics[scale=0.33]{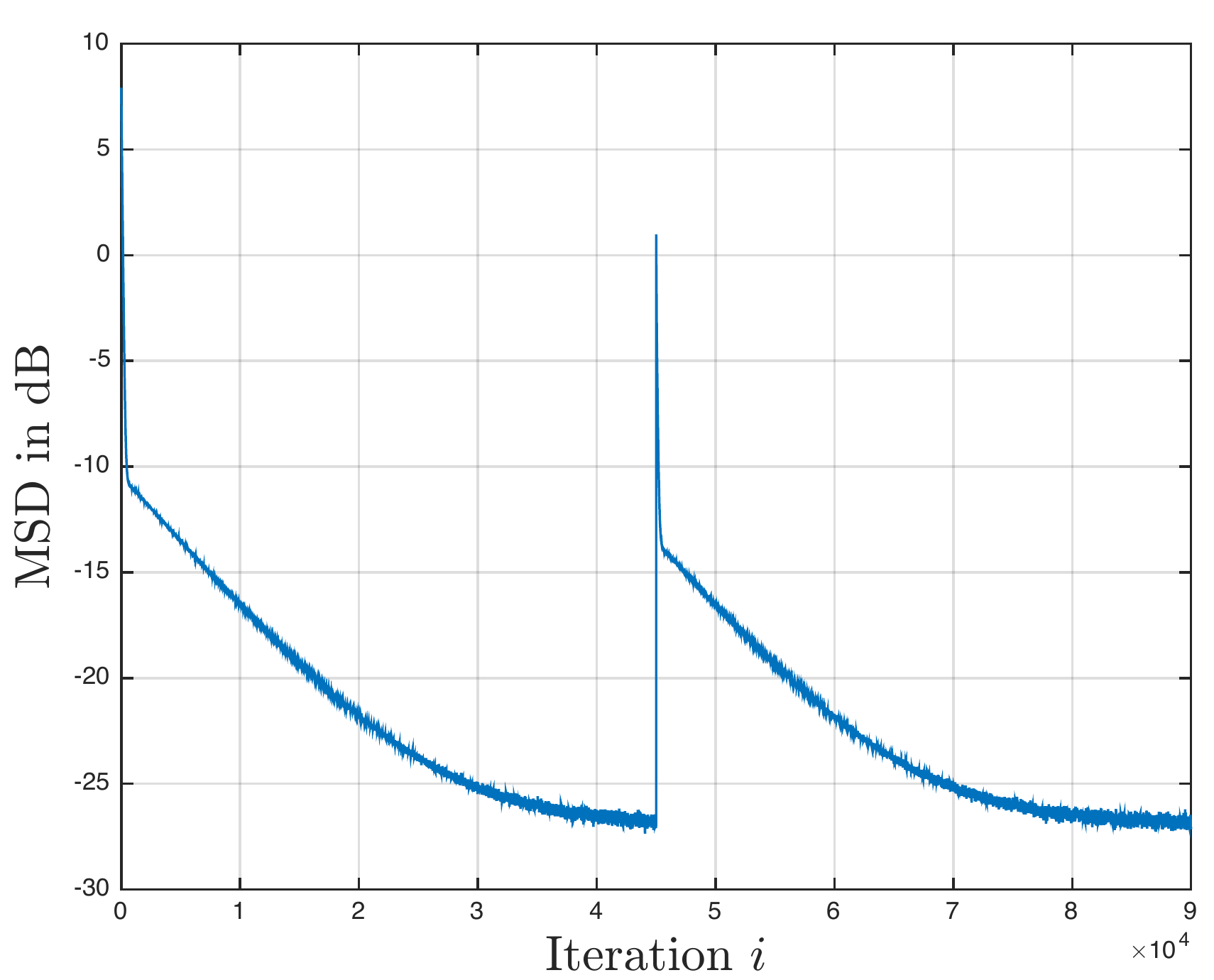}
	\caption{MSD performance and tracking ability of algorithm~\eqref{eq: multitask algorithm} for the minimum cost network flow problem.}
	\label{fig: network flow problem MSD}}
\end{figure}

\begin{figure*}[h]
	{\centering
	\includegraphics[scale=0.3]{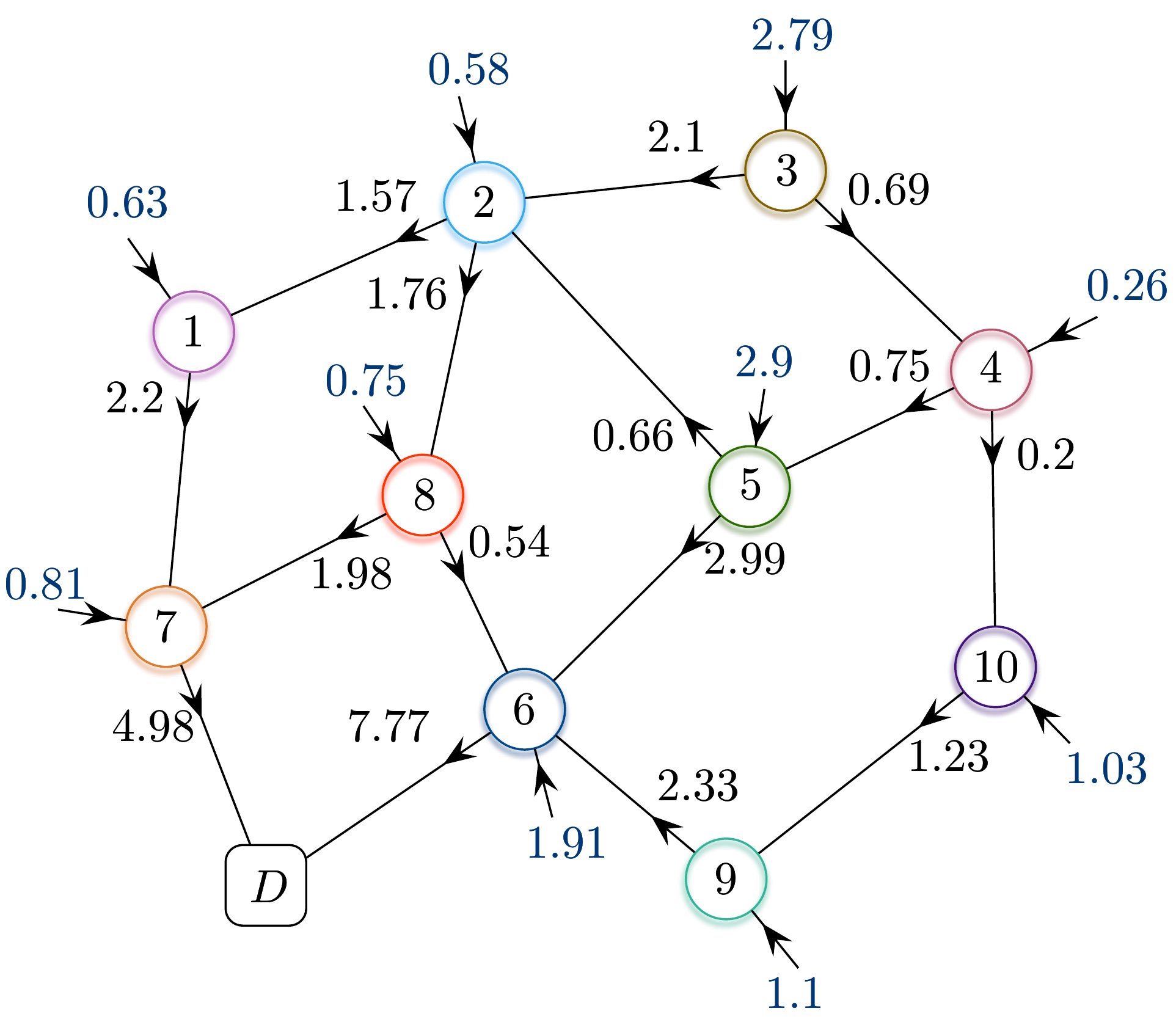}
	\includegraphics[scale=0.3]{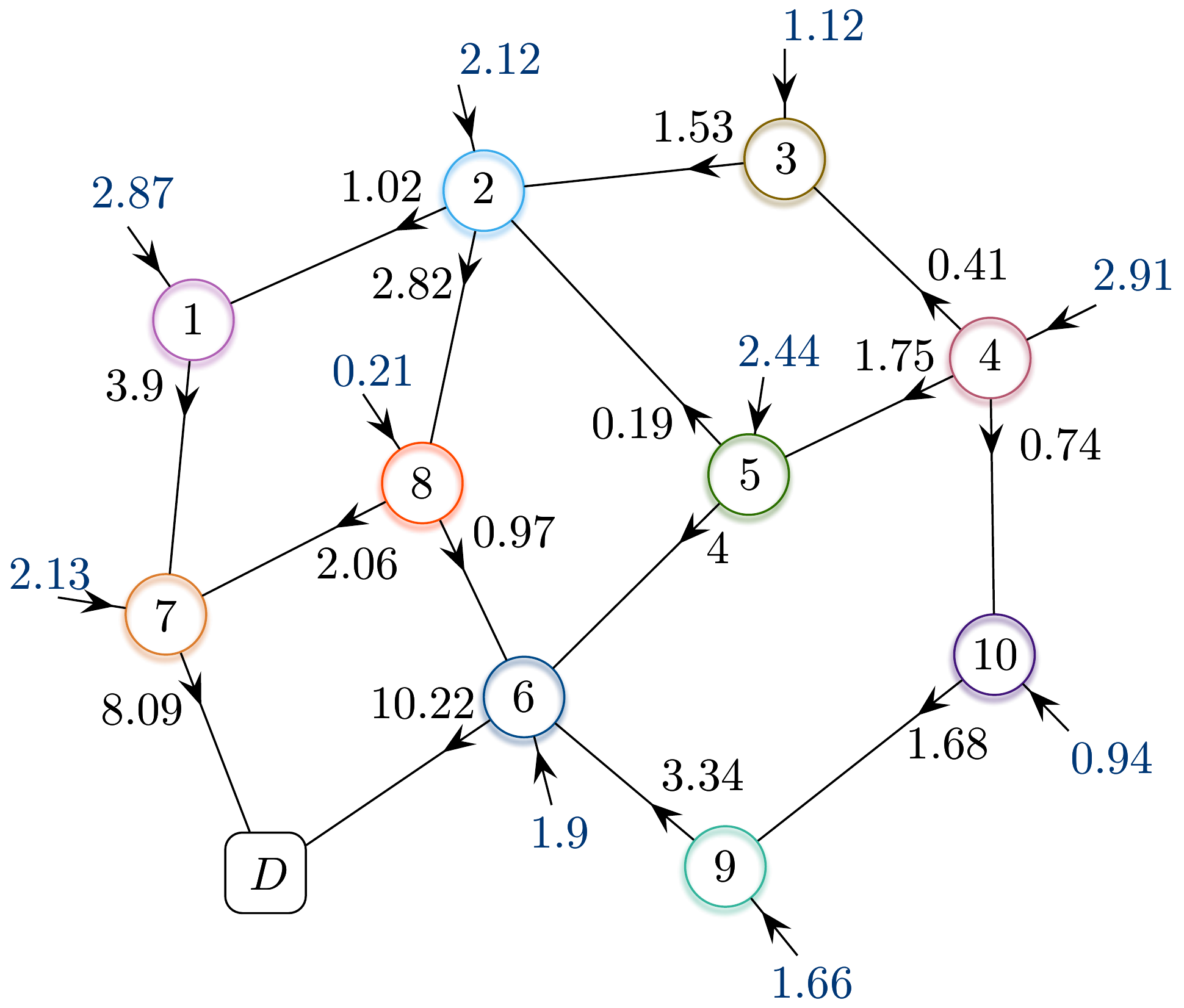}
	\includegraphics[scale=0.3]{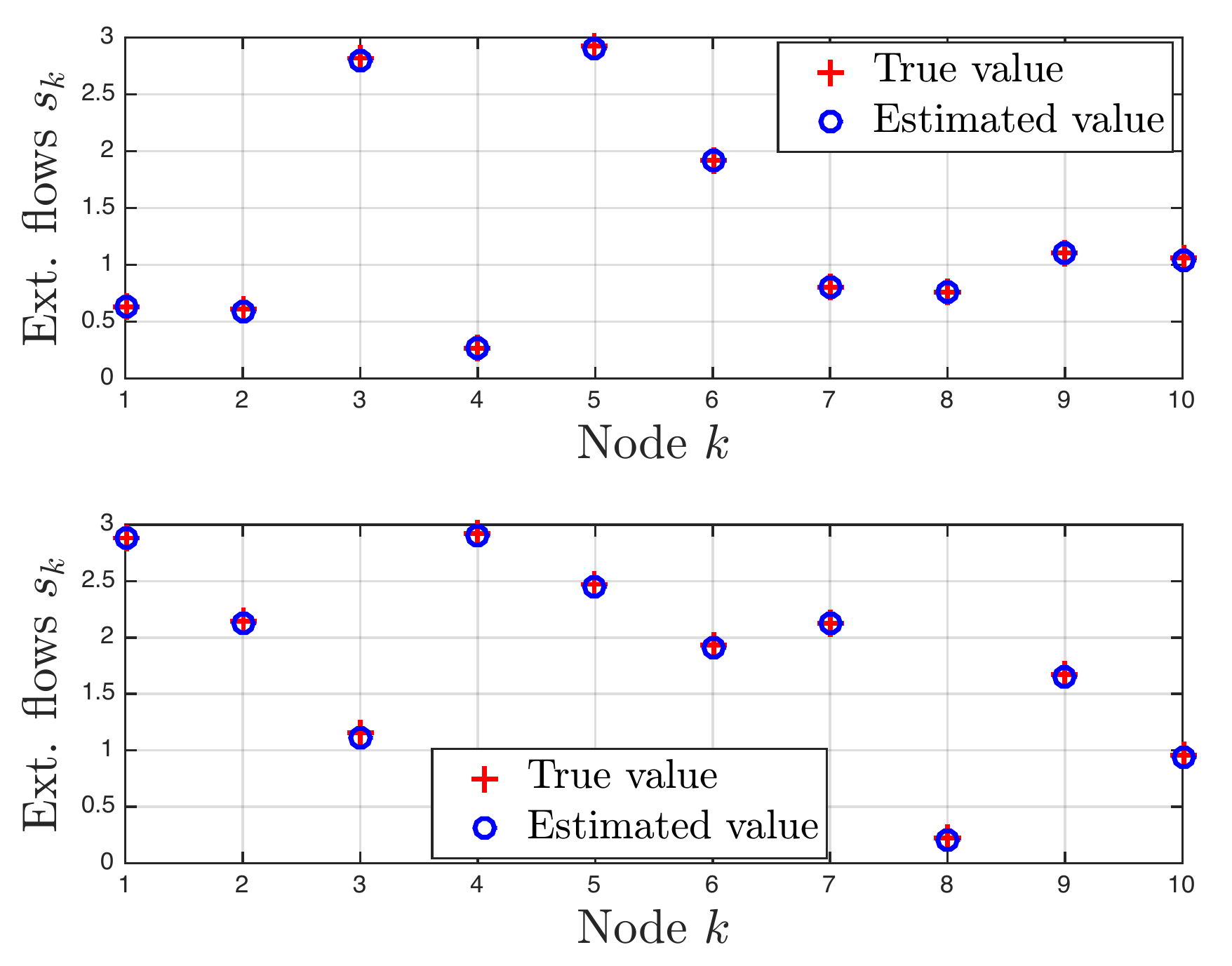}
	\caption{Estimated network flows. ({\em Left}) First experiment. ({\em Middle}) Second experiment. A rounding to 2 decimal places is adopted when visualizing the estimated flows.
	({\em Right}) Comparison of the true and estimated flows $s_k$ (top: first experiment, bottom: second experiment).}
	\label{fig: network flow results}}
\end{figure*}

For each agent $k$, the external flow $s_k$ and the variance $\sigma^2_{z,k}$ of the Gaussian noise $z_k(i)$ were randomly generated from the uniform distributions $\mathcal{U}(0,3)$ and $\mathcal{U}(0.1,0.14)$, respectively. In order to solve the multitask problem~\eqref{eq: optimization problem of the network flow} in a fully distributed manner, we applied algorithm~\eqref{eq: multitask algorithm} by modifying the adaptation step according to:
\begin{equation}
\label{eq: leaky LMS}
\begin{split}
	\bpsi_{k_m}(i+1)&=\bw_{k_m}(i)+\mu\,c_{k_m}\boldsymbol{1}_{M_k \times 1}[s_k(i)-\boldsymbol{1}_{M_k \times 1}^\top\bw_{k_m}(i)]\\
	&\qquad-\frac{\mu}{2} c_{km}\eta\,\bw_{k_m}(i),
\end{split}
\end{equation}
and setting $\mu=0.2$ and $\eta=0.002$. \cblue{Note that equation~\eqref{eq: leaky LMS} leads to a leaky-LMS version of the proposed algorithm. It is well-known that the leaky-LMS algorithm introduces a bias compared to the LMS, but improves its robustness against the so-called weight-drift problem of the LMS algorithm~\cite{sayed2011adaptive}.} In order to test the tracking ability of the algorithm, the external flow $s_k$ at each node $k$ was re-generated from $\mathcal{U}(0,3)$ after $45000$ iterations. The MSD learning curve with respect to the solution of problem~\eqref{eq: optimization problem of the network flow} is reported in Fig.~\ref{fig: network flow problem MSD}. This result was obtained by averaging over 150 Monte-Carlo runs. This figure shows that our strategy was able to solve the minimum-cost flow problem in a fully distributed manner. The estimated flows over the network for both settings considered in the tracking experiment are showed in Fig.~\ref{fig: network flow results} (left and middle). Note that the direction of the estimated flow between nodes $3$ and $4$ is reversed. The true and estimated flows are reported in Fig.~\ref{fig: network flow results} (right) for both settings.
\vspace{-3mm}
\subsection{Numerical solution of a two-dimensional process}

\begin{figure*}[t]
	{\centering
	\subfigure[An $n\times n$ grid network for the solution of Poisson's equation]{\includegraphics[scale=0.3]{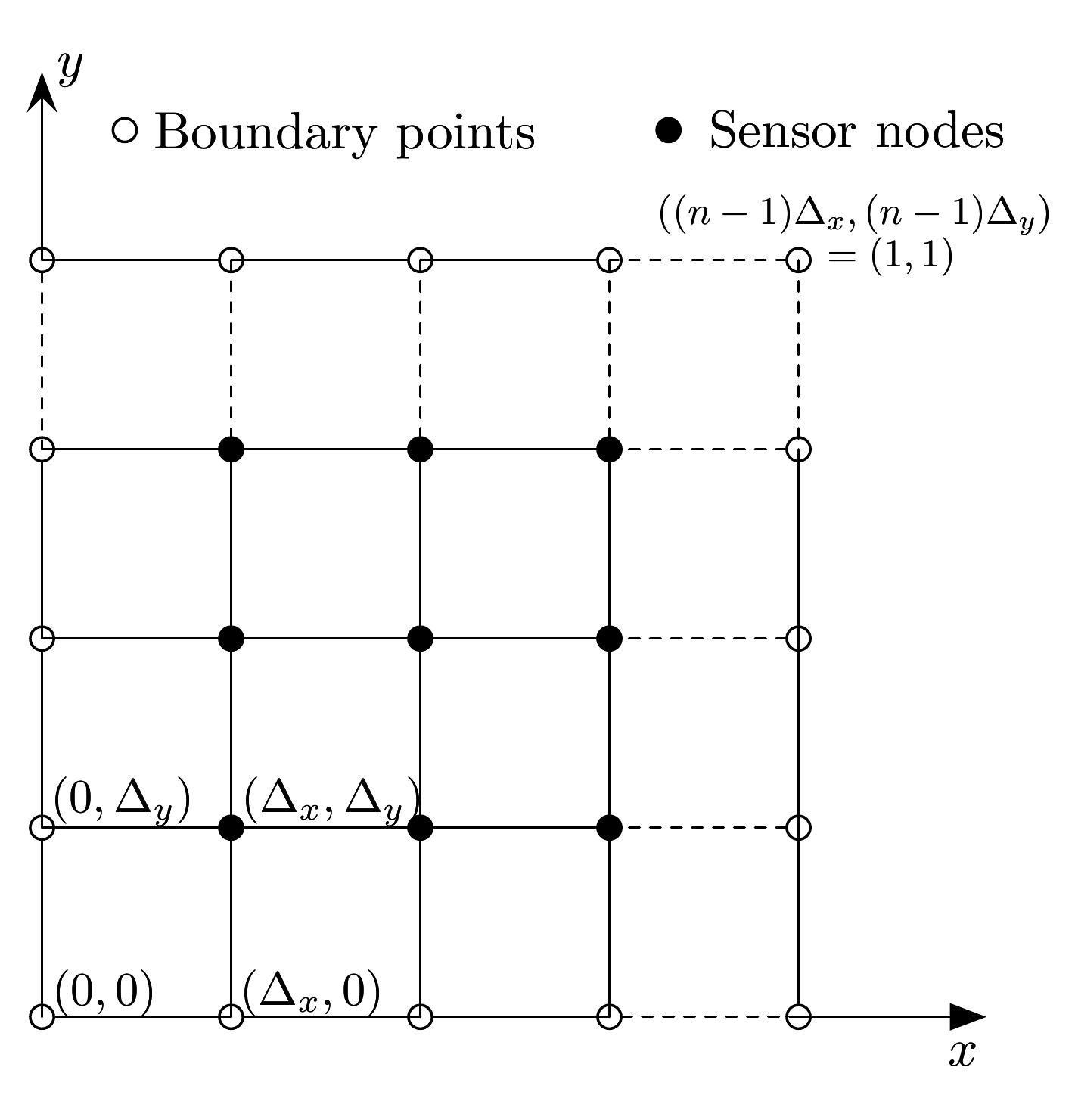}}
	\subfigure[$f(x,y)=(1-x^2)(2y^3-3y^2+1)$]{\includegraphics[scale=0.3]
	{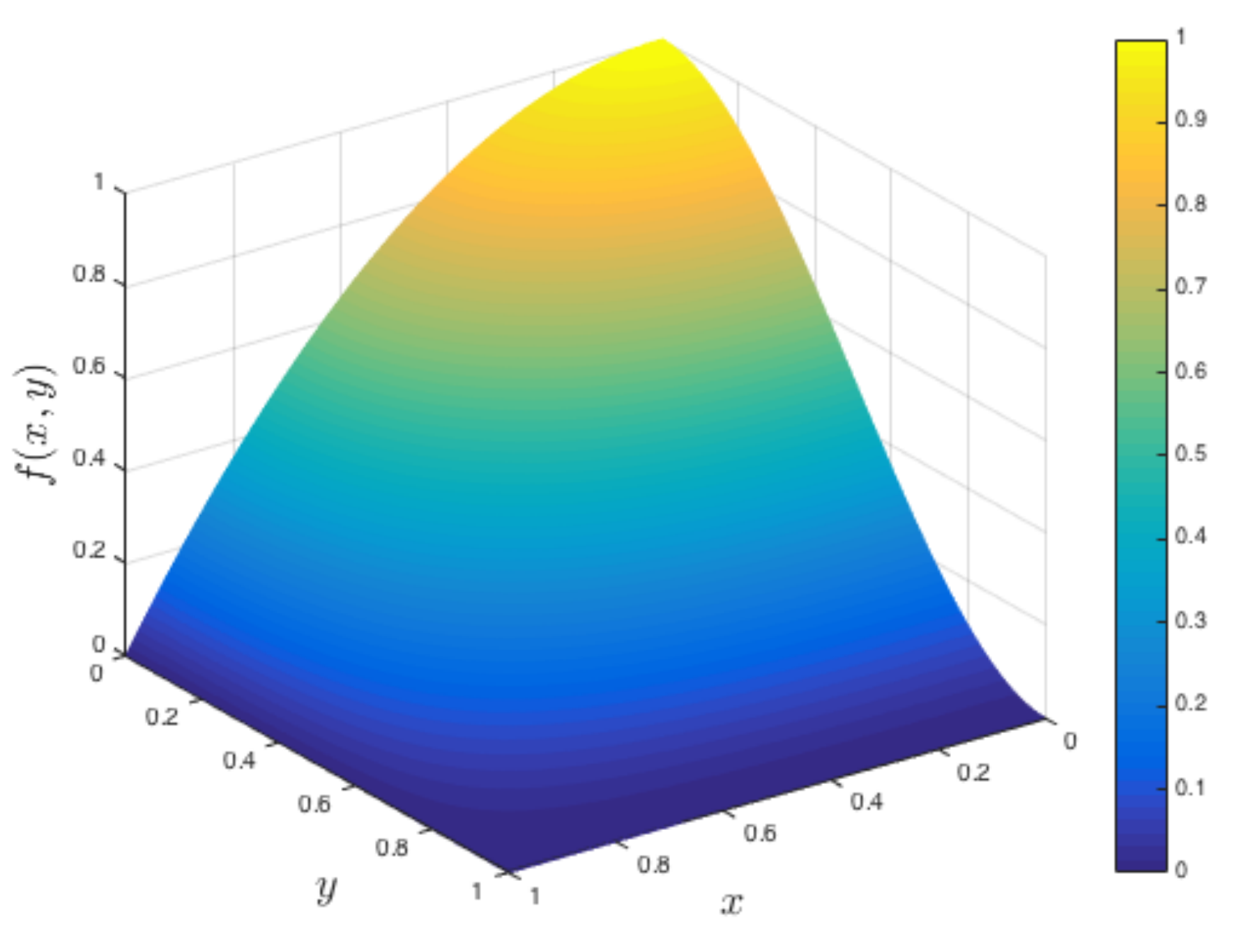}}
	\subfigure[$g(x,y)=-2(2y^3-3y^2+1)+6(1-x^2)(2y-1)$]	{\includegraphics[scale=0.3]{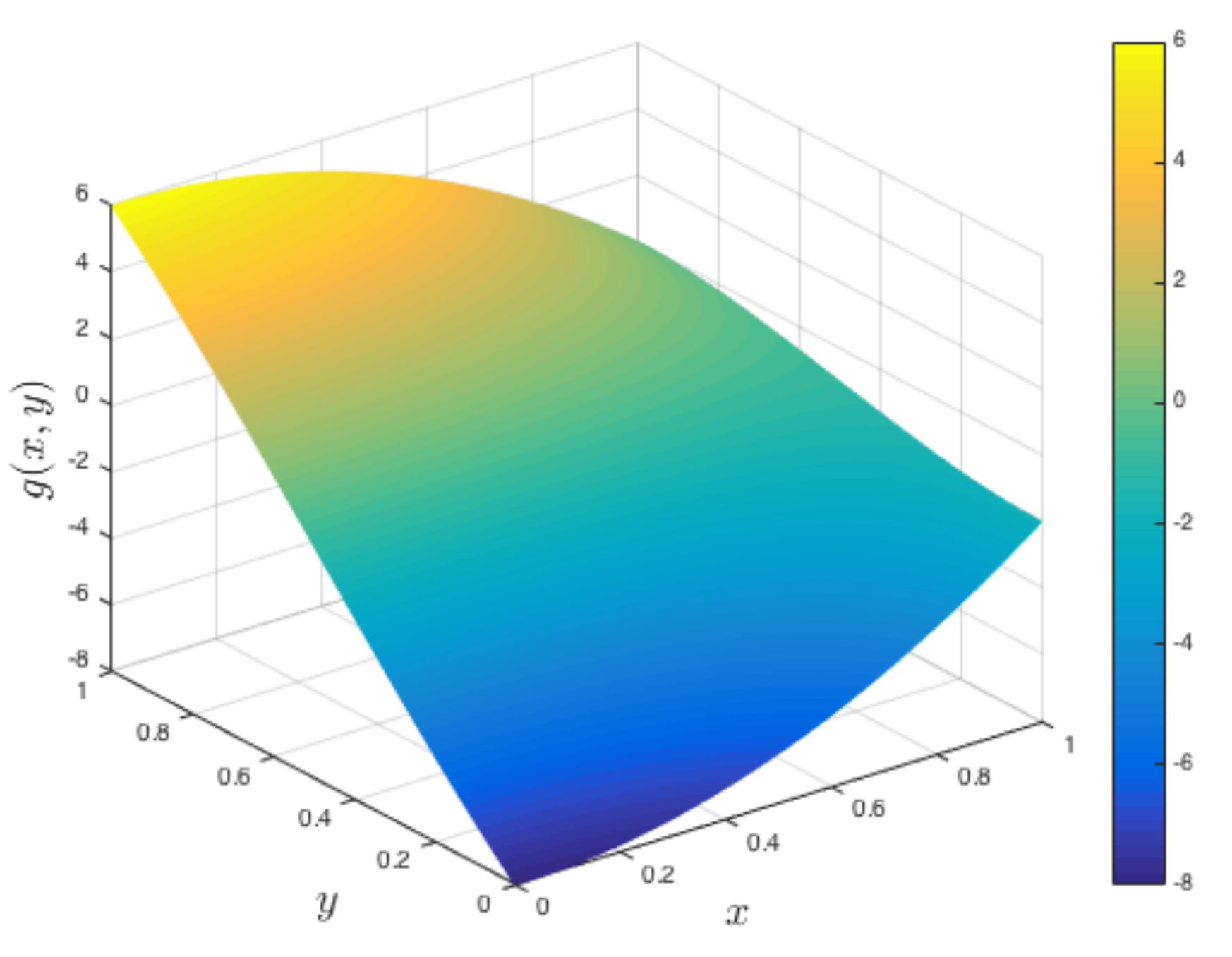}}
	\caption{Network topology, function $f(x,y)$ to estimate over the interior grid points, and input function $g(x,y)$.} 
	\label{fig: grid network}}
\end{figure*}

Consider now the problem of estimating a two-dimensional process driven by a partial differential equation (PDE) with a sensor network. To see how our distributed algorithm can be tuned to address this issue, we shall focus on the Poisson's PDE defined by:
\begin{equation}
	\label{eq:poisson}
	\frac{\partial^2 f(x,y)}{\partial x^2}+\frac{\partial^2 f(x,y)}{\partial y^2}=g(x,y), \qquad (x,y)\in[0,1]^2,
\end{equation}
with $g:[0,1]^2\rightarrow\mathbb{R}$ an input function, and on a two dimensional network of $(n-2)^2$ sensor nodes and $4(n-1)$ boundary points equally spaced over the unit square $(x,y)\in[0,1]^2$ with $\Delta_x=\Delta_y=\Delta=\frac{1}{n-1}$, as illustrated in Fig.~\ref{fig: grid network} (a).

We introduce the grid point $(x_k,y_\ell)\triangleq (k\Delta,\ell\Delta)$ and the sampled values at this point $f_{k,\ell}\triangleq f(k\Delta,\ell\Delta)$ and $g_{k,\ell}\triangleq g(k\Delta,\ell\Delta)$ with $0\leq k,\ell \leq n-1$. We use the central difference approximation for the second derivative~\cite{bertsekas1989parallel}:
\begin{align}
	\frac{\partial^2 f(k\Delta,\ell\Delta)}{\partial x^2}&\approx\frac{1}{\Delta^2}\Big(f_{k+1,\ell}-2 f_{k,\ell}+f_{k-1,\ell}\Big)\\
	\frac{\partial^2 f(k\Delta,\ell\Delta)}{\partial y^2}&\approx\frac{1}{\Delta^2}\Big(f_{k,\ell+1}-2 f_{k,\ell}+f_{k,\ell-1}\Big)
\end{align}
which leads to:
\begin{equation}
	\label{eq: linear relation between f and g}
	\frac{1}{\Delta^2}\Big(-4 f_{k,\ell}+f_{k-1,\ell}+f_{k,\ell-1}+f_{k,\ell+1}+f_{k+1,\ell}\Big)=g_{k,\ell}.
\end{equation}
In this experiment, we shall consider the unknown physical process $f$ and the input function $g$ given by:
\begin{align}
	&f(x,y)=(1-x^2)(2y^3-3y^2+1),\\
	&g(x,y)=-2(2y^3-3y^2+1)+6(1-x^2)(2y-1),
\end{align}
for $(x,y)\in[0,1]^2$ with boundary conditions 
$f(0,y)=2y^3-3y^2+1$, $f(x,0)=1-x^2$, and $f(1,y)=f(x,1)=0$. These functions are illustrated in Fig.~\ref{fig: grid network} (b), (c).

The objective is to estimate $f(x,y)$ at the interior grid points $(x_k,y_\ell)$ with $0<k,\ell<n-1$, given noisy measurements $g_{k\ell}(i)=g_{k\ell}+z_{k\ell}(i)$ of $g(x,y)$ collected by the sensors located at these interior grid points. The noise process $z_{k\ell}(i)$ is assumed to be zero mean, temporally white, and spatially independent. The values of $f(x,y)$ at the boundary points are known a priori as they correspond to boundary conditions. We denote by $f^o_{k\ell}$ the value at $(x_k,y_{\ell})$ of the function $f(x,y)$ that satisfies \eqref{eq:poisson}, and by $f_{k\ell}$ the estimated value of $f^o_{k\ell}$. To each node $(k,\ell)$ we associate an $M_{k\ell}\times1$ parameter vector $\bw_{k\ell}$ to estimate, an $M_{k\ell} \times 1$ regression vector $\bx_{k\ell}$ and a scalar $v^o_{k\ell}$, defined in Table~\ref{tab:parameter vectors and regression vectors} depending on the node location on the grid. 

\cblue{Given the values of $f(x,y)$ at the boundary points, and} according to~\eqref{eq: linear relation between f and g}, the linear regression model can be written as follows:
\begin{equation}
	g_{k\ell}(i)=\bx_{k\ell}^\top\bw_{k\ell}+v^o_{k\ell}+z_{k\ell}(i).
\end{equation}
As can be seen in Table~\ref{tab:parameter vectors and regression vectors}, equality constraints of the form~\eqref{eq: multitask problem (b)} need to be imposed on the parameter vectors of neighboring sensor nodes in order to achieve equality between common entries. For instance, let us consider neighboring nodes $(k,\ell)$ and $(k+1,\ell)$ with $2\leq k \leq n-4$ and $2\leq \ell \leq n-3$. Since these nodes are jointly estimating $f_{k,\ell}$ and $f_{k+1,\ell}$, the following equality constraint is required:
\begin{equation}
	\left[ \setlength\arraycolsep{2pt}
	\begin{array}{ccccc} 
		1 & 0 &0 & 0&0 \\
		0&0&0&0&1
	\end{array}
 	\right]\bw_{k\ell}+
 	\left[  \setlength\arraycolsep{2pt}
	\begin{array}{ccccc} 
		0 & -1 &0 & 0&0 \\
		-1&0&0&0&0
	\end{array}
 	\right]\bw_{(k+1)\ell}=\Zero.
\end{equation} 

\begin{figure}[h]
{\centering
\includegraphics[scale=0.3]{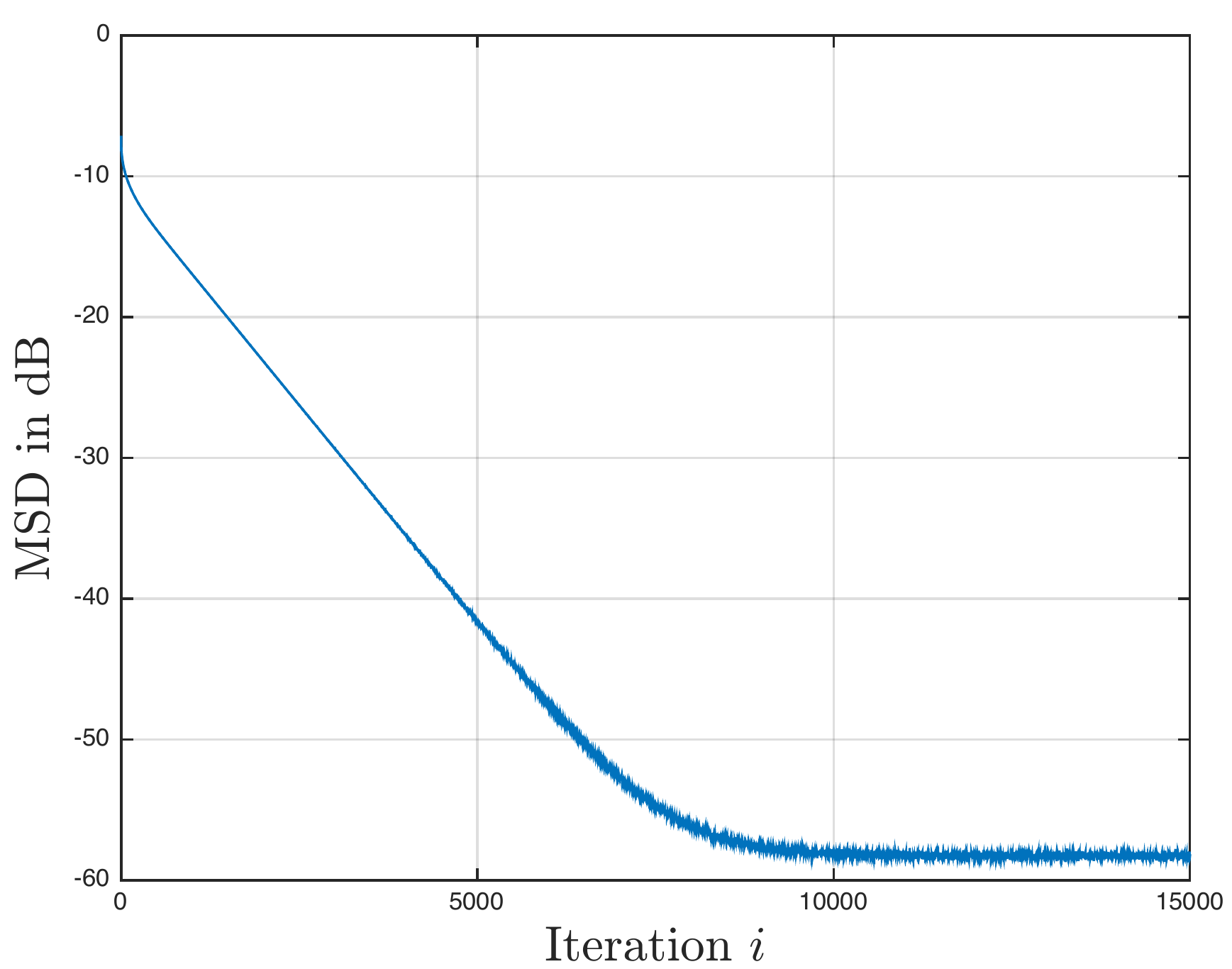}
\caption{Network MSD performance for $n=9$.}
\label{fig: poisson process problem MSD}}
\end{figure}

\begin{figure}[h]
{\centering
\includegraphics[scale=0.24]{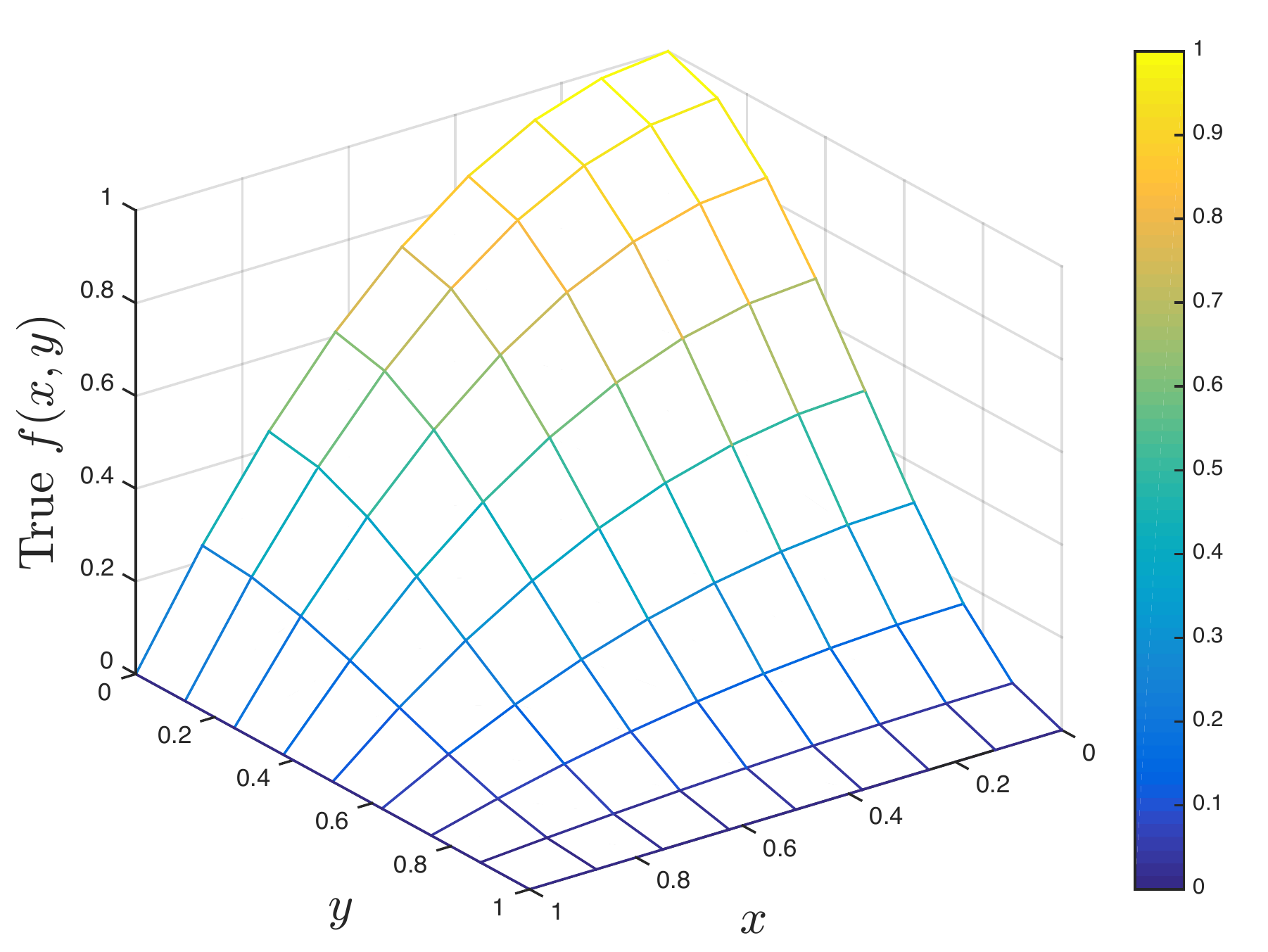}
\includegraphics[scale=0.24]{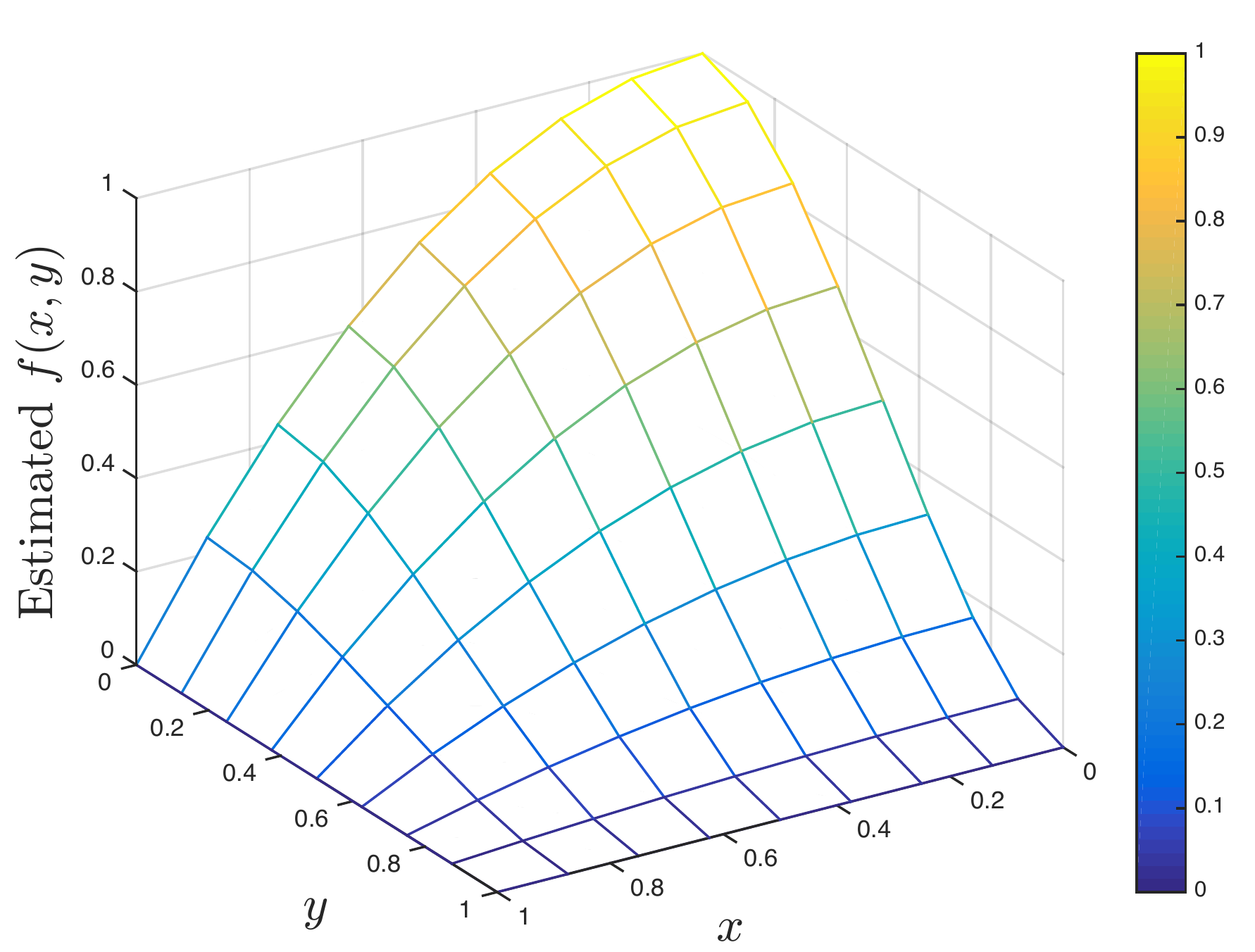}
\caption{Poisson process $f(x,y)$ over the network grid. ({\em Left}) True process. ({\em Right}) Estimated process.}
\label{fig: poisson process estimation}}
\end{figure}

\begin {table*}[t]
\begin{center}
   \begin{tabular}{ |c|c|c|c| }
    \hline
    \diagbox[]{$\ell$}{$k$}& $1$ & $2,\ldots, n-3$ & $n-2$ \\ \hline
    \multirow{3}{*}
    {$1$} & $[f_{k,\ell},f_{k,\ell+1},f_{k+1,\ell}]^\top$&  $[f_{k,\ell},f_{k-1,\ell},f_{k,\ell+1},f_{k+1,\ell}]^\top$ &$[f_{k,\ell},f_{k-1,\ell},f_{k,\ell+1}]^\top$ \\
              & $[-4,1,1]^\top$ &$[-4,1,1,1]^\top$&$[-4,1,1]^\top$ \\
              &$f^o_{1,0}+f^o_{0,1}$&$f^o_{k,0}$&$f^o_{n-2,0}+f^o_{n-1,1}$\\ \hline
    $2$ & $[f_{k,\ell},f_{k,\ell-1},f_{k,\ell+1},f_{k+1,\ell}]^\top$  & $[f_{k,\ell},f_{k-1,\ell},f_{k,\ell-1},f_{k,\ell+1},f_{k+1,\ell}]^\top$ 
    &$[f_{k,\ell},f_{k-1,\ell},f_{k,\ell-1},f_{k,\ell+1}]^\top$  \\
    $ \vdots$ & $[-4,1,1,1]^\top$ &$[-4,1,1,1,1]^\top$ &$[-4,1,1,1]^\top$ \\
    $n-3$ &$f^o_{0,\ell}$ & 0&  $f^o_{n-1,\ell}$\\ \hline
    \multirow{3}{*}
    {$n-2$} & $[f_{k,\ell},f_{k,\ell-1},f_{k+1,\ell}]^\top$ &  $[f_{k,\ell},f_{k-1,\ell},f_{k,\ell-1},f_{k+1,\ell}]^\top$ 
    &$[f_{k,\ell},f_{k-1,\ell},f_{k,\ell-1}]^\top$ \\
                 & $[-4,1,1]^\top$ &$[-4,1,1,1]^\top$&$[-4,1,1]^\top$ \\
                 &$f^o_{0,n-2}+f^o_{1,n-1}$ & $f^o_{k,n-1}$&  $f^o_{n-2,n-1}+f^o_{n-1,n-2}$ \\  \hline
    \end{tabular}
\end{center}
\caption{Parameter vector $\bw_{k\ell}$ (first row of each cell), regression vector $\Delta^2\bx_{k\ell}$ (second row of each cell), and scalar value $\Delta^2v^o_{k\ell}$ (last row of each cell) at each node $(k,\ell)$.} 
\label{tab:parameter vectors and regression vectors}
\end{table*}
Algorithm~\eqref{eq: multitask algorithm} can be used to address this problem by replacing the adaptation step~\eqref{eq: 1 adaptation step} by:
\begin{equation}
	\bpsi_{k\ell_m}(i+1)=\bw_{k\ell_m}(i)+\mu\,c_{k\ell_m}\bx_{k\ell}\big[g_{k\ell}(i)-\bx_{k\ell}^\top\bw_{k\ell_m}(i)-v^o_{k\ell}\big],
\end{equation}
where $\bw_{k\ell_m}(i)$ denotes the estimate of $\bw_{k\ell}$ at the $m$-th sub-node of $(k,\ell)$. The noises $z_{k,\ell}(i)$ were zero-mean i.i.d. Gaussian distributed with variances $\sigma^2_{z,k\ell}$ randomly generated from the uniform distribution $\mathcal{U}(0.1,0.14)$. We used a constant step-size $\mu=7\cdot10^{-5}$ for all nodes. Figure~\ref{fig: poisson process problem MSD} shows the network MSD learning curves for $n=9$. The simulated curves were obtained by averaging over $100$ independent runs. Figure~\ref{fig: poisson process estimation} shows the true (left) and estimated (right) process after convergence of our algorithm.

\section{Conclusion and Perspectives}
In this work, we proposed a multitask LMS algorithm for solving problems that require {the} simultaneous estimation of multiple parameter vectors that are related {locally} via linear constraints. Our primal technique was based on the stochastic gradient projection algorithm with constant step-sizes. The behavior of the algorithm in the mean and mean-square-error sense was studied. \cblue{We checked with simulations that the} agents are able to reach the optimal solution with good precision. In future {work}, we shall extend our approach to other {types} of constraints and also consider other constraints distribution over networks.
\begin{appendices}
\section{Projection matrix structure}
\label{sec: Projection matrix structure}
We denote by $\cD_{e,p}$ the $p$-th block row in $\cD_e$ and by $[\cD_{e}]_{p,k_m}$ the $L_p\times M_k$ block of $\cD_{e,p}$ corresponding to the $k_m$-th sub-node. First, we show that the $M_k\times M_\ell$ $(k_m,\ell_n)$-th block of the $N_e\times N_e$ block matrix $\cP_e$ in~\eqref{eq: cP e} is equal to:
\begin{equation}
	\label{eq: result 1}
	[\cP_e]_{k_m,\ell_n}=
	\left\lbrace
	\begin{array}{lr}
	\bI_{M_k}-[\cD_{e}]_{p,k_m}^\top(\cD_{e,p}\cD^\top_{e,p})^{-1}[\cD_{e}]_{p,k_m},\\
	\qquad\qquad\text{if $k_m=\ell_n$ and $k_m\in\cI_{e,p}$,} \\
	-[\cD_{e}]_{p,k_m}^\top(\cD_{e,p}\cD^\top_{e,p})^{-1}[\cD_{e}]_{p,\ell_n},\\
	\qquad\qquad~\text{if $k_m\neq\ell_n$ and $k_m,\ell_n\in\cI_{e,p}$,}\\
\Zero_{M_k\times M_\ell},\quad\qquad\quad\qquad~\quad\text{otherwise.}
\end{array}
\right.
\end{equation}
Furthermore, we show that the $k_m$-th block of the $N_e\times 1$ block column vector $\boldf_e$ in~\eqref{eq: boldf e} is equal to:
\begin{equation}
	\label{eq: result 2}
	[\boldf_e]_{k_m}=
	\left\lbrace
	\begin{array}{lr}
		[\cD_{e}]_{p,k_m}^\top(\cD_{e,p}\cD^\top_{e,p})^{-1}\bb_p,\text{ if $k_m\in\cI_{e,p}$,}\\
		\Zero_{M_k\times 1}, \qquad\qquad\qquad\qquad\text{otherwise.}
	\end{array}
	\right.
\end{equation}

It can be verified that $\cD_e\cD_e^\top$ is a $P\times P$ block diagonal matrix whose $(p,p)$-th block is of dimension $L_p\times L_p$ and is given by:
\begin{equation}
	[\cD_e\cD_e^\top]_{p,p}=\cD_{e,p}\cD^\top_{e,p}=\cD_{p}\cD_{p}^\top.
\end{equation}
The inverse of the block diagonal matrix $\cD_e\cD_e^\top$ is:
\begin{equation}
	(\cD_e\cD_e^\top)^{-1}=\diag\Big\{(\cD_{e,p}\cD^\top_{e,p})^{-1}\Big\}_{p=1}^P.
\end{equation}
By multiplying the matrix $(\cD_e\cD_e^\top)^{-1}$ from the left by $\cD_e^\top$ we obtain an $N_e\times P$ block matrix whose $(k_m,p)$-th block is of dimension $M_k\times L_p$ given by:
\begin{equation}
\label{eq: pinvD}
\begin{split}
&[\cD_e^\top(\cD_e\cD_e^\top)^{-1}]_{k_m,p}\\
&=\left\lbrace
\begin{array}{lr}
[\cD_{e}]^\top_{p,k_m}(\cD_{e,p}\cD^\top_{e,p})^{-1},~ \text{if} ~k_m\in\cI_{e,p}\\
\Zero_{M_k\times L_p},\qquad\qquad\qquad~\text{otherwise}.
\end{array}
\right.
\end{split}
\end{equation}
When we multiply the matrix $\cD_e^\top(\cD_e\cD_e^\top)^{-1}$ from the right by $\cD_e$, we obtain an $N_e\times N_e$ block matrix whose $(k_m,\ell_n)$-th block corresponding to sub-nodes $k_m$, $\ell_n$ is of dimension $M_k\times M_\ell$ and is given by:
\begin{equation}
\label{eq: pinvDD}
\begin{split}
&[\cD_e^\top(\cD_e\cD_e^\top)^{-1}\cD_e]_{k_m,\ell_n}\\
&=
\left\lbrace
\begin{array}{lr}
[\cD_{e}]^\top_{p,k_m}(\cD_{e,p}\cD^\top_{e,p})^{-1}[\cD_e]_{p,\ell_n},\quad\text{if $k_m,\ell_n\in\cI_{e,p}$,}\\
\Zero_{M_k\times M_\ell},\qquad\qquad\qquad\qquad\qquad~\text{otherwise.}
\end{array}
\right.
\end{split}
\end{equation}
From~\eqref{eq: boldf e} and~\eqref{eq: pinvD}, we obtain~\eqref{eq: result 2}.


\section{Evaluation of the matrix $\cF$}
\label{sec: matrix F}

{Without loss of generality, we assume in the following that $M_k$ is uniform across the network, i.e., $M_k=M_0$ for all $k$.} We note that for any {symmetric} matrix $\bT$, we have~\cite{isserlis1918formula}:
\begin{equation}
\begin{split}
&\expec\{\bx_k(i)\bx_k^\top(i)\bT\bx_{\ell}(i)\bx_{\ell}^\top(i)\}\\
&=\bR_{x,k}\bT\bR_{x,\ell}+\delta_{k,\ell}\big(\bR_{x,k}\bT\bR_{x,k}+\bR_{x,k}\tr(\bR_{x,k}\bT)\big).
\end{split}
\end{equation}
From~\eqref{eq: B(i)} and~\eqref{eq: sigma'}, we obtain:
\begin{equation}
\begin{split}
&\bSig'=\cP_e\cA\bSig\cA^\top\cP_e-\mu\cP_e\cA\bSig\cA^\top\cP_e\cR_{x,e}-\\
&\mu\cR_{x,e}\cP_e\cA\bSig\cA^\top\cP_e+\mu^2\expec\{\cR_{x,e}(i)\cP_e\cA\bSig\cA^\top\cP_e\cR_{x,e}(i)\}.
\end{split}
\end{equation}
In order to evaluate $\bSig'$ we need to evaluate the fourth term on the RHS of the above equation. Let:
{\begin{eqnarray}
\cK&\triangleq&\expec\{\cR_{x,e}(i)\cP_e\cA\bSig\cA^\top\cP_e\cR_{x,e}(i)\},\label{eq: matrix K} \\
\cT&\triangleq&\cP_e\cA\bSig\cA^\top\cP_e.
\end{eqnarray}}
It can be verified that the $(k_m,\ell_n)$-th block of the matrix $\cK$ corresponding to the $(k_m,\ell_n)$-th sub-node is given by:
\begin{equation}
\begin{split}
&[\cK]_{k_m,\ell_n}\\
&=c_{k_m}c_{\ell_n}\expec\big\{\bx_k(i)\bx_{k}^\top(i)[\cT]_{k_m,\ell_n}\bx_{\ell}(i)\bx_{\ell}^\top(i)\big\}\\
&=c_{k_m}c_{\ell_n}\bR_{x,k}[\cT]_{k_m,\ell_n}\bR_{x,\ell}+\\
&~ ~\delta_{k,\ell}c_{k_m}c_{\ell_n}\big(\bR_{x,k}[\cT]_{k_m,\ell_n}\bR_{x,k}+\bR_{x,k}\tr(\bR_{x,k}[\cT]_{k_m,\ell_n})\big),
\end{split}
\end{equation}
where the $M_0\times M_0$ matrix $[\cT]_{k_m,\ell_n}$ is the $(k_m,\ell_n)$-th block of the matrix $\cT$. The matrix $\cK$ in~\eqref{eq: matrix K} can be written as:
\begin{equation}
\begin{split}
\cK=&\cR_{x,e}\cT\cR_{x,e}+\sum_{k=1}^N\cS_k(\bI_{N_e}\otimes\bR_{x,k})\cT(\bI_{N_e}\otimes\bR_{x,k})\cS_k+\\
&\quad\sum_{k=1}^N\cS_k(\bI_{N_e}\otimes\bR_{x,k})\cZ_k\cS_k,
\end{split}
\end{equation}
where $\cS_k$ is the $N\times N$ block diagonal matrix whose $(k,k)$-th block is equal to $\bC_k\otimes\bI_{M_0}$, and $\cZ_k$ is the $N_e\times N_e$ block matrix whose $(k_m,\ell_n)$-th block is given by:
\begin{equation}
\big[\cZ_k\big]_{h_m,\ell_n}=\bI_{M_0}[\vc(\bR_{x,k})]^\top\vc([\cT]_{k_m,\ell_n}).
\end{equation}
Applying the block-vectorization operator to $\cK$ and using the property $\bvc(\bA\bB\bC)=(\bC^\top\otimes_b\bA)\bvc(\bB)$, we obtain:
\begin{equation}
	\begin{split}
	&\bvc(\cK)=(\cR_{x,e}\otimes_b\cR_{x,e})\bvc(\cT)+\\
	&\sum_{k=1}^N\Big(\cS_k^\top(\bI_{N_e}\otimes\bR_{x,k})\otimes_b\cS_k(\bI_{N_e}\otimes	\bR_{x,k})\Big)\bvc(\cT)+\\
	&\sum_{k=1}^N\Big(\cS_k^\top\otimes_b\big(\cS_k[\bI_{N_e}\otimes\bR_{x,k}]\big)\Big)\bvc(\cZ_k),
	\end{split}
\end{equation}
where $\bvc(\cZ_k)$ can be expressed as:
\begin{equation}
\bvc(\cZ_k)=\Big(\bI_{N_e^2}\otimes\vc(\bI_{M_0})\otimes[\vc(\bR_{x,k})]^\top\Big)\bvc(\cT),
\end{equation}
where {$\bvc(\cT)=(\cP_e\cA\otimes_b\cP_e\cA)\bsig$}. Finally, we conclude that the matrix $\cF$ in~\eqref{eq: matrix F} can be written as:
{\begin{equation}
\begin{split}
&\cF=\cB^\top\otimes_b\cB^\top+\\
&\mu^2\sum_{k=1}^N\Big(\cS_k^\top(\bI_{N_e}\otimes\bR_{x,k})\otimes_b\cS_k(\bI_{N_e}\otimes\bR_{x,k})\Big)(\cP_e\cA\otimes_b\cP_e\cA)\\
&+\mu^2\sum_{k=1}^N\Big(\cS_k^\top\otimes_b\big(\cS_k(\bI_{N_e}\otimes\bR_{x,k})\big)\Big)\cdot\\
&\qquad\qquad\Big(\bI_{N_e^2}\otimes\vc(\bI_{M_0})\otimes[\vc(\bR_{x,k})]^\top\Big)(\cP_e\cA\otimes_b\cP_e\cA).
\end{split}
\end{equation}}
\section{Performance of competing algorithms}
\label{app: Performance of competing algorithms}
We compare in the simulation section algorithm~\eqref{eq: multitask algorithm} with the non-cooperative LMS algorithm (obtained from~\eqref{eq: CLMS} by setting $\cP=\bI_{M}$ and $\boldf=\Zero$), the centralized CLMS algorithm~\eqref{eq: CLMS}, and the following algorithm:
\begin{subequations}
		\label{eq: multitask algorithm with steps inversion}
		\begin{eqnarray}
			\bpsi_{k_m}(i+1)&=&\bw_{k_m}(i)+\nonumber
			\\
			&&\mu\,c_{k_m}\bx_k(i)[d_k(i)-\bx_k^\top(i)\bw_{k_m}(i)],\label{eq: 1 adaptation step}\\
			\bphi_{k_m}(i+1)&=&\sum_{k_n\in{\N_{k_m}\cap\,\C_k}}a_{k_n,k_m}\bpsi_{k_n}(i+1),\label{eq: 2 combination step}\\
			\bw_{k_m}(i+1)&= &[\cP_{p}]_{k_m,\bullet}\cdot\col\big\{\bphi_{\ell_n}(i+1)\big\}_{\ell_n\in\cI_{e,p}}-[\boldf_{p}]_{k_m},\nonumber\\
			\label{eq: 3 projection step}
		\end{eqnarray}
\end{subequations}
where the sub-nodes ``combine-then-project'' instead of ``project-then-combine''. In the following, we show how the theoretical learning curves of these algorithms can be obtained from the analysis in Sections~\ref{sec: Performance analysis I}  and \ref{sec: Performance analysis II}.
Consider the centralized CLMS algorithm~\eqref{eq: CLMS}. Let $\bwt(i)$ and $\bwt'(i)$ denote the $N\times 1$ block error vectors at the fusion center given by:
\begin{equation}
	\bwt(i)\triangleq\bw^o-\bw(i),\qquad
	\bwt'(i)\triangleq\bw^\star-\bw(i).
\end{equation}
Subtracting $\bw^o$ from both sides of recursion~\eqref{eq: CLMS} and using the linear data model~\eqref{eq: linear data model}, we obtain:
\begin{equation}
\label{eq: evolution of w}
\begin{split}
\bwt(i+1)&=\cP\left(\bI_{M}-\mu\cR_x(i)\right)\bwt(i)-\mu\cP\bp_{xz}(i)+\\
&\qquad\left(\bI_{M}-\cP\right)\bw^o+\boldf,
\end{split}
\end{equation}
where $\cR_x(i)$ and $\bp_{xz}(i)$ are given by:
\begin{align}
&\cR_x(i)\triangleq\diag\left\{\bx_k(i)\bx_k^\top(i)\right\}_{k=1}^N, \\
&\bp_{xz}(i)\triangleq\col\left\{d_k(i)\bx_k(i)\right\}_{k=1}^N.
\end{align}
Let $\bw^\delta\triangleq\bw^o-\bw^\star$. Using $\bwt'(i)=\bwt(i)-\bw^\delta$ with~\eqref{eq: evolution of w} and the fact that $\bw^\star$ satisfies $\cP\bw^\star-\boldf=\bw^\star$, we obtain:
\begin{equation}
\label{eq: evolution of w'}
\bwt'_b(i+1)=\cP\left(\bI_{M}-\mu\cR_x(i)\right)\bwt'(i)-\mu\cP\bp_{xz}(i)-\mu\cP\cR_x(i)\bw^\delta.
\end{equation}
Comparing recursions~\eqref{eq: evolution of w} and \eqref{eq: evolution of w'} with recursions~\eqref{eq: weight error vector recursion 0} and \eqref{eq: weight error vector recursion 0'}, we observe that the learning curves of the centralized solution~\eqref{eq: CLMS} can be deduced from those of the decentralized solution~\eqref{eq: multitask algorithm} by properly modifying the coefficient matrices and vectors. Note that, the centralized solution is unbiased with respect to $\bw^\star$ since $\mu\cP\expec\cR_x(i)\bw^\delta=\Zero$.

Next, consider the distributed solution~\eqref{eq: multitask algorithm with steps inversion}. Following the same line of reasoning as in Subsection~\ref{subsec: network error vector recursion}, we obtain the following recursions for the block error vectors~\eqref{eq: definition of wt_e} and~\eqref{eq: definition of wt'_e}:
\begin{equation}
	\label{eq: weight error vector recursion 0 with step inverse}
	\begin{split}
	\bwt_e(i+1)&=\cP_e\cA^\top\left[\bI_{M_e}-\mu\cR_{x,e}(i)\right]\bwt_e(i)-\\
	&\mu\cP_e\cA^\top\bp_{xz,e}(i)+\left(\bI_{M_e}-\cP_e\right)\bw^o_e+\boldf_e,
	\end{split}
\end{equation}
\begin{equation}
	\label{eq: weight error vector recursion 0' with step inverse}
	\begin{split}
	\bwt_e'(i+1)&=\cP_e\cA^\top\left[\bI_{M_e}-\mu\cR_{x,e}(i)\right]\bwt_e'(i)-\\
	&\mu\cP_e\cA^\top\bp_{xz,e}(i)-\mu\cP_e\cA^\top\cR_{x,e}(i)\bw^\delta_e.
	\end{split}
\end{equation}
Comparing recursions~\eqref{eq: weight error vector recursion 0 with step inverse} and \eqref{eq: weight error vector recursion 0' with step inverse} with recursions~\eqref{eq: weight error vector recursion 0} and \eqref{eq: weight error vector recursion 0'}, we observe that the learning curves of the distributed solution~\eqref{eq: multitask algorithm with steps inversion} can be deduced from the theoretical curves of the decentralized solution~\eqref{eq: multitask algorithm} by properly replacing the product $\cA^\top\cP_e$ in the analysis of Sections~\ref{sec: Performance analysis I}  and \ref{sec: Performance analysis II} by the product $\cP_e\,\cA^\top$ and the vector $\br$ in~\eqref{eq: br} by $\br\triangleq\left(\bI_{M_e}-\cP_e\right)\bw^o_e+\boldf_e$. 

\section{Performance comparison}
\label{app: Performance comparison}

As explained after algorithm (35), this permutation of the projection and the aggregation steps is very useful since it allows to simplify the algorithm.  We show hereafter that, for the perfect model scenario (i.e., $\bw^o=\bw^\star$), this permutation enhances the steady-state mean-square-error performance. 

First, let us consider the algorithm where agent $k$ performs the projection step before the aggregation step. In the case of the perfect model scenario, we know that $\br=0$ (see~(56)), and $\expec\,\bwt_e(\infty)=0$ (see (59)). When matrix $\cF$ is stable, from (75)--(77), we obtain the following steady-state network performance with metric $\bsig_{ss}\triangleq\bvc(\bSig_{ss})$:
\begin{equation}
\begin{split}
\zeta^\star&=[\bvc(\cY(\infty))]^\top(\bI-\cF)^{-1}\bsig_{ss}\\
&=[\bvc(\cY(\infty))]^\top\sum_{j=0}^{+\infty}[(\cB^{\top})^j\otimes_b(\cB^{\top})^j]\bsig_{ss}\\
&=[\bvc(\cY(\infty))]^\top\sum_{j=0}^{+\infty}\bvc\left((\cB^{\top})^j\bSig_{ss}\cB^j\right)\\
&=\sum_{j=0}^{+\infty}\tr\left(\cY(\infty)(\cB^{\top})^j\bSig_{ss}\cB^j\right)\\
&=\sum_{j=0}^{+\infty}\tr\left(\bSig_{ss}\cB^j\cY(\infty)(\cB^{\top})^j\right)
\end{split}
\end{equation} 
Replacing $\cB$ and $\cY(\infty)$ in the above expression by (58) and (77), we obtain:
\begin{equation}
\begin{split}
\zeta^\star=\mu^2\sum_{j=0}^{+\infty}\tr\Big(\bSig_{ss}&\left(\cA^\top\cP_e(\bI-\mu\cR_{x,e})\right)^j\cA^\top\cP_e\cS\cP_e\cA\cdot\\
&\qquad\qquad\Big((\bI-\mu\cR_{x,e})\cP_e\cA\Big)^j\Big),
\end{split}
\end{equation} 
where $\cS\triangleq\diag\left\{\bc_k\bc_k^\top\otimes\sigma^2_{z,k}\bR_{x,k}\right\}_{k=1}^N$.

Let us consider now that agent $k$ performs the aggregation step before projecting. Let $\zeta^\star_1$ be the steady-state network performance with metric $\bsig_{ss}\triangleq\bvc(\bSig_{ss})$. Following the same line of reasoning, we obtain:
\begin{equation}
\begin{split}
\zeta^\star_1=\mu^2\sum_{j=0}^{+\infty}\tr\Big(\bSig_{ss}&\left(\cP_e\cA^\top(\bI-\mu\cR_{x,e})\right)^j\cP_e\cA^\top\cS\cA\cP_e\\
&\qquad\qquad\Big((\bI-\mu\cR_{x,e})\cA\cP_e\Big)^j\Big).
\end{split}
\end{equation} 
Let us assume that the factors $c_{k_m}$ are set to $\frac{1}{j_k}$, and $\N_{k_m}\cap\C_k=\C_k$ for all $m$. We further assume that $a_{k_n,k_m}$ are set to $\frac{1}{j_k}$ for all $n$. In this case, it can be verified that $\cA=\cA^\top$, $\cA^\top(\bI-\mu\cR_{x,e})=(\bI-\mu\cR_{x,e})\cA^\top$, and $\cA^\top\cS\cA=\cS$. Thus, $\zeta^\star$ can be written alternatively as:
\begin{equation}
\begin{split}
\zeta^\star=\mu^2\sum_{j=0}^{+\infty}\tr\Big(\bSig_{ss}&\cA^\top\underbrace{\left(\cP_e\cA^\top(\bI-\mu\cR_{x,e})\right)^j}_{\cW_j}\cP_e\cdot\\
&~~\cS\cP_e\underbrace{\Big((\bI-\mu\cR_{x,e})\cA\cP_e\Big)^j}_{\cW_j^\top}\cA\Big).
\end{split}
\end{equation}
Using $\cA^\top\cS\cA=\cS$, $\zeta^\star_1$ can be written alternatively as:
\begin{equation}
\begin{split}
\zeta^\star_1=\mu^2\sum_{j=0}^{+\infty}\tr\Big(\bSig_{ss}&\underbrace{\left(\cP_e\cA^\top(\bI-\mu\cR_{x,e})\right)^j}_{\cW_j}\cP_e\cdot\\
&~~\cS\cP_e\underbrace{\Big((\bI-\mu\cR_{x,e})\cA\cP_e\Big)^j}_{\cW_j^\top}\Big).
\end{split}
\end{equation} 
Hence, we obtain:
\begin{equation}
\begin{split}
&\zeta^\star-\zeta^\star_1\\
&=\mu^2\sum_{j=0}^{+\infty}\tr\left(\bSig_{ss}\left(\cA^\top\cW_j\cP_e\cS\cP_e\cW_j^\top\cA-\cW_j\cP_e\cS\cP_e\cW_j^\top\right)\right).
\end{split}
\end{equation}
When $\bSig_{ss}=\frac{1}{N_e}\bI$, we obtain:
\begin{equation*}
\begin{split}
&\zeta^\star-\zeta^\star_1\\
&=\frac{\mu^2}{N_e}\sum_{j=0}^{+\infty}\tr\left(\cA^\top\cW_j\cP_e\cS\cP_e\cW_j^\top\cA-\cW_j\cP_e\cS\cP_e\cW_j^\top\right)\leq 0,
\end{split}
\end{equation*}
where we used the fact that $\tr(\cA^\top\cH\cA)\leq\tr(\cH)$ for any doubly-stochastic matrix $\cA$ and any non-negative matrix $\cH$ of compatible dimensions (see Theorem C.3 in [11]).  

\end{appendices}
\bibliographystyle{IEEEbib}
\bibliography{bibliography}


\begin{IEEEbiography}[{\includegraphics[width=1in,height=1.25in,clip,keepaspectratio]{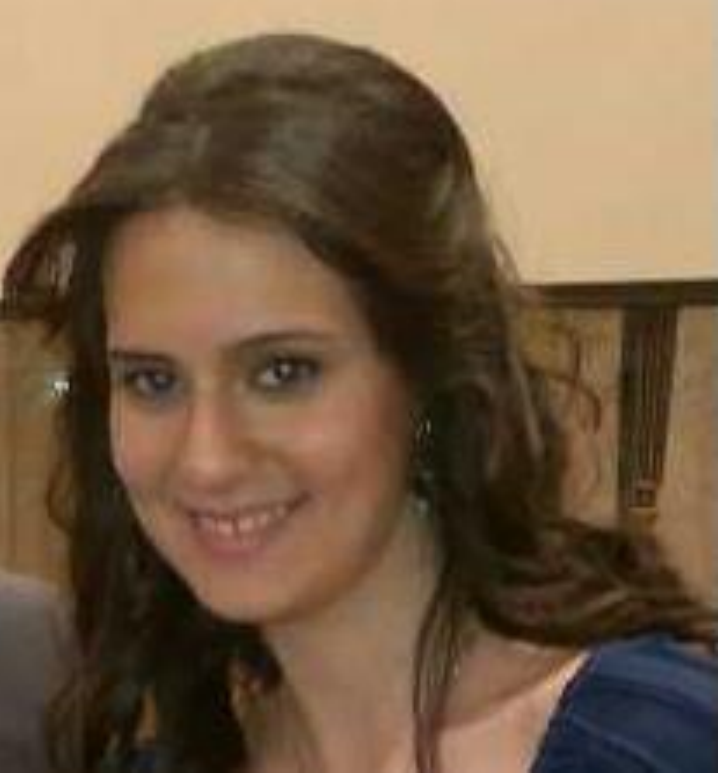}}
]{Roula Nassif}

received the bachelor's degree in Electrical Engineering from the Lebanese University, Lebanon, in 2013. She received the M.S. degrees in Industrial Control and Intelligent Systems for Transport from the Lebanese University, Lebanon, and from Compi\`egne University of Technology, France, in 2013. She received the Ph.D. degree in 2016, from the University of C\^ote d'Azur (UCA), France. She is currently a researcher and a teaching assistant at UCA. Her current research interests include adaptation and learning over networks.

\end{IEEEbiography}

\begin{IEEEbiography}[{\includegraphics[width=1in,height=1.25in,trim = 50mm 50mm 50mm 50mm, clip,keepaspectratio]{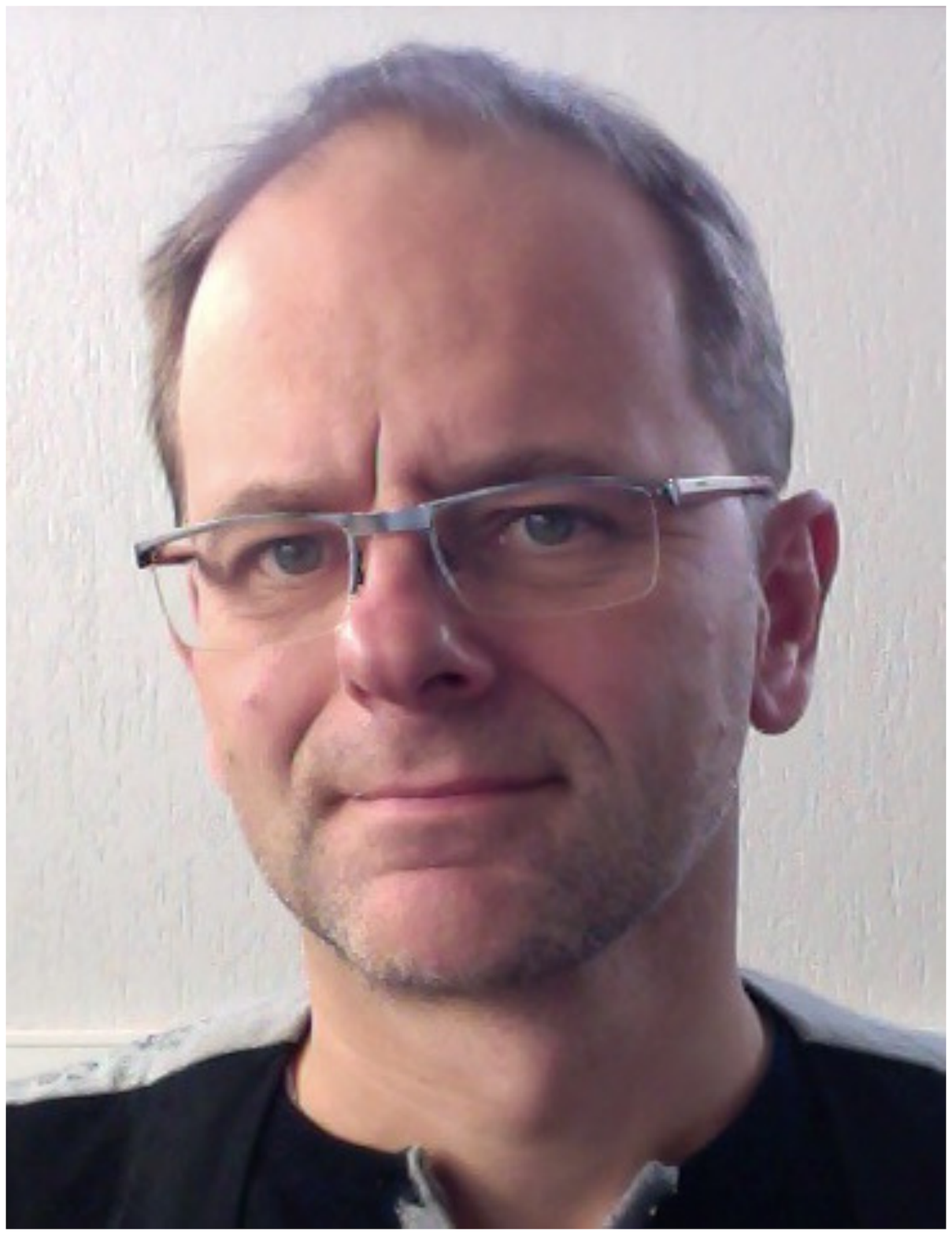}}]{C{\'e}dric Richard}
(S'98--M'01--SM'07)  received the Dipl.-Ing. and the M.S. degrees in 1994, and the Ph.D. degree in 1998, from Compi\`egne University of Technology, France.  He is a Full Professor at the University of Nice Sophia Antipolis, France. He was a junior member of the Institut Universitaire de France in 2010-2015.

His current research interests include statistical signal processing and machine learning. Prof. Richard is the author of over 250 papers. He was the General Co-Chair of the IEEE SSP'11 Workshop that was held in Nice, France. He was the Technical Co-Chair of EUSIPCO'15 that was held in Nice, France, and of the IEEE CAMSAP'15 Workshop that was held in Cancun, Mexico. Since 2015, he serves as a Senior Area Editor of the IEEE Transactions on Signal Processing, and as an Associate Editor of the IEEE Transactions on Signal and Information Processing over Networks. He is an Associate Editor of Signal Processing Elsevier since 2009. Prof. Richard is member of the IEEE Machine Learning for Signal Processing Technical Committee, and served as member of the IEEE Signal Processing Theory and Methods Technical Committee in 2009-2014.
\end{IEEEbiography}

\begin{IEEEbiography}[{\includegraphics[width=1in,height=1.25in,clip,keepaspectratio]{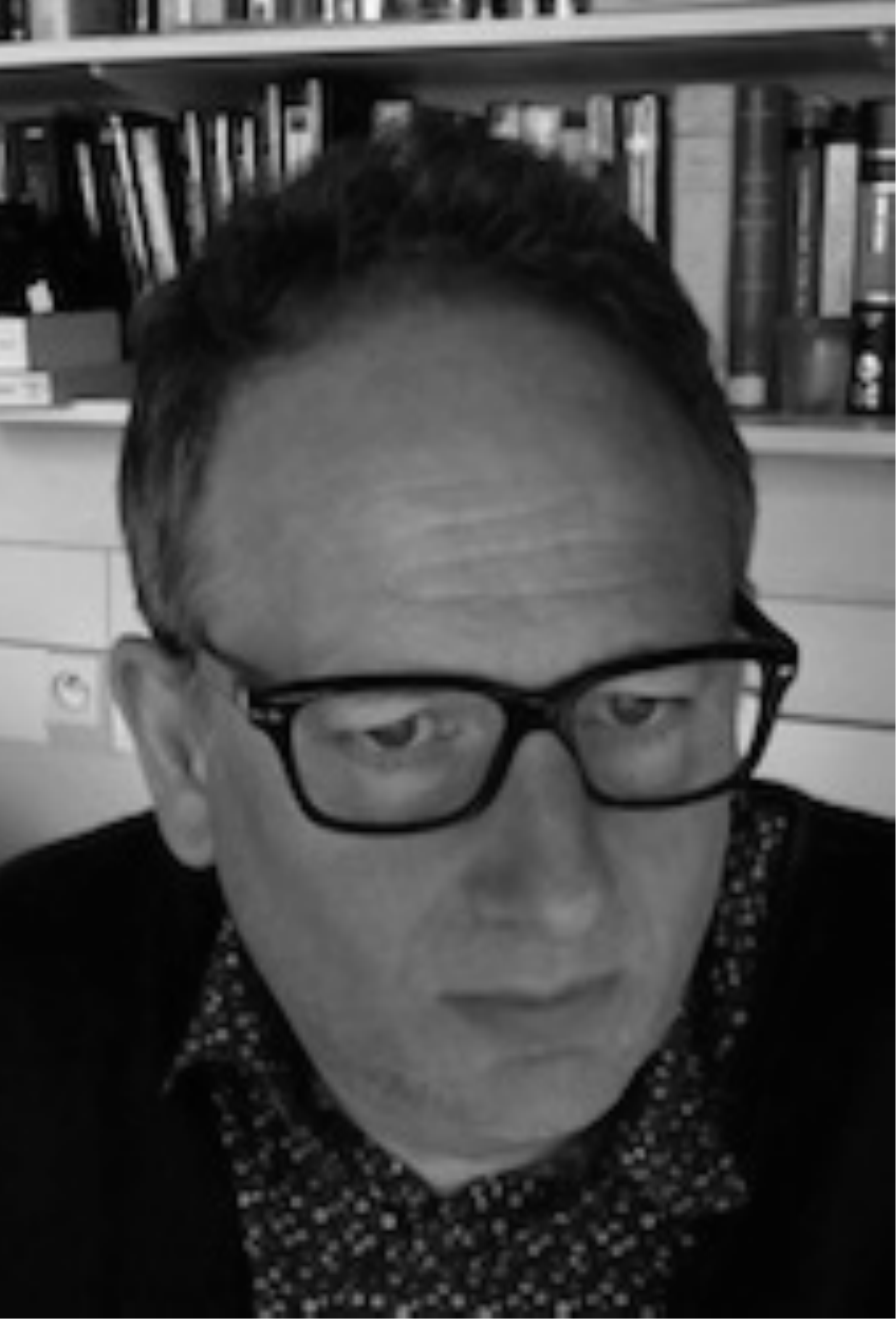}}
]{Andr{\'e} Ferrari}

(SM'91-M'93) received the Ing\'enieur degree from \'Ecole Centrale de Lyon, Lyon, France, in 1988 and the M.Sc. and Ph.D. degrees from the University of Nice Sophia Antipolis (UNS), France, in 1989 and 1992, respectively, all in electrical and computer engineering. 

He is currently a Professor at UNS. He is a member of the Joseph-Louis Lagrange Laboratory (CNRS, OCA), where his research activity is centered around statistical signal processing and modeling, with a particular interest in applications to astrophysics.

\end{IEEEbiography}

\begin{IEEEbiography}[{\includegraphics[width=1in,height=1.25in,clip,keepaspectratio]{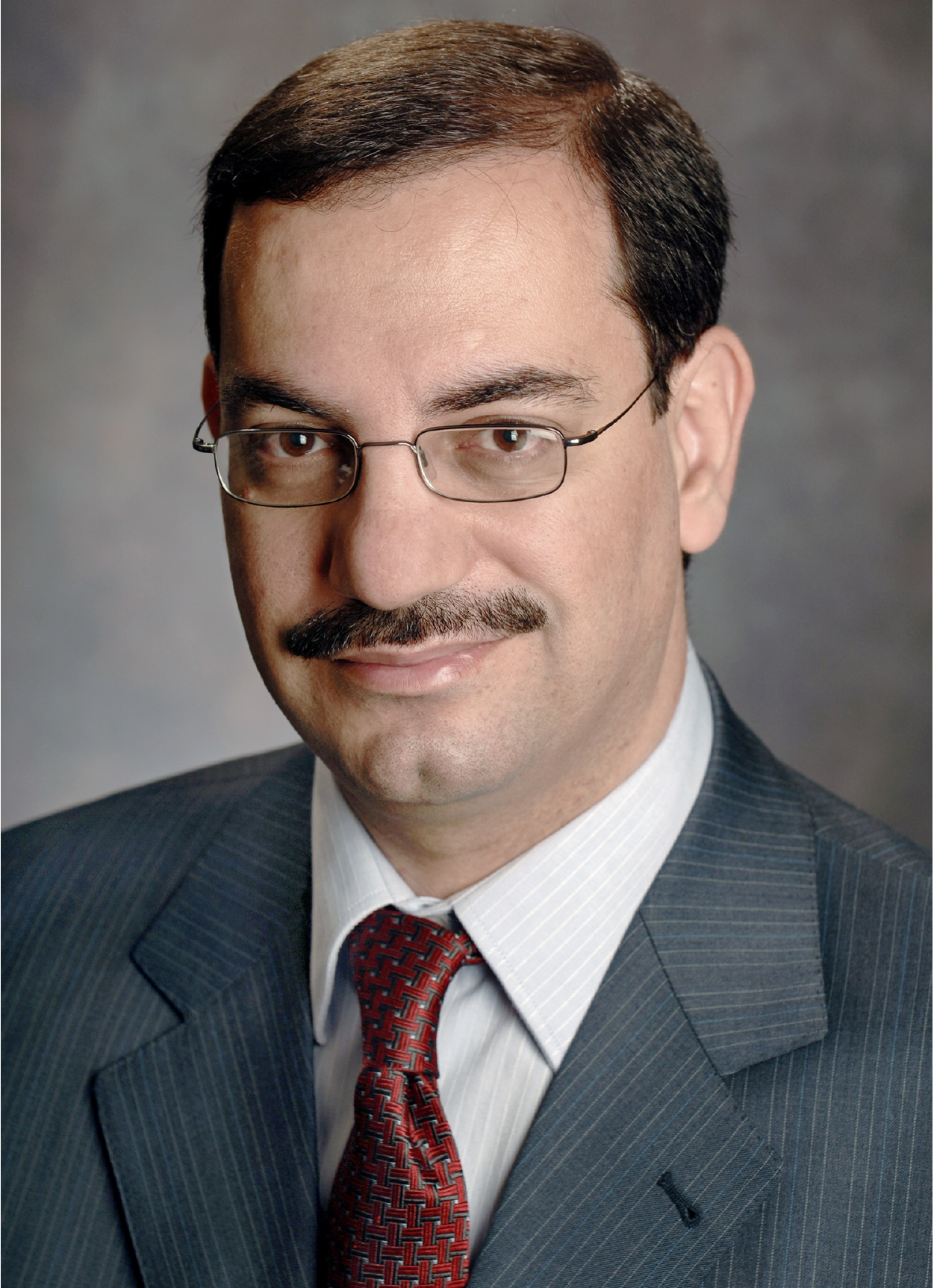}}
]{Ali H. Sayed}

(S'90-M'92-SM'99-F'01) is a professor and former chairman of electrical engineering at the University of California, Los Angeles, where
he directs the UCLA Adaptive Systems Laboratory. An author of over 480 scholarly publications and six books, his research involves several areas
including adaptation and learning, statistical signal processing, distributed processing, network and data sciences, and biologically-inspired designs. Dr. Sayed has received several awards including the 2015 Education Award from the IEEE Signal Processing Society, the 2014 Athanasios Papoulis Award from the European Association for Signal Processing, the 2013 Meritorious Service Award, and the 2012 Technical Achievement Award from the IEEE Signal Processing Society. Also, the 2005 Terman Award from the American Society for Engineering Education, the 2003 Kuwait Prize, and the 1996 IEEE Donald G. Fink Prize. He served as Distinguished Lecturer for the IEEE Signal Processing Society in 2005 and as Editor-in-Chief of the IEEE TRANSACTIONS ON SIGNAL PROCESSING (2003?2005). His articles received several Best Paper Awards from the IEEE Signal Processing Society (2002, 2005, 2012, 2014). He is a Fellow of the American Association for the Advancement of Science (AAAS). He is recognized as a Highly Cited Researcher by Thomson Reuters. He is serving as President-Elect of the IEEE Signal Processing Society.

\end{IEEEbiography}

\end{document}